\documentclass[
 aps,
 prl,
 twocolumn,
 floats,
 epsf,
 amsmath,amssymb,
 superscriptaddress,
]{revtex4-2}

\usepackage{graphicx}
\usepackage{dcolumn}
\usepackage{bm}
\usepackage{hyperref}
\usepackage[capitalise]{cleveref}
\usepackage{bbold}
\usepackage{MnSymbol}
\usepackage{physics}
\usepackage{extarrows}
\usepackage[normalem]{ulem}
\usepackage{cancel}
\usepackage{placeins}
\usepackage{booktabs}
\usepackage{soul}
\usepackage{multirow}
\usepackage{orcidlink}

\usepackage{tikz}
\usetikzlibrary{tikzmark, arrows.meta}
\usepackage{tikz-cd}
\usetikzlibrary{decorations.pathmorphing,decorations.markings,shapes.geometric}
\usepackage{colortbl}
\usepackage{xcolor}
\usepackage{microtype}
\crefrangeformat{equation}{Eqs.~(#3#1#4) to (#5#2#6)}
\crefformat{equation}{Eq.~(#2#1#3)}
\crefmultiformat{equation}{Eqs.~(#2#1#3)}
{,~(#2#1#3)}{,(#2#1#3)}{,~(#2#1#3)}
\crefformat{table}{Tab.~#2#1#3}
\crefmultiformat{table}{Tabs.~#2#1#3}
{,~#2#1#3}{,#2#1#3}{,~#2#1#3}
\crefrangeformat{table}{Tabs.~#3#1#4 to #5#2#6}
\crefformat{figure}{Fig.~#2#1#3}
\crefmultiformat{figure}{Figs.~#2#1#3}
{,~#2#1#3}{,#2#1#3}{~#2#1#3}
\crefrangeformat{figure}{Figs.~#3#1#4 to #5#2#6}

\newcommand{\numcircled}[1]{{\textcircled{\raisebox{-0.6pt}{#1}}}}

\newcommand{\vc}{\ensuremath{\mathbf{\hat{c}}}}
\newcommand{\vR}{\ensuremath{\mathbf{R}}}
\newcommand{\e}{\ensuremath{\mathrm{e}}}

\newcommand{\vq}{\ensuremath{\mathbf{q}}}
\newcommand{\vtq}{\ensuremath{\mathbf{\tilde{q}}}}
\newcommand{\vtk}{\ensuremath{\mathbf{\tilde{k}}}}
\newcommand{\vk}{\ensuremath{\mathbf{k}}}
\newcommand{\vnull}{\ensuremath{\mathbf{0}}}
\newcommand{\vpi}{\ensuremath{\boldsymbol{\pi}}}
\renewcommand{\i}{{\mathrm{i}}}
\renewcommand{\j}{{\mathrm{j}}}
\newcommand{\up}{\uparrow}
\newcommand{\dn}{\downarrow}
\newcommand{\down}{\downarrow}
\newcommand{\h}{{\mathrm{h}}}
\renewcommand{\l}{{\mathrm{l}}}
\newcommand{\ch}{{\mathrm{ch}}}
\renewcommand{\sp}{{\mathrm{sp}}}
\newcommand{\pp}{{\mathrm{pp}}}
\newcommand{\ph}{{\mathrm{ph}}}
\definecolor{TUgreen}{RGB}{130,158,155}

\usepackage{comment}

\newcommand{\TUVienna}
{\affiliation{Institute of Solid State Physics, TU Wien, 1040 Vienna, Austria}}
\newcommand{\MPI}
{\affiliation{Max-Planck-Institut für Festkörperforschung, Heisenbergstraße 1, 70569 Stuttgart, Germany}}
\newcommand{\Trieste}
{\affiliation{Dipartimento di Fisica, Università di Trieste, Strada Costiera 11, I-34151 Trieste, Italy}}

\begin{document}

\title{
On Degeneracies of Density, Magnetic, and Pairing Responses:\\How Competing Orders Echo Underlying Symmetries in the Hubbard Model
}

\author{Michael Meixner}
\thanks{Both authors contributed equally.}
\MPI
\author{Herbert E{\ss}l\orcidlink{0009-0005-9883-8104}}
\thanks{Both authors contributed equally.}
\TUVienna
\author{Matthias Reitner\orcidlink{0000-0002-2529-0847}}
\TUVienna
\author{Alessandro Toschi\orcidlink{0000-0001-5669-3377}}
\TUVienna
\author{Thomas Schäfer\orcidlink{0000-0002-7550-4807}}
\Trieste
\MPI

\date{\today}

\begin{abstract} 
Strongly correlated electron systems often display competing or even intertwined ordering tendencies, hinting to extremely close or degenerate many-electron energies. While degeneracies are directly rooted in the underlying symmetries of the problem under investigation, their multifaceted effects on different response functions and their mutual relations often remain elusive. Here we put this subject on a rigorous basis by investigating the degeneracies of charge, spin, and pairing susceptibilities for the unfrustrated, bipartite Hubbard model. Exploiting its pseudospin symmetry, we analytically derive the mutual relations between these response functions for generic spatial modulations, highly relevant, e.g., for the competition of stripe and superconducting orders. By means of two-particle numerical simulations we demonstrate the occurrence of a simultaneous $d$-wave pairing/$d$-density wave (loop current) instability in the vicinity of the metal-insulator transition, driven by short-ranged spin fluctuations for the two-dimensional case. We show how this degeneracy is gradually lifted by geometrical frustration, which favors superconductivity. Our study provides a general tool for revealing symmetry relations in correlated electron systems and establishes a unifying perspective on the nature of their intermingled charge/loop current, pairing, and spin orders.
\end{abstract}

\maketitle
\textsl{Introduction.} Quantum materials, in which the effects of the mutual interaction of electrons cannot be neglected, arguably exhibit the most fascinating phenomena of contemporary condensed matter physics. Famous examples are transition metal oxides (such as the celebrated cuprates and nickelates) \cite{Bednorz1986,Keimer2015,Keng2023,Puphal2026}, organic charge-transfer salts \cite{Kanoda2011,Powell2006,Riedl2022}, as well as artificial quantum materials such as moir{\'e} transition metal dichalcogenides \cite{Li2021,Ghiotto2021}, and cold atoms \cite{Esslinger2010,Chalopin2026,Kendrick2025}. What makes these systems so appealing is the richness of their phase diagrams, hosting, for instance, unconventional superconductivity \cite{Stewart2017}, (Mott or charge-transfer) metal-to-insulator transitions \cite{Imada1998}, as well as pseudogaps \cite{Norman2005,Damascelli2003}, quantum criticality \cite{Sachdev1999,Loehneysen2007,Brando2016}, and quantum magnetism \cite{Savary2017,Sachdev1999}. These multifaceted phases hint towards many competing energy scales at play, which make these materials highly sensitive to small parameter changes (such as doping and pressure), and, therefore, potentially very useful for technological applications.

From a theoretical point of view, the rivalry (or potential coexistence) of phases is deeply rooted in the (exact or approximate) degeneracies of the systems' ground states, not only making the energetically most favorable phase hard to resolve, but also possibly leading to competing or intertwined orders \cite{Kaminski2002,Fradkin2015}, and thereby posing a formidable challenge for contemporary quantum many-body theory \cite{Viteritti2026}. This is particularly true in the most important model for electronic correlations, the Hubbard model \cite{Hubbard1963,Qin2022,Arovas2022}, believed to qualitatively describe large parts of the phase diagrams of strongly correlated systems. For instance, concerning the model's putative superconducting regime in two dimensions, careful numerical studies have revealed a subtle competition of stripe ordering (a concomitant modulation of charge and spin degrees of freedom) with pairing tendencies, both for the ground state and for non-zero temperature (e.g.,~\cite{Huang2018,Qin2020,Xu2024,Qin2022,Chakravarty2011,Jiang2019,Wang2015,Lao2025,Wietek2020}). More specifically, given the dependence on minuscule parameter changes \cite{Scholle2026,Reitner2024B,Adler2024B,Rampon2025}, the outcome of simulations can strongly depend on the approximation employed \cite{Leblanc2015,Schaefer2021}.

A fruitful path for understanding the emergence of competing correlated phases is the search for the fundamental \textit{roots} of energetic degeneracies, rather than the analysis of their \textit{implications}. Clearly, degeneracies must emerge from symmetries, whose concrete manifestations can be, however, elusive, depending on the observable under scrutiny. The arguably most direct probes for ordering tendencies in interacting systems are susceptibilities, i.e., response functions to external perturbations, such as charge, spin, and pairing susceptibilities, with different spatial modulations. In contrast to \textit{single-particle} Green functions, generalized susceptibilities represent \textit{two-particle} Green functions with multiple space and time variables. In these more involved objects, the multifaceted echoes of symmetries can be difficult to capture at first sight.

\begin{figure*}[t!]
    \centering
    \includegraphics[width=\linewidth]{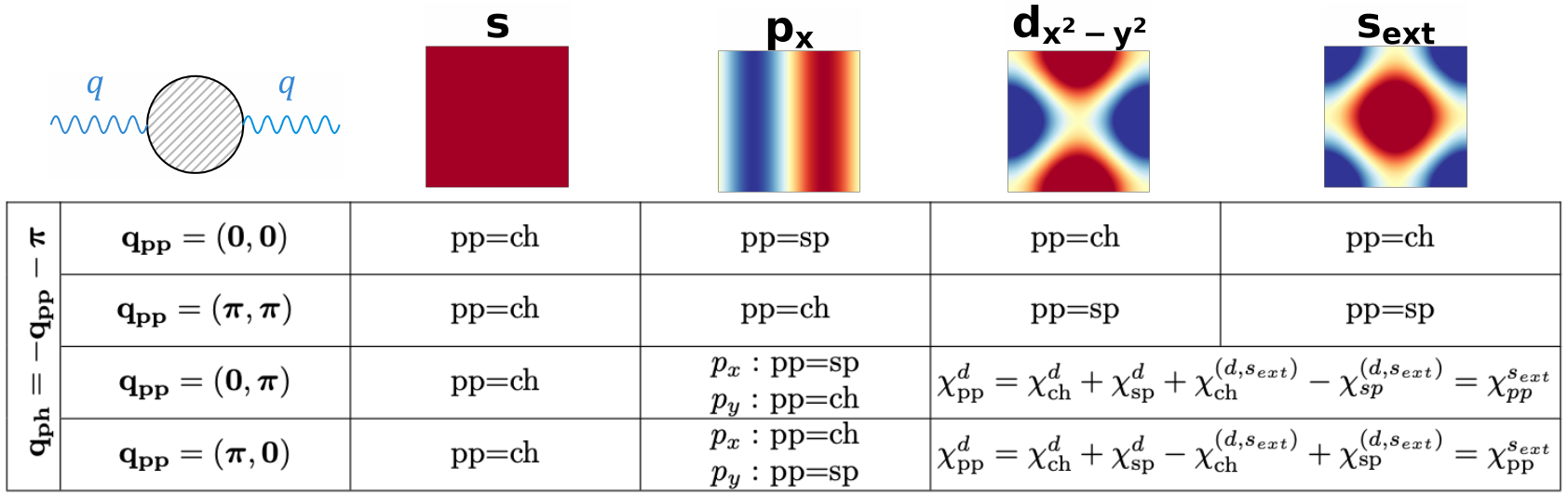}
    \caption{Exact symmetry relations between pairing, charge and spin response functions with $s,p,d,s_{ext}$ form-factors (columns) and exemplary transfer momenta $\vq_\pp=-\vq_\mathrm{ph}-\vpi$ (rows) for square lattice models with 
    $SU(2)_P$-symmetry. The relations hold not only for the static case ($\omega=0$) but for all transfer-frequencies, with $\omega_{\pp}= -\omega_{\ph}$ and non-zero external magnetic field \cite{sup}.}
    \label{fig:table}
\end{figure*}

Here, we address this problem on a rigorous and general basis. We first exploit the $SU(2)_P$ pseudospin symmetry of the Hubbard model on a simple square lattice to analytically demonstrate the equivalence of response functions in the charge, spin, and particle-particle (pairing) sectors, and, hence, the close competition of their respective ordering tendencies in adjacent regions of the phase diagrams. In particular, as opposed to previous studies \cite{Zhang1990,Keller2001,Chakravarty2001}, we do not restrict ourselves to selected form factors in space (like uniform ones), but provide a framework applicable to general momentum transfers. Since spacial dependencies occur in many strongly correlated materials, these momentum modulations are crucial for their description. 

Secondly, we investigate the implications of our analytical results, by numerically computing the susceptibilities for a charge-density wave with $d_{x^2-y^2}$-wave form factor (`$d$-density wave', $d$-DW) and $d$-wave pairing ($d$-SC), in close proximity to a Mott metal-insulator transition with a $2\times 2$ cellular dynamical mean-field theory (CDMFT) \cite{Kotliar2001} at the two-particle level \cite{Meixner2026a}. We show that the significant enhancement in \textit{both} responses is crucially driven by repeated non-local spin fluctuations on nearest-neighbor sites. Eventually, we break the particle-hole (and pseudospin) symmetry by introducing a non-zero nearest-neighbor hopping amplitude $t'$. Thereby we numerically demonstrate the \textit{gradual} lifting of the degeneracies of $d$-DW and $d$-SC, systematically favoring the superconducting instability.

\textsl{Degeneracies of the Hubbard Hamiltonian.}\,
To describe the interplay of competing orders, we employ the fundamental model for correlated electrons, the [particle-hole (ph)-symmetric] Hubbard model \cite{Hubbard1963,Qin2022,Arovas2022}. Its Hamiltonian on a n-dimensional cubic lattice reads
\begin{equation} \label{eq:Hamiltonian}
\begin{split}
\hat{H}_U=-&t\sum_{\langle \mathrm{i,j} \rangle,\sigma}\hat{c}^\dagger_{\mathrm{\i},\sigma}\hat{c}_{\mathrm{j},\sigma}+U\sum_{\mathrm{i}}\left(\hat{n}_{\i,\uparrow}-\frac{1}{2}\right)\left(\hat{n}_{i,\downarrow}-\frac{1}{2}\right),
\end{split}
\end{equation} 
which includes nearest-neighbor hopping $t$ for electrons with spin flavor $\sigma$, on-site repulsion $U$ and $\hat{n}_{\i},\hat{c}^{\dagger}_\i,\hat{c}_\i$ are density, creation and annihilation operators on lattice site $\i$.  Additionally, next-nearest-neighbor hopping $t'$, a magnetic field $B_z$, or doping via a chemical potential $\delta\mu$ may be defined, which break particle-hole symmetry.
Setting the lattice spacing to $1$, we define the partial particle-hole mapping by Shiba \cite{Shib1972}:
\begin{equation}
\label{Eq:Shiba}
\hat{c}^{\dagger}_{\i\downarrow}\rightarrow\hat{c}^{\dagger}_{P,\i\downarrow}=\e^{\i\vpi\vR_i}\hat{c}_{\i\downarrow}; \qquad \hat{c}_{\i\downarrow}\rightarrow\hat{c}_{P,\i\downarrow}=\e^{-\i\vpi\vR_i}\hat{c}_{\i\downarrow}^\dagger,
\end{equation} where $\vpi=(\pi,\pi,\dots)$ and $\vR_\i$ the location of the $\i^\text{th}$ lattice site.
The Shiba mapping flips the sign of the interaction $\hat{H}(U)\xleftrightarrow{U\rightarrow -U}\hat{H}_{P}=\hat{H}(-U)$ and $\delta\mu$ to $B_z$ in the Hubbard model \cite{Singh1991,Essl2024}.
For the spin operators $\hat{S}_{\alpha,\i}=\vc_\i^\dagger \sigma_\alpha\vc_\i$\cite{Shen1996}, defined via spinors $\vc_\i\xleftrightarrow{U\rightarrow -U} \vc_{P,\i}$ and the Pauli matrices $\sigma_\alpha$, the Shiba mapping results in \textit{pseudospin} operators $\hat{S}_{P,\alpha,\i}=\vc_{P,\i}^\dagger \sigma_\alpha\vc_{P,\i}$, which are generators of an $SU(2)_P$ symmetry \cite{Yang1990,Pu1994,Scaletter2024,Carmelo2026}.
 For two exemplary operators $A,B$, which can be written in terms of fermionic creation- and annihilation operators, we exploit the mapping as follows:\begin{equation}
\label{Eq:mapping}
\begin{split}
    \tikzmark{mapstart} \hat{A} 
    \underset{\mathrm{Eq.(\ref{Eq:Shiba})}}{\xleftrightarrow{U\rightarrow -U}}
     \hat{A}_{P} 
    \xleftrightarrow{\mathrm{SU(2)}}
     \hat{B}_{P} 
    \underset{\mathrm{Eq.(\ref{Eq:Shiba})}}{\xleftrightarrow{-U\rightarrow U}}
     \hat{B} 
    \tikzmark{mapend},\\
    \begin{tikzpicture}[overlay, remember picture]
    \draw[{Glyph[glyph math command=rightarrow]}-{Glyph[glyph math command=rightarrow]}, bend right]
        ([xshift=10pt, yshift=-6pt]{pic cs:mapstart})
        to[bend right=25]
        node[midway, above, font=\scriptsize]{\,SU(2)$_p$}
        ([xshift=-10pt, yshift=-6pt]{pic cs:mapend});
\end{tikzpicture}\\[4pt]
\end{split}
\end{equation} establishing $ \langle A\rangle = \langle B\rangle $. 
Relating the two operators $A,B$ by exploiting the $SU(2)_P$ pseudospin symmetry of $H(U)$ is equivalent to exploiting the $SU(2)$ symmetry of $H(-U)$.
This holds if the initial Hubbard model is defined on a bipartite lattice with $t'=0$ and $\delta\mu=0$. For simplicity, we set $B_z=0$, which results in a particle-hole symmetric model $SO(4)=[SU(2)\cross SU(2)]/  \mathrm{Z}_2$ \cite{Yang1990,Carmelo2026}.

To obtain a relation between particle-hole and particle-particle (pp) response functions, we define the generalized susceptibilities 
\begin{equation}
\begin{split}
\chi^{12|34}_{\ph,\sigma\sigma^\prime}=&\langle c^\dagger_{1\sigma}c_{2\sigma} c^\dagger_{3\sigma^\prime} c_{4\sigma^\prime}\rangle-G_{41}G_{23},\\
\chi^{13|42}_{\pp,\up\dn}=&\langle c^\dagger_{1\up} c_{2\up} c^\dagger_{3\dn} c_{4\dn}\rangle\\
\end{split}
\end{equation} with the combined space-time indices $1=(\vR_1,\tau_1)$ and the charge, spin and particle-particle susceptibilities
\begin{equation}
\begin{split}
    \chi^{12|34}_{\ch}=&\chi^{12|34}_{\ph,\up\up}+\chi^{12|34}_{\ph,\up\dn},\\
    \chi^{12|34}_{\sp}=&\chi^{12|34}_{\ph,\up\up}-\chi^{12|34}_{\ph,\up\dn},\\
    \chi^{13|42}_{\pp}=&\chi^{13|42}_{\pp,\up\dn},
\end{split}
\end{equation}
see Sec.~II in \cite{sup} for detailed frequency definitions. The physical responses can be obtained by pairwise contractions of incoming and outgoing indices, e.g.~$\chi^{11|33}$. 
Applying the mapping Eq.~(\ref{Eq:mapping}) to $\chi^{12|34}_{\ph,\up\up}$ results in the following real-space identity:
\begin{equation}
\begin{split}
\label{Eq:real-space}
   2 \e^{\i\vpi(\vR_2-\vR_3)}\chi_\text{pp}^{12|34}=&\chi_\text{ch}^{12|34}+\chi_\text{ch}^{12|43}\e^{\i\vpi(\vR_3-\vR_4)}\\
    &+\chi_\text{sp}^{12|34}-\chi_\text{sp}^{12|43}\e^{\i\vpi(\vR_3-\vR_4)},
\end{split}
\end{equation} 
the transformation to Matsubara frequencies is given in Sec.~III of \cite{sup}. This relates the particle-particle response function to the particle-hole sector, including their real-space dependence. To clarify the physical content of this relation, we present a momentum-space equivalent 
\begin{equation}
\label{Eq:momentumspace}
\begin{split}
    \chi_\text{pp}^{k,k^\prime,-q-\Pi}=\frac{1}{2}\left(\chi_\text{ch}^{k,k^\prime, q}+\chi_\text{ch}^{k,\overline{k^\prime}, q}+\chi_\text{sp}^{k,k^\prime, q}-\chi_\text{sp}^{k,\overline{k^\prime}, q}\right),
\end{split}
\end{equation}
\begin{figure*}
    \centering
    \begin{tikzpicture}
    \node[anchor=south west, inner sep=0] at (0,0)
    {\includegraphics[width=0.54285\linewidth]{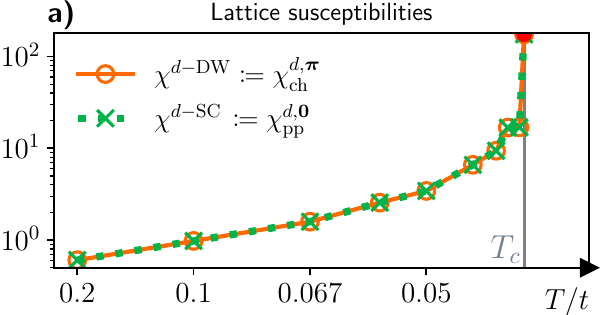}};
    \node[anchor=south west, inner sep=0] at (10.4,0.0) {\includegraphics[width=0.4071428\linewidth]{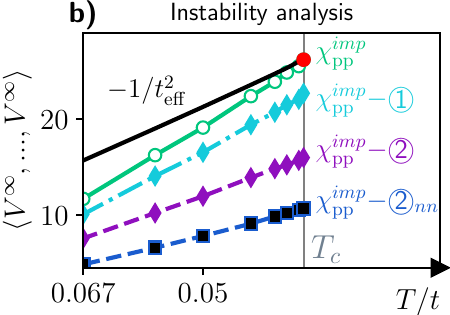}};
    \node[anchor=south west, inner sep=0] at (5.5,3) {\includegraphics[width=0.07\linewidth]{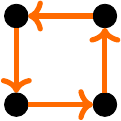}};
    \end{tikzpicture}
    \begin{tikzpicture}
    \node[anchor=south west, inner sep=0] at (0,-0.1)
    {\includegraphics[width=0.75\linewidth]{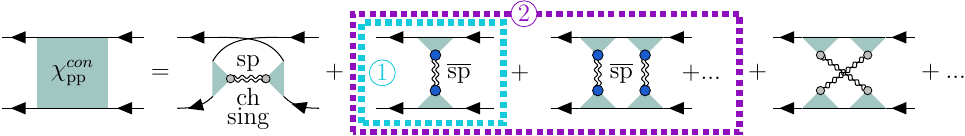}};
    \node at (-0.5,1.5) {\fontsize{14pt}{15pt}\textsf{\bfseries c)}};
    \end{tikzpicture}
    \caption{a) $d$-density wave $\chi^{d-\mathrm{DW}}$ and $d$-wave superconductivity $\chi^{d-\mathrm{SC}}$ susceptibilities over temperature in CDMFT, which show a concomitant divergence at $T_{c}\approx 0.0413t$ for $U=5.6t$. b) Projection on the instability eigenvector $V_\infty$ of the two terms of Eq.~(\ref{Eq:BSE_pp}): the $d-$SC generalized susceptibility of the impurity cluster (green circles) and corresponding inverse-bubble contributions  $-1/t_\text{eff}^2$ (black bold line). The thermodynamic instability occurs at their intersection (red dot). c) Diagrams which are subtracted from the impurity response in b), see text: \numcircled{1} the single boson exchange spin-transverse diagram $\overline{\mathrm{sp}}$ of the generalized susceptibility, \numcircled{2} the single-boson $\overline{\mathrm{sp}}$ of the two-particle irreducible vertex and \numcircled{2}$_{nn}$, the same diagrams subtracted, with the $\overline{\mathrm{sp}}$ boson exchange  restricted to the nearest-neighbor distance.}
\label{fig:CDMFT_ph_symmetry}\end{figure*}
where a combined momentum- and frequency notation $k=(\nu,\vk),k^\prime=(\nu^\prime,\vk^\prime)$ and $q=(\omega,\vq)$ is employed and
$\overline{k}=-k-\Pi-q=(-\nu-\omega,-\vk-\vpi-\vq)$ with $\Pi=(0,\vpi)$. 
Similar equations may be found for non-zero magnetic field, see Sec.~V of \cite{sup}\footnote{This momentum space equation is related to the real-space Eq.~(\ref{Eq:real-space}) by Fourier transforms, where pp and ph sectors are subject to different Fourier conventions \cite{Pu1994,Rohringer2013a}(see Sec.~II of \cite{sup}) to implement the adequate operator ordering.}. The exact symmetry relations (\ref{Eq:real-space})-(\ref{Eq:momentumspace}), which link the generalized susceptibilities in different scattering channels, hold for general bipartite lattices if the vector $\vpi$ is changed accordingly, and get reflected in a series of important degeneracies of the corresponding physical susceptibilities (see Sec.~V of \cite{sup}). Note, that the static limit of the latter controls the onset of (possibly degenerate) thermodynamic instabilities. 

In the following, we will proceed by projecting Eq.~(\ref{Eq:momentumspace}) onto square lattice form factors $f_r(\vk)$, and perform all internal summations over fermionic momenta and frequencies ($k,k'$). The analytical findings of this procedure are reported in Fig.~\ref{fig:table} for the most relevant form factors [isotropic $s$-wave ($r\!=\!s$), $p$-wave ($r\!=\!p_x,p_y$), $d_{x^2-y^2}$-wave ($r\!=\!d$), extended $s$-wave ($r\!=\!s_{ext}$)\footnote{corresponding to the truncated unity expansion \cite{Lichtenstein2017}}], and for representative external/transfer momenta of the particle-hole response $\vq_\pp$ (and its symmetry-related 
$\ph$ one $\vq_\pp = -\vq_\mathrm{ph}- \vpi $). 
In particular, when $\vq_\pp=\mathbf{0}$ and for the isotropic $s$-wave sector, we immediately recover the famous ``super-solid'' three-fold degeneracy between the $s$-wave pairing and the staggered density (often dubbed charge density wave, CDW) responses \cite{Zhang1990,Chakravarty2001}. 
Remarkably, this equivalence can indeed be generalized for a multitude of form factors. For instance, for $\vq_\pp= \vnull$ an \emph{analogous} equivalence holds in the case of \emph{even} parity form factors, yielding a degeneracy of the $d$-wave (or $s_{ext}$) pairing with the staggered
$d-$DW \cite{Affleck1988,Nayak2000b,Chakravarty2002,Chakravarty2011,Leeb2026} (or $s_{ext}-$DW) response. Needless to say, the degeneracies in the $d-$wave sector are of particular importance in the case of repulsive Hubbard interaction considered here \footnote{We note that an alike relation of $d$-DW and pairing instability has been discussed in the different context of the $t-J$ model \cite{Kotliar1988,Cappelluti1999} and, recently, the $\sigma_z$-Hubbard model \cite{Zhu2025}}.

While the generalization of the isotropic $s$-wave result to the $d$/$s_{ext}$-wave sectors may appear somewhat intuitive, and, thus, easily generalizeable, it represents a rather special case. In fact, the degeneracy relations between different response functions do depend in a highly non-trivial way on both the momentum transfer and the form factor considered. As a result, for $\vq_\pp= \vnull$, within the \emph{odd} $p$-wave sector, quite different degeneracy relations are found: the $p$-wave pairing response is coincident with the $p$-spin DW ($p-$SDW) one. 
Further, except for the isotropic $s$-wave case, the degeneracy relations explicitly depend on the transfer vector considered. For instance, if $\vq_\pp =\vpi$, one finds a degeneracy between the staggered pair-DW and the uniform SDW responses in both (even) $d-$ and $s_{ext}$-wave sectors, while in the (odd) $p$-wave sector the staggered $p$-pair-DW susceptibility becomes degenerate with the uniform $p$-DW one.

For arbitrary choices of momenta, the symmetry relations of the physical responses become more involved. In Fig.~\ref{fig:table} we explicitly report the interesting case of a stripe-like momentum transfer, i.e.,~$\vq=(0,\pi)/(\pi,0)$: Within the $p$-wave sector, a plain juxtaposition of the previous results (depending on the stripe orientation) occurs, while more complex combinations of pairing, spin, and charge DW responses hold for $d$- and $s_{ext}$-wave responses. 
Finally, we emphasize that Eq.~(\ref{Eq:momentumspace}) entails degeneracies for any $\vq$-vector and corresponding form factor combination.

It is important to note that, apart from the lattice Hubbard Hamiltonian, the relations presented also hold for isolated clusters, and (cluster) Anderson impurity models \cite{Anderson1961} with a pseudospin-symmetric bath and bipartite boundary conditions. Furthermore, by using the dual-BSE \cite{vanLoon2024-2}, we explicitly demonstrate that the relations remain fully valid also for the results of quantum embedding methods for the Hubbard model such as the dynamical cluster approximation \cite{Jarrell2001} and the (cellular) dynamical mean field theory (C)DMFT \cite{Georges1992,Kotliar2000}, see Sec.~VI of \cite{sup}.

\textsl{Results for real space embedding theories.}\, For an illustration of these degeneracies and their physical significance,  we computed the temperature dependence of the $d-$wave lattice susceptibilities by means of a Bethe-Salpeter equation (BSE) treatment in four-site CDMFT \cite{Meixner2026a} for $U=5.6t$ and pseudo-spin symmetry (see the End Matter for computational details, \footnote{The $2\times2$ CDMFT cell is sufficient to capture the $d-$wave instability, consequently, the BSE-superlattice vector is set to 0.}) An instability in the particle-particle channel with $d-$wave symmetry occurs in close proximity to the CDMFT Mott metal-insulator transition (MIT) \cite{Park2008}, indicated in Fig.~\ref{fig:CDMFT_ph_symmetry}a) by a divergence of the corresponding susceptibility (green dashed line) at the critical temperature $T_c$. This matches results reported by \cite{Fratino2016} obtained from applied (uniform) pairing field calculations, see Sec.~VI of \cite{sup}. Fulfilling the symmetry relations of Fig.~\ref{fig:table}, the values of the $d-$DW response function for $\vq_\ph=\vpi$ (orange bold line) are \textit{numerically identical} to those of the $d-$wave pairing one.
We note here that the related $d$-DW fluctuations, which essentially correspond to vertex/loop currents [cf.~the inset of Fig.~\ref{fig:CDMFT_ph_symmetry}a)], 
 have been discussed in previous literature as a possible origin of the pseudogap \cite{Varma1997,Varma1999,Chakravarty2001,Goswami2013}. Yet, no CDMFT study of their behavior, to the best of our knowledge, has been been reported so far.

The degeneracy of $d$-SC and $d$-DW responses has evident consequences. 
The three-component order parameter \cite{Nayak2000} associated to the $d$-SC and $d$-DW fluctuations should then fall under the Mermin-Wagner-Hohenberg theorem conditions \cite{Mermin1966,Hohenberg1967}, featuring an \emph{exponential} $T-$dependence \cite{Dare1996,Vilk1997,Schaefer2015b,Schaefer2021} of the corresponding susceptibilities. Such an exponential behavior is indeed observed in the intermediate-to-high-$T$ regime. At lower $T\lesssim0.067t$, when the correlation lengths of both (degenerate) $d-$SC and $d-$DW fluctuations exceed the size of the CDMFT cluster, the mean-field nature of CDMFT allows for a divergence of the susceptibilities at a critical temperature $T_c$, which is \emph{identical} for both $d-$SC and $d-$DW sectors.

We now investigate the microscopic origin of the observed $T$-behavior of the degenerate $d-$SC/$d$-DW responses, and of their low-$T$ divergence in CDMFT. To this aim, we recall that a generalized (static) susceptibility (e.g.~$\chi_\pp$) can be viewed as a matrix in the combined fermionic frequencies and cluster site indices \cite{Musshoff2021}. 
When working in the corresponding eigenbasis, a divergence of the associated physical response is driven by a divergence of \textit{one} of its eigenvalues ($\lambda_{\infty}$). The frequency/momentum structure of the corresponding eigenvector $V^{\infty}$ entails, then, crucial information about the underlying symmetries of the phase transitions.

Beyond these general considerations, we note that in the CDMFT BSE~(\ref{Eq:BSE_pp}), the generalized susceptibility can be separated into two terms: $-t_\text{eff}^2$ (stemming from the dressed bubble) and the cluster-impurity susceptibility $\left[\chi_\pp^{imp}\right]^{-1}$ (containing vertex corrections). 
As we observe that $V^{\infty}$ is approximately an eigenvector of both the lattice and the impurity susceptibility over a wide range of temperatures, see Sec.~VIII of \cite{sup}, the BSE acquires a diagonal block structure, with $V^{\infty}$ spanning an independent subspace. In this fortunate situation, both terms in the BSE can be projected onto $V^\infty$ to obtain their respective contributions to the instability in a scalar form, see Fig.~\ref{fig:CDMFT_ph_symmetry}b). 
In this situation, the term $1/t_\text{eff}^2$ (black line) acts as a critical threshold for the instability. In fact, when it gets exactly canceled by the corresponding projection of the inverse generalized susceptibility of the impurity (green $o$-marker) at $T_c$, it triggers the divergence of the physical lattice susceptibility (red dot) and, thus, the associated thermodynamic phase transition.

Within this framework, it is particularly insightful to identify the physical nature of the fluctuations driving the enhanced thermodynamic response and, ultimately, its instability. To this aim, we exploit the single-boson exchange (SBE) decomposition of $\chi_\pp^{imp}$ \cite{Krien2019c,Bonetti2022,Adler2024,Gievers2025,Meixner2026a} in Fig.~\ref{fig:CDMFT_ph_symmetry}c), i.e.,~we evaluate how different scattering processes defined by the exchange of single or multiple boson fluctuations contribute to the value of the lattice $d-$SC susceptibility $\chi_\pp$. In particular, the first term of this decomposition shown in Fig.~\ref{fig:CDMFT_ph_symmetry}c) entails the exchange of a single longitudinal bosonic mode of either spin, charge or singlet (pairing) type. The second term describes, instead, the exchange of a single spin-boson mode in the transversal direction $\overline{\sp}$. Eventually, all remaining processes involving multiple boson exchanges are encoded in the other diagrams on the r.h.s.~of Fig.~\ref{fig:CDMFT_ph_symmetry}c). When subtracting the transversal diagram \numcircled{1}, the impurity response's contribution to the instability [Fig.~\ref{fig:CDMFT_ph_symmetry}b), turquoise diamond markers] significantly drops. From the slope of $\chi_\pp^{imp}$- \numcircled{1}, one infers that the crossing with the $-t_\text{eff}^2$ line would then occur at much lower temperatures, yielding a significant suppression of $T_c$.
Beyond that, a full disappearance of the instability can be obtained by subtracting \emph{all} two-particle irreducible scattering contributions (subset \numcircled{2}) featuring transversal single spin-boson exchanges \footnote{This corresponds to applying the SBE decomposition to the two-particle irreducible vertex $\Gamma_\pp$ \cite{Krien2020b,Meixner2026a}.}. Indeed, subtracting \numcircled{2} from $\chi_\pp^{imp}$ (purple line) results in such a marked slope-reduction of the impurity contribution, that prohibits the occurrence of the instability \textit{at any} finite temperature. Remarkably, the subtraction of a \emph{much smaller} diagrammatic subset \numcircled{2}$_{nn}$, removing only diagrams containing nearest-neighbor ($nn$) transversal spin-boson modes, features an \emph{even stronger} suppression of the instability. This identifies, on a microscopic level, repeated transversal spin-boson exchanges on $nn$ sites as the main scattering process driving the observed degenerate $d-$SC/$d-$DW instability. We note that the very same scattering processes, at larger $U$ values, are the ones mostly responsible for the occurrence of the Mott MIT \cite{Park2008,Meixner2025,Meixner2026a}.

\begin{figure}
    \centering
    \includegraphics[width=0.95\linewidth]{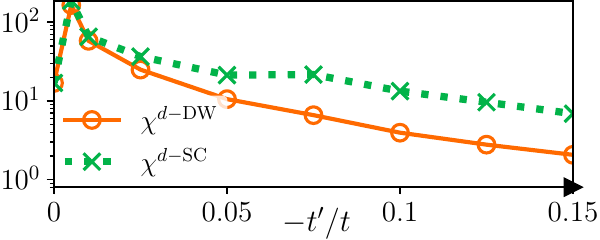}
    \caption{$d$-density wave $\chi^{d-\mathrm{DW}}$ and $d$-wave superconductivity $\chi^{d-\mathrm{SC}}$ susceptibilities over second-next neighbor hopping $t'$ at $U=5.6t$ and $T=0.0417t$, slightly above  $T_c$.}
    \label{fig:CDMFT_tp}
\end{figure}

\textsl{Breaking of pseudospin symmetry.}\, To study an experimentally relevant situation, an additional next-nearest-neighbor hopping $t'$ is introduced. This breaking of the pseudospin symmetry formally lifts the degeneracy between $d$-DW and the $d$-SC, thereby splitting the order parameter into a one-component charge and a two-component SC one. The corresponding response functions at $T/t=1/24$ and as a function of $t'/t$, remarkably follow the same qualitative behavior in Fig.~\ref{fig:CDMFT_tp}. For small values of $t'\neq 0$, both response functions increase essentially by the same amount. This trend can be explained by the degree of geometric frustration of the AFM fluctuations controlled by $t'$, which counteracts the $d$-DW and $d$-SC instability in the particle-hole symmetric case \cite{Kyung2003,Kancharla2008}. By further increasing $t'$, both susceptibilities, while remaining quite large, are gradually reduced, with superconducting fluctuations prevailing over the $d$-DW ones. The observed similarity of both response functions out of $\mathrm{SU(2)}_P$ symmetry suggests a strong intertwining between d-wave superconductivity and vertex currents, previously indicated by vertex model \cite{Varma1997} and Hartree-Fock considerations \cite{Kotliar1988,Chakravarty2001}.

The gradual lifting of the degeneracy observed here appears at odds w.r.t.~the famous example of the (C)DMFT Mott transition \cite{Georges1992a,Kotliar2001,Parcollet2004,Park2008}. There, an observation of a divergent charge susceptibility at perfect particle-hole symmetry would fall into the $s$-wave category of a supersolid: $\chi_\ch^{s,\vpi}=\chi_\pp^{s,\vnull}$. However, in contrast to the previously discussed $d$-wave instability, the Mott instability at $t'=0$ displays a divergent generalized charge susceptibility, whose matrix elements in $(\nu,\nu')$ are \textit{perfectly antisymmetric} \cite{Reitner2020,Meixner2026a}. Hence, these divergent contributions cancel when summing for the \textit{physical} response function $\chi_\ch^{s,\vnull}$. The argument of cancellation, however, does not hold at the critical endpoint for non-zero $t'$, where the physical charge response $\chi_\ch^{s,\vnull}$ diverges instantly \cite{Sordi2012,Sordi2012b}.

The abrupt differentiation between a divergent charge and a non-divergent pairing physical response upon lifting $SU(2)_P$-symmetry is very transparent at the respective two-particle level [\cref{Eq:momentumspace}]:
The antisymmetric matrix elements driving the Mott transition are \emph{not} mapped to the pairing channel. This contrasts the $d-$wave case, whose frequency-symmetric  
instability \emph{does} get mapped onto the paring channel, yielding a similar response in two sectors, when the $SU(2)_P$-symmetry is lifted. This example nicely illustrates how studying the two-particle level allows for assessing the robustness of  degeneracies in physical response functions, even upon particle-hole symmetry breaking (see End Matter \ref{sec:PT}).

\textsl{Conclusions and Outlook.} We established a general framework for the derivation of degeneracy relations between pairing and particle-hole response functions at the two-particle level. In the particle-hole symmetric Hubbard model these degeneracies arise from its pseudospin symmetry and hold, in different forms, for general form factors and transfer momenta. As an important example, we numerically investigate the symmetry-related $d$-SC and $d$-DW channels. These exhibit a concomitant instability in the vicinity of a Mott MIT, driven by short-ranged spin fluctuations. Breaking the symmetry by introducing a non-zero $t'$ only gradually lifts the degeneracy and favors $d$-wave SC. Our results pave the way to a deeper understanding of the highly non-trivial competition of exotic ordering tendencies in the Hubbard model. In particular, our calculations offer in-depth explanations of previous works such as the observed degeneracies in the Schwinger-Dyson equation within the truncated unity expansion \cite{Patricolo2025}, and Yukawa couplings in multiloop functional renormalization group formulations \cite{Fraboulet2025}. More broadly, the generic frequency and momentum dependence on the two-particle level can be linked to the behavior of physical response functions, thus serving as powerful analysis tools for understanding complex phase diagrams. In this light, extending our approach to more material-linked settings may help to clarify the microscopic links between superconductivity, pseudogap physics, and even more unconventional orders, such as the loop currents recently connected to exotic collinear odd-wave magnetism \cite{Leeb2026}.

\begin{acknowledgments}
\textsl{Acknowledgements.} We thank Georg Rohringer, Sergio Ciuchi, Giorgio Sangiovanni, Sergio Caprara, and Massimo Capone for inspiring comments, and Pietro Bonetti for carefully reading the manuscript. We appreciate the collaboration with Nils Wentzell of CCQ at the Flatiron Institute and the computer service facility of the MPI-FKF. We acknowledge funding from the Deutsche Forschungsgemeinschaft (DFG) through research unit FOR5249 ``QUAST'' project No.~449872909 (Projects P1 and P4), partially funded by the FWF as (sub)project with Grant-DOI 10.55776/KIN2563725. HE also acknowledge support from the FWF project with Grant-DOI 10.55776/I5487.
\end{acknowledgments}
\vspace{-0.5cm}

\bibliography{main}

\begin{thebibliography}{137}%
\makeatletter
\providecommand \@ifxundefined [1]{%
 \@ifx{#1\undefined}
}%
\providecommand \@ifnum [1]{%
 \ifnum #1\expandafter \@firstoftwo
 \else \expandafter \@secondoftwo
 \fi
}%
\providecommand \@ifx [1]{%
 \ifx #1\expandafter \@firstoftwo
 \else \expandafter \@secondoftwo
 \fi
}%
\providecommand \natexlab [1]{#1}%
\providecommand \enquote  [1]{``#1''}%
\providecommand \bibnamefont  [1]{#1}%
\providecommand \bibfnamefont [1]{#1}%
\providecommand \citenamefont [1]{#1}%
\providecommand \href@noop [0]{\@secondoftwo}%
\providecommand \href [0]{\begingroup \@sanitize@url \@href}%
\providecommand \@href[1]{\@@startlink{#1}\@@href}%
\providecommand \@@href[1]{\endgroup#1\@@endlink}%
\providecommand \@sanitize@url [0]{\catcode `\\12\catcode `\$12\catcode
  `\&12\catcode `\#12\catcode `\^12\catcode `\_12\catcode `\%12\relax}%
\providecommand \@@startlink[1]{}%
\providecommand \@@endlink[0]{}%
\providecommand \url  [0]{\begingroup\@sanitize@url \@url }%
\providecommand \@url [1]{\endgroup\@href {#1}{\urlprefix }}%
\providecommand \urlprefix  [0]{URL }%
\providecommand \Eprint [0]{\href }%
\providecommand \doibase [0]{https://doi.org/}%
\providecommand \selectlanguage [0]{\@gobble}%
\providecommand \bibinfo  [0]{\@secondoftwo}%
\providecommand \bibfield  [0]{\@secondoftwo}%
\providecommand \translation [1]{[#1]}%
\providecommand \BibitemOpen [0]{}%
\providecommand \bibitemStop [0]{}%
\providecommand \bibitemNoStop [0]{.\EOS\space}%
\providecommand \EOS [0]{\spacefactor3000\relax}%
\providecommand \BibitemShut  [1]{\csname bibitem#1\endcsname}%
\let\auto@bib@innerbib\@empty
%</preamble>
\bibitem [{\citenamefont {Bednorz}\ and\ \citenamefont
  {M\"uller}(1986)}]{Bednorz1986}%
  \BibitemOpen
  \bibfield  {author} {\bibinfo {author} {\bibfnamefont {J.~G.}\ \bibnamefont
  {Bednorz}}\ and\ \bibinfo {author} {\bibfnamefont {K.~A.}\ \bibnamefont
  {M\"uller}},\ }\bibfield  {title} {\bibinfo {title} {Possible high t$_{\rm
  c}$ superconductivity in the ba--la--cu--o system},\ }\href@noop {}
  {\bibfield  {journal} {\bibinfo  {journal} {Zeitschrift f\"ur Physik B
  Condensed Matter}\ }\textbf {\bibinfo {volume} {64}},\ \bibinfo {pages} {189}
  (\bibinfo {year} {1986})}\BibitemShut {NoStop}%
\bibitem [{\citenamefont {{Keimer}}\ \emph {et~al.}(2015)\citenamefont
  {{Keimer}}, \citenamefont {{Kivelson}}, \citenamefont {{Norman}},
  \citenamefont {{Uchida}},\ and\ \citenamefont {{Zaanen}}}]{Keimer2015}%
  \BibitemOpen
  \bibfield  {author} {\bibinfo {author} {\bibfnamefont {B.}~\bibnamefont
  {{Keimer}}}, \bibinfo {author} {\bibfnamefont {S.~A.}\ \bibnamefont
  {{Kivelson}}}, \bibinfo {author} {\bibfnamefont {M.~R.}\ \bibnamefont
  {{Norman}}}, \bibinfo {author} {\bibfnamefont {S.}~\bibnamefont {{Uchida}}},\
  and\ \bibinfo {author} {\bibfnamefont {J.}~\bibnamefont {{Zaanen}}},\
  }\bibfield  {title} {\bibinfo {title} {{{From quantum matter to
  high-temperature superconductivity in copper oxides}}},\ }\href
  {https://doi.org/10.1038/nature14165} {\bibfield  {journal} {\bibinfo
  {journal} {\nat}\ }\textbf {\bibinfo {volume} {518}},\ \bibinfo {pages} {179}
  (\bibinfo {year} {2015})}\BibitemShut {NoStop}%
\bibitem [{\citenamefont {Kang}\ \emph {et~al.}(2023)\citenamefont {Kang},
  \citenamefont {Zhang}, \citenamefont {Schierle}, \citenamefont {McCoy},
  \citenamefont {Li}, \citenamefont {Sutarto}, \citenamefont {Suter},
  \citenamefont {Prokscha}, \citenamefont {Salman}, \citenamefont {Weschke},
  \citenamefont {Cybart}, \citenamefont {Wei},\ and\ \citenamefont
  {Comin}}]{Keng2023}%
  \BibitemOpen
  \bibfield  {author} {\bibinfo {author} {\bibfnamefont {M.}~\bibnamefont
  {Kang}}, \bibinfo {author} {\bibfnamefont {C.~C.}\ \bibnamefont {Zhang}},
  \bibinfo {author} {\bibfnamefont {E.}~\bibnamefont {Schierle}}, \bibinfo
  {author} {\bibfnamefont {S.}~\bibnamefont {McCoy}}, \bibinfo {author}
  {\bibfnamefont {J.}~\bibnamefont {Li}}, \bibinfo {author} {\bibfnamefont
  {R.}~\bibnamefont {Sutarto}}, \bibinfo {author} {\bibfnamefont
  {A.}~\bibnamefont {Suter}}, \bibinfo {author} {\bibfnamefont
  {T.}~\bibnamefont {Prokscha}}, \bibinfo {author} {\bibfnamefont
  {Z.}~\bibnamefont {Salman}}, \bibinfo {author} {\bibfnamefont
  {E.}~\bibnamefont {Weschke}}, \bibinfo {author} {\bibfnamefont
  {S.}~\bibnamefont {Cybart}}, \bibinfo {author} {\bibfnamefont {J.~Y.~T.}\
  \bibnamefont {Wei}},\ and\ \bibinfo {author} {\bibfnamefont {R.}~\bibnamefont
  {Comin}},\ }\bibfield  {title} {\bibinfo {title} {{Discovery of charge order
  in a cuprate Mott insulator}},\ }\href
  {https://doi.org/10.1073/pnas.2302099120} {\bibfield  {journal} {\bibinfo
  {journal} {Proceedings of the National Academy of Sciences}\ }\textbf
  {\bibinfo {volume} {120}},\ \bibinfo {pages} {e2302099120} (\bibinfo {year}
  {2023})}\BibitemShut {NoStop}%
\bibitem [{\citenamefont {{Puphal}}\ \emph {et~al.}(2026)\citenamefont
  {{Puphal}}, \citenamefont {{Sch{\"a}fer}}, \citenamefont {{Keimer}},\ and\
  \citenamefont {{Hepting}}}]{Puphal2026}%
  \BibitemOpen
  \bibfield  {author} {\bibinfo {author} {\bibfnamefont {P.}~\bibnamefont
  {{Puphal}}}, \bibinfo {author} {\bibfnamefont {T.}~\bibnamefont
  {{Sch{\"a}fer}}}, \bibinfo {author} {\bibfnamefont {B.}~\bibnamefont
  {{Keimer}}},\ and\ \bibinfo {author} {\bibfnamefont {M.}~\bibnamefont
  {{Hepting}}},\ }\bibfield  {title} {\bibinfo {title} {{Superconductivity in
  infinite-layer and Ruddlesden-Popper nickelates}},\ }\href
  {https://doi.org/10.1038/s42254-025-00898-2} {\bibfield  {journal} {\bibinfo
  {journal} {Nature Reviews Physics}\ }\textbf {\bibinfo {volume} {8}},\
  \bibinfo {pages} {70} (\bibinfo {year} {2026})}\BibitemShut {NoStop}%
\bibitem [{\citenamefont {{Kanoda}}\ and\ \citenamefont
  {{Kato}}(2011)}]{Kanoda2011}%
  \BibitemOpen
  \bibfield  {author} {\bibinfo {author} {\bibfnamefont {K.}~\bibnamefont
  {{Kanoda}}}\ and\ \bibinfo {author} {\bibfnamefont {R.}~\bibnamefont
  {{Kato}}},\ }\bibfield  {title} {\bibinfo {title} {{Mott Physics in Organic
  Conductors with Triangular Lattices}},\ }\href
  {https://doi.org/10.1146/annurev-conmatphys-062910-140521} {\bibfield
  {journal} {\bibinfo  {journal} {Annual Review of Condensed Matter Physics}\
  }\textbf {\bibinfo {volume} {2}},\ \bibinfo {pages} {167} (\bibinfo {year}
  {2011})}\BibitemShut {NoStop}%
\bibitem [{\citenamefont {{Powell}}\ and\ \citenamefont
  {{McKenzie}}(2006)}]{Powell2006}%
  \BibitemOpen
  \bibfield  {author} {\bibinfo {author} {\bibfnamefont {B.~J.}\ \bibnamefont
  {{Powell}}}\ and\ \bibinfo {author} {\bibfnamefont {R.~H.}\ \bibnamefont
  {{McKenzie}}},\ }\bibfield  {title} {\bibinfo {title} {{TOPICAL REVIEW:
  Strong electronic correlations in superconducting organic charge transfer
  salts}},\ }\href {https://doi.org/10.1088/0953-8984/18/45/R03} {\bibfield
  {journal} {\bibinfo  {journal} {Journal of Physics Condensed Matter}\
  }\textbf {\bibinfo {volume} {18}},\ \bibinfo {pages} {R827} (\bibinfo {year}
  {2006})},\ \Eprint {https://arxiv.org/abs/cond-mat/0607078}
  {arXiv:cond-mat/0607078 [cond-mat.str-el]} \BibitemShut {NoStop}%
\bibitem [{\citenamefont {Riedl}\ \emph {et~al.}(2022)\citenamefont {Riedl},
  \citenamefont {Gati},\ and\ \citenamefont {Valenti}}]{Riedl2022}%
  \BibitemOpen
  \bibfield  {author} {\bibinfo {author} {\bibfnamefont {K.}~\bibnamefont
  {Riedl}}, \bibinfo {author} {\bibfnamefont {E.}~\bibnamefont {Gati}},\ and\
  \bibinfo {author} {\bibfnamefont {R.}~\bibnamefont {Valenti}},\ }\bibfield
  {title} {\bibinfo {title} {{Ingredients for Generalized Models of
  $\kappa$-Phase Organic Charge-Transfer Salts: A Review}},\ }\href
  {https://doi.org/10.3390/cryst12121689} {\bibfield  {journal} {\bibinfo
  {journal} {Crystals}\ }\textbf {\bibinfo {volume} {12}},\ \bibinfo {pages}
  {1689} (\bibinfo {year} {2022})}\BibitemShut {NoStop}%
\bibitem [{\citenamefont {{Li}}\ \emph {et~al.}(2021)\citenamefont {{Li}},
  \citenamefont {{Jiang}}, \citenamefont {{Li}}, \citenamefont {{Zhang}},
  \citenamefont {{Kang}}, \citenamefont {{Zhu}}, \citenamefont {{Watanabe}},
  \citenamefont {{Taniguchi}}, \citenamefont {{Chowdhury}}, \citenamefont
  {{Fu}}, \citenamefont {{Shan}},\ and\ \citenamefont {{Mak}}}]{Li2021}%
  \BibitemOpen
  \bibfield  {author} {\bibinfo {author} {\bibfnamefont {T.}~\bibnamefont
  {{Li}}}, \bibinfo {author} {\bibfnamefont {S.}~\bibnamefont {{Jiang}}},
  \bibinfo {author} {\bibfnamefont {L.}~\bibnamefont {{Li}}}, \bibinfo {author}
  {\bibfnamefont {Y.}~\bibnamefont {{Zhang}}}, \bibinfo {author} {\bibfnamefont
  {K.}~\bibnamefont {{Kang}}}, \bibinfo {author} {\bibfnamefont
  {J.}~\bibnamefont {{Zhu}}}, \bibinfo {author} {\bibfnamefont
  {K.}~\bibnamefont {{Watanabe}}}, \bibinfo {author} {\bibfnamefont
  {T.}~\bibnamefont {{Taniguchi}}}, \bibinfo {author} {\bibfnamefont
  {D.}~\bibnamefont {{Chowdhury}}}, \bibinfo {author} {\bibfnamefont
  {L.}~\bibnamefont {{Fu}}}, \bibinfo {author} {\bibfnamefont {J.}~\bibnamefont
  {{Shan}}},\ and\ \bibinfo {author} {\bibfnamefont {K.~F.}\ \bibnamefont
  {{Mak}}},\ }\bibfield  {title} {\bibinfo {title} {{Continuous Mott transition
  in semiconductor moir{\'e} superlattices}},\ }\href
  {https://doi.org/10.1038/s41586-021-03853-0} {\bibfield  {journal} {\bibinfo
  {journal} {Nature}\ }\textbf {\bibinfo {volume} {597}},\ \bibinfo {pages}
  {350} (\bibinfo {year} {2021})},\ \Eprint {https://arxiv.org/abs/2103.09779}
  {arXiv:2103.09779 [cond-mat.str-el]} \BibitemShut {NoStop}%
\bibitem [{\citenamefont {{Ghiotto}}\ \emph {et~al.}(2021)\citenamefont
  {{Ghiotto}}, \citenamefont {{Shih}}, \citenamefont {{Pereira}}, \citenamefont
  {{Rhodes}}, \citenamefont {{Kim}}, \citenamefont {{Zang}}, \citenamefont
  {{Millis}}, \citenamefont {{Watanabe}}, \citenamefont {{Taniguchi}},
  \citenamefont {{Hone}}, \citenamefont {{Wang}}, \citenamefont {{Dean}},\ and\
  \citenamefont {{Pasupathy}}}]{Ghiotto2021}%
  \BibitemOpen
  \bibfield  {author} {\bibinfo {author} {\bibfnamefont {A.}~\bibnamefont
  {{Ghiotto}}}, \bibinfo {author} {\bibfnamefont {E.-M.}\ \bibnamefont
  {{Shih}}}, \bibinfo {author} {\bibfnamefont {G.~S.~S.~G.}\ \bibnamefont
  {{Pereira}}}, \bibinfo {author} {\bibfnamefont {D.~A.}\ \bibnamefont
  {{Rhodes}}}, \bibinfo {author} {\bibfnamefont {B.}~\bibnamefont {{Kim}}},
  \bibinfo {author} {\bibfnamefont {J.}~\bibnamefont {{Zang}}}, \bibinfo
  {author} {\bibfnamefont {A.~J.}\ \bibnamefont {{Millis}}}, \bibinfo {author}
  {\bibfnamefont {K.}~\bibnamefont {{Watanabe}}}, \bibinfo {author}
  {\bibfnamefont {T.}~\bibnamefont {{Taniguchi}}}, \bibinfo {author}
  {\bibfnamefont {J.~C.}\ \bibnamefont {{Hone}}}, \bibinfo {author}
  {\bibfnamefont {L.}~\bibnamefont {{Wang}}}, \bibinfo {author} {\bibfnamefont
  {C.~R.}\ \bibnamefont {{Dean}}},\ and\ \bibinfo {author} {\bibfnamefont
  {A.~N.}\ \bibnamefont {{Pasupathy}}},\ }\bibfield  {title} {\bibinfo {title}
  {{Quantum criticality in twisted transition metal dichalcogenides}},\ }\href
  {https://doi.org/10.1038/s41586-021-03815-6} {\bibfield  {journal} {\bibinfo
  {journal} {Nature}\ }\textbf {\bibinfo {volume} {597}},\ \bibinfo {pages}
  {345} (\bibinfo {year} {2021})},\ \Eprint {https://arxiv.org/abs/2103.09796}
  {arXiv:2103.09796 [cond-mat.mes-hall]} \BibitemShut {NoStop}%
\bibitem [{\citenamefont {Esslinger}(2010)}]{Esslinger2010}%
  \BibitemOpen
  \bibfield  {author} {\bibinfo {author} {\bibfnamefont {T.}~\bibnamefont
  {Esslinger}},\ }\bibfield  {title} {\bibinfo {title} {{Fermi-Hubbard Physics
  with Atoms in an Optical Lattice}},\ }\href
  {https://doi.org/https://doi.org/10.1146/annurev-conmatphys-070909-104059}
  {\bibfield  {journal} {\bibinfo  {journal} {Annual Review of Condensed Matter
  Physics}\ }\textbf {\bibinfo {volume} {1}},\ \bibinfo {pages} {129} (\bibinfo
  {year} {2010})}\BibitemShut {NoStop}%
\bibitem [{\citenamefont {Chalopin}\ \emph {et~al.}(2026)\citenamefont
  {Chalopin}, \citenamefont {Bojović}, \citenamefont {Wang}, \citenamefont
  {Franz}, \citenamefont {Sinha}, \citenamefont {Wang}, \citenamefont
  {Bourgund}, \citenamefont {Obermeyer}, \citenamefont {Grusdt}, \citenamefont
  {Bohrdt}, \citenamefont {Pollet}, \citenamefont {Wietek}, \citenamefont
  {Georges}, \citenamefont {Hilker},\ and\ \citenamefont
  {Bloch}}]{Chalopin2026}%
  \BibitemOpen
  \bibfield  {author} {\bibinfo {author} {\bibfnamefont {T.}~\bibnamefont
  {Chalopin}}, \bibinfo {author} {\bibfnamefont {P.}~\bibnamefont {Bojović}},
  \bibinfo {author} {\bibfnamefont {S.}~\bibnamefont {Wang}}, \bibinfo {author}
  {\bibfnamefont {T.}~\bibnamefont {Franz}}, \bibinfo {author} {\bibfnamefont
  {A.}~\bibnamefont {Sinha}}, \bibinfo {author} {\bibfnamefont
  {Z.}~\bibnamefont {Wang}}, \bibinfo {author} {\bibfnamefont {D.}~\bibnamefont
  {Bourgund}}, \bibinfo {author} {\bibfnamefont {J.}~\bibnamefont {Obermeyer}},
  \bibinfo {author} {\bibfnamefont {F.}~\bibnamefont {Grusdt}}, \bibinfo
  {author} {\bibfnamefont {A.}~\bibnamefont {Bohrdt}}, \bibinfo {author}
  {\bibfnamefont {L.}~\bibnamefont {Pollet}}, \bibinfo {author} {\bibfnamefont
  {A.}~\bibnamefont {Wietek}}, \bibinfo {author} {\bibfnamefont
  {A.}~\bibnamefont {Georges}}, \bibinfo {author} {\bibfnamefont
  {T.}~\bibnamefont {Hilker}},\ and\ \bibinfo {author} {\bibfnamefont
  {I.}~\bibnamefont {Bloch}},\ }\bibfield  {title} {\bibinfo {title}
  {{Observation of emergent scaling of spin–charge correlations at the onset
  of the pseudogap}},\ }\bibfield  {journal} {\bibinfo  {journal} {Proceedings
  of the National Academy of Sciences}\ }\textbf {\bibinfo {volume} {123}},\
  \href {https://doi.org/10.1073/pnas.2525539123} {10.1073/pnas.2525539123}
  (\bibinfo {year} {2026})\BibitemShut {NoStop}%
\bibitem [{\citenamefont {Kendrick}\ \emph {et~al.}(2025)\citenamefont
  {Kendrick}, \citenamefont {Kale}, \citenamefont {Gang}, \citenamefont
  {Deters}, \citenamefont {Lebrat}, \citenamefont {Young},\ and\ \citenamefont
  {Greiner}}]{Kendrick2025}%
  \BibitemOpen
  \bibfield  {author} {\bibinfo {author} {\bibfnamefont {L.~H.}\ \bibnamefont
  {Kendrick}}, \bibinfo {author} {\bibfnamefont {A.}~\bibnamefont {Kale}},
  \bibinfo {author} {\bibfnamefont {Y.}~\bibnamefont {Gang}}, \bibinfo {author}
  {\bibfnamefont {A.~D.}\ \bibnamefont {Deters}}, \bibinfo {author}
  {\bibfnamefont {M.}~\bibnamefont {Lebrat}}, \bibinfo {author} {\bibfnamefont
  {A.~W.}\ \bibnamefont {Young}},\ and\ \bibinfo {author} {\bibfnamefont
  {M.}~\bibnamefont {Greiner}},\ }\href {https://arxiv.org/abs/2509.18075}
  {\bibinfo {title} {{Pseudogap in a Fermi-Hubbard quantum simulator}}}
  (\bibinfo {year} {2025}),\ \Eprint {https://arxiv.org/abs/2509.18075}
  {arXiv:2509.18075 [cond-mat.quant-gas]} \BibitemShut {NoStop}%
\bibitem [{\citenamefont {{Stewart}}(2017)}]{Stewart2017}%
  \BibitemOpen
  \bibfield  {author} {\bibinfo {author} {\bibfnamefont {G.~R.}\ \bibnamefont
  {{Stewart}}},\ }\bibfield  {title} {\bibinfo {title} {{Unconventional
  superconductivity}},\ }\href {https://doi.org/10.1080/00018732.2017.1331615}
  {\bibfield  {journal} {\bibinfo  {journal} {Advances in Physics}\ }\textbf
  {\bibinfo {volume} {66}},\ \bibinfo {pages} {75} (\bibinfo {year} {2017})},\
  \Eprint {https://arxiv.org/abs/1705.05593} {arXiv:1705.05593
  [cond-mat.supr-con]} \BibitemShut {NoStop}%
\bibitem [{\citenamefont {Imada}\ \emph {et~al.}(1998)\citenamefont {Imada},
  \citenamefont {Fujimori},\ and\ \citenamefont {Tokura}}]{Imada1998}%
  \BibitemOpen
  \bibfield  {author} {\bibinfo {author} {\bibfnamefont {M.}~\bibnamefont
  {Imada}}, \bibinfo {author} {\bibfnamefont {A.}~\bibnamefont {Fujimori}},\
  and\ \bibinfo {author} {\bibfnamefont {Y.}~\bibnamefont {Tokura}},\
  }\bibfield  {title} {\bibinfo {title} {Metal-insulator transitions},\ }\href
  {https://doi.org/10.1103/RevModPhys.70.1039} {\bibfield  {journal} {\bibinfo
  {journal} {Rev. Mod. Phys.}\ }\textbf {\bibinfo {volume} {70}},\ \bibinfo
  {pages} {1039} (\bibinfo {year} {1998})}\BibitemShut {NoStop}%
\bibitem [{\citenamefont {{Norman}}\ \emph {et~al.}(2005)\citenamefont
  {{Norman}}, \citenamefont {{Pines}},\ and\ \citenamefont
  {{Kallin}}}]{Norman2005}%
  \BibitemOpen
  \bibfield  {author} {\bibinfo {author} {\bibfnamefont {M.~R.}\ \bibnamefont
  {{Norman}}}, \bibinfo {author} {\bibfnamefont {D.}~\bibnamefont {{Pines}}},\
  and\ \bibinfo {author} {\bibfnamefont {C.}~\bibnamefont {{Kallin}}},\
  }\bibfield  {title} {\bibinfo {title} {{The pseudogap: friend or foe of high
  Tc?}},\ }\href {https://doi.org/10.1080/00018730500459906} {\bibfield
  {journal} {\bibinfo  {journal} {Advances in Physics}\ }\textbf {\bibinfo
  {volume} {54}},\ \bibinfo {pages} {715} (\bibinfo {year} {2005})},\ \Eprint
  {https://arxiv.org/abs/cond-mat/0507031} {arXiv:cond-mat/0507031
  [cond-mat.supr-con]} \BibitemShut {NoStop}%
\bibitem [{\citenamefont {Damascelli}\ \emph {et~al.}(2003)\citenamefont
  {Damascelli}, \citenamefont {Hussain},\ and\ \citenamefont
  {Shen}}]{Damascelli2003}%
  \BibitemOpen
  \bibfield  {author} {\bibinfo {author} {\bibfnamefont {A.}~\bibnamefont
  {Damascelli}}, \bibinfo {author} {\bibfnamefont {Z.}~\bibnamefont
  {Hussain}},\ and\ \bibinfo {author} {\bibfnamefont {Z.-X.}\ \bibnamefont
  {Shen}},\ }\bibfield  {title} {\bibinfo {title} {Angle-resolved photoemission
  studies of the cuprate superconductors},\ }\href
  {https://doi.org/10.1103/RevModPhys.75.473} {\bibfield  {journal} {\bibinfo
  {journal} {Rev. Mod. Phys.}\ }\textbf {\bibinfo {volume} {75}},\ \bibinfo
  {pages} {473} (\bibinfo {year} {2003})}\BibitemShut {NoStop}%
\bibitem [{\citenamefont {Sachdev}(1999)}]{Sachdev1999}%
  \BibitemOpen
  \bibfield  {author} {\bibinfo {author} {\bibfnamefont {S.}~\bibnamefont
  {Sachdev}},\ }\href@noop {} {\emph {\bibinfo {title} {{Quantum Phase
  Transitions}}}}\ (\bibinfo  {publisher} {Cambridge University Press},\
  \bibinfo {year} {1999})\BibitemShut {NoStop}%
\bibitem [{\citenamefont {v.~L\"ohneysen}\ \emph {et~al.}(2007)\citenamefont
  {v.~L\"ohneysen}, \citenamefont {Rosch}, \citenamefont {Vojta},\ and\
  \citenamefont {W\"olfle}}]{Loehneysen2007}%
  \BibitemOpen
  \bibfield  {author} {\bibinfo {author} {\bibfnamefont {H.}~\bibnamefont
  {v.~L\"ohneysen}}, \bibinfo {author} {\bibfnamefont {A.}~\bibnamefont
  {Rosch}}, \bibinfo {author} {\bibfnamefont {M.}~\bibnamefont {Vojta}},\ and\
  \bibinfo {author} {\bibfnamefont {P.}~\bibnamefont {W\"olfle}},\ }\bibfield
  {title} {\bibinfo {title} {Fermi-liquid instabilities at magnetic quantum
  phase transitions},\ }\href {https://doi.org/10.1103/RevModPhys.79.1015}
  {\bibfield  {journal} {\bibinfo  {journal} {Rev. Mod. Phys.}\ }\textbf
  {\bibinfo {volume} {79}},\ \bibinfo {pages} {1015} (\bibinfo {year}
  {2007})}\BibitemShut {NoStop}%
\bibitem [{\citenamefont {Brando}\ \emph {et~al.}(2016)\citenamefont {Brando},
  \citenamefont {Belitz}, \citenamefont {Grosche},\ and\ \citenamefont
  {Kirkpatrick}}]{Brando2016}%
  \BibitemOpen
  \bibfield  {author} {\bibinfo {author} {\bibfnamefont {M.}~\bibnamefont
  {Brando}}, \bibinfo {author} {\bibfnamefont {D.}~\bibnamefont {Belitz}},
  \bibinfo {author} {\bibfnamefont {F.~M.}\ \bibnamefont {Grosche}},\ and\
  \bibinfo {author} {\bibfnamefont {T.~R.}\ \bibnamefont {Kirkpatrick}},\
  }\bibfield  {title} {\bibinfo {title} {Metallic quantum ferromagnets},\
  }\href {https://doi.org/10.1103/RevModPhys.88.025006} {\bibfield  {journal}
  {\bibinfo  {journal} {Rev. Mod. Phys.}\ }\textbf {\bibinfo {volume} {88}},\
  \bibinfo {pages} {025006} (\bibinfo {year} {2016})}\BibitemShut {NoStop}%
\bibitem [{\citenamefont {{Savary}}\ and\ \citenamefont
  {{Balents}}(2017)}]{Savary2017}%
  \BibitemOpen
  \bibfield  {author} {\bibinfo {author} {\bibfnamefont {L.}~\bibnamefont
  {{Savary}}}\ and\ \bibinfo {author} {\bibfnamefont {L.}~\bibnamefont
  {{Balents}}},\ }\bibfield  {title} {\bibinfo {title} {{Quantum spin liquids:
  a review}},\ }\href {https://doi.org/10.1088/0034-4885/80/1/016502}
  {\bibfield  {journal} {\bibinfo  {journal} {Reports on Progress in Physics}\
  }\textbf {\bibinfo {volume} {80}},\ \bibinfo {eid} {016502} (\bibinfo {year}
  {2017})},\ \Eprint {https://arxiv.org/abs/1601.03742} {arXiv:1601.03742
  [cond-mat.str-el]} \BibitemShut {NoStop}%
\bibitem [{\citenamefont {Kaminski}\ \emph {et~al.}(2002)\citenamefont
  {Kaminski}, \citenamefont {Rosenkranz}, \citenamefont {Fretwell},
  \citenamefont {Campuzano}, \citenamefont {Li}, \citenamefont {Raffy},
  \citenamefont {Cullen}, \citenamefont {You}, \citenamefont {Olson},
  \citenamefont {Varma},\ and\ \citenamefont {Höchst}}]{Kaminski2002}%
  \BibitemOpen
  \bibfield  {author} {\bibinfo {author} {\bibfnamefont {A.}~\bibnamefont
  {Kaminski}}, \bibinfo {author} {\bibfnamefont {S.}~\bibnamefont
  {Rosenkranz}}, \bibinfo {author} {\bibfnamefont {H.}~\bibnamefont
  {Fretwell}}, \bibinfo {author} {\bibfnamefont {J.}~\bibnamefont {Campuzano}},
  \bibinfo {author} {\bibfnamefont {Z.}~\bibnamefont {Li}}, \bibinfo {author}
  {\bibfnamefont {H.}~\bibnamefont {Raffy}}, \bibinfo {author} {\bibfnamefont
  {W.}~\bibnamefont {Cullen}}, \bibinfo {author} {\bibfnamefont
  {H.}~\bibnamefont {You}}, \bibinfo {author} {\bibfnamefont {C.}~\bibnamefont
  {Olson}}, \bibinfo {author} {\bibfnamefont {C.}~\bibnamefont {Varma}},\ and\
  \bibinfo {author} {\bibfnamefont {H.}~\bibnamefont {Höchst}},\ }\bibfield
  {title} {\bibinfo {title} {Spontaneous breaking of time-reversal symmetry in
  the pseudogap state of a high-tc superconductor},\ }\href
  {https://www.nature.com/articles/416610a} {\bibfield  {journal} {\bibinfo
  {journal} {Nature}\ }\textbf {\bibinfo {volume} {416}},\ \bibinfo {pages}
  {610–613} (\bibinfo {year} {2002})}\BibitemShut {NoStop}%
\bibitem [{\citenamefont {Fradkin}\ \emph {et~al.}(2015)\citenamefont
  {Fradkin}, \citenamefont {Kivelson},\ and\ \citenamefont
  {Tranquada}}]{Fradkin2015}%
  \BibitemOpen
  \bibfield  {author} {\bibinfo {author} {\bibfnamefont {E.}~\bibnamefont
  {Fradkin}}, \bibinfo {author} {\bibfnamefont {S.~A.}\ \bibnamefont
  {Kivelson}},\ and\ \bibinfo {author} {\bibfnamefont {J.~M.}\ \bibnamefont
  {Tranquada}},\ }\bibfield  {title} {\bibinfo {title} {{Colloquium: Theory of
  intertwined orders in high temperature superconductors}},\ }\href
  {https://doi.org/10.1103/RevModPhys.87.457} {\bibfield  {journal} {\bibinfo
  {journal} {Rev. Mod. Phys.}\ }\textbf {\bibinfo {volume} {87}},\ \bibinfo
  {pages} {457} (\bibinfo {year} {2015})}\BibitemShut {NoStop}%
\bibitem [{\citenamefont {Viteritti}\ \emph {et~al.}(2026)\citenamefont
  {Viteritti}, \citenamefont {Rende}, \citenamefont {Roth}, \citenamefont
  {Sengupta}, \citenamefont {Carleo},\ and\ \citenamefont
  {Georges}}]{Viteritti2026}%
  \BibitemOpen
  \bibfield  {author} {\bibinfo {author} {\bibfnamefont {L.~L.}\ \bibnamefont
  {Viteritti}}, \bibinfo {author} {\bibfnamefont {R.}~\bibnamefont {Rende}},
  \bibinfo {author} {\bibfnamefont {C.}~\bibnamefont {Roth}}, \bibinfo {author}
  {\bibfnamefont {A.}~\bibnamefont {Sengupta}}, \bibinfo {author}
  {\bibfnamefont {G.}~\bibnamefont {Carleo}},\ and\ \bibinfo {author}
  {\bibfnamefont {A.}~\bibnamefont {Georges}},\ }\href
  {https://arxiv.org/abs/2604.21978} {\bibinfo {title} {{Beyond Variational
  Bias: Resolving Intertwined Orders in the Hubbard Model}}} (\bibinfo {year}
  {2026}),\ \Eprint {https://arxiv.org/abs/2604.21978} {arXiv:2604.21978
  [cond-mat.str-el]} \BibitemShut {NoStop}%
\bibitem [{\citenamefont {Hubbard}(1963)}]{Hubbard1963}%
  \BibitemOpen
  \bibfield  {author} {\bibinfo {author} {\bibfnamefont {J.}~\bibnamefont
  {Hubbard}},\ }\bibfield  {title} {\bibinfo {title} {{Electron Correlations in
  Narrow Energy Bands}},\ }\href {https://doi.org/10.1098/rspa.1963.0204}
  {\bibfield  {journal} {\bibinfo  {journal} {Proceedings of the Royal Society
  of London. Series A, Mathematical and Physical Sciences}\ }\textbf {\bibinfo
  {volume} {276}},\ \bibinfo {pages} {238} (\bibinfo {year}
  {1963})}\BibitemShut {NoStop}%
\bibitem [{\citenamefont {Qin}\ \emph {et~al.}(2022)\citenamefont {Qin},
  \citenamefont {Schäfer}, \citenamefont {Andergassen}, \citenamefont
  {Corboz},\ and\ \citenamefont {Gull}}]{Qin2022}%
  \BibitemOpen
  \bibfield  {author} {\bibinfo {author} {\bibfnamefont {M.}~\bibnamefont
  {Qin}}, \bibinfo {author} {\bibfnamefont {T.}~\bibnamefont {Schäfer}},
  \bibinfo {author} {\bibfnamefont {S.}~\bibnamefont {Andergassen}}, \bibinfo
  {author} {\bibfnamefont {P.}~\bibnamefont {Corboz}},\ and\ \bibinfo {author}
  {\bibfnamefont {E.}~\bibnamefont {Gull}},\ }\bibfield  {title} {\bibinfo
  {title} {{The Hubbard Model: A Computational Perspective}},\ }\bibfield
  {journal} {\bibinfo  {journal} {{Annual Review of Condensed Matter Physics}}\
  }\textbf {\bibinfo {volume} {13}},\ \href
  {https://doi.org/10.1146/annurev-conmatphys-090921-033948}
  {10.1146/annurev-conmatphys-090921-033948} (\bibinfo {year} {2022}),\ \Eprint
  {https://arxiv.org/abs/https://doi.org/10.1146/annurev-conmatphys-090921-033948}
  {https://doi.org/10.1146/annurev-conmatphys-090921-033948} \BibitemShut
  {NoStop}%
\bibitem [{\citenamefont {Arovas}\ \emph {et~al.}(2022)\citenamefont {Arovas},
  \citenamefont {Berg}, \citenamefont {Kivelson},\ and\ \citenamefont
  {Raghu}}]{Arovas2022}%
  \BibitemOpen
  \bibfield  {author} {\bibinfo {author} {\bibfnamefont {D.~P.}\ \bibnamefont
  {Arovas}}, \bibinfo {author} {\bibfnamefont {E.}~\bibnamefont {Berg}},
  \bibinfo {author} {\bibfnamefont {S.~A.}\ \bibnamefont {Kivelson}},\ and\
  \bibinfo {author} {\bibfnamefont {S.}~\bibnamefont {Raghu}},\ }\bibfield
  {title} {\bibinfo {title} {{The Hubbard Model}},\ }\bibfield  {journal}
  {\bibinfo  {journal} {{Annual Review of Condensed Matter Physics}}\ }\textbf
  {\bibinfo {volume} {13}},\ \href
  {https://doi.org/10.1146/annurev-conmatphys-031620-102024}
  {10.1146/annurev-conmatphys-031620-102024} (\bibinfo {year} {2022}),\ \Eprint
  {https://arxiv.org/abs/https://doi.org/10.1146/annurev-conmatphys-031620-102024}
  {https://doi.org/10.1146/annurev-conmatphys-031620-102024} \BibitemShut
  {NoStop}%
\bibitem [{\citenamefont {Huang}\ \emph {et~al.}(2018)\citenamefont {Huang},
  \citenamefont {Mendl}, \citenamefont {Jiang}, \citenamefont {Moritz},\ and\
  \citenamefont {Devereaux}}]{Huang2018}%
  \BibitemOpen
  \bibfield  {author} {\bibinfo {author} {\bibfnamefont {E.~W.}\ \bibnamefont
  {Huang}}, \bibinfo {author} {\bibfnamefont {C.~B.}\ \bibnamefont {Mendl}},
  \bibinfo {author} {\bibfnamefont {H.-C.}\ \bibnamefont {Jiang}}, \bibinfo
  {author} {\bibfnamefont {B.}~\bibnamefont {Moritz}},\ and\ \bibinfo {author}
  {\bibfnamefont {T.~P.}\ \bibnamefont {Devereaux}},\ }\bibfield  {title}
  {\bibinfo {title} {{Stripe order from the perspective of the Hubbard
  model}},\ }\bibfield  {journal} {\bibinfo  {journal} {npj Quantum Materials}\
  }\textbf {\bibinfo {volume} {3}},\ \href
  {https://doi.org/10.1038/s41535-018-0097-0} {10.1038/s41535-018-0097-0}
  (\bibinfo {year} {2018})\BibitemShut {NoStop}%
\bibitem [{\citenamefont {Qin}\ \emph {et~al.}(2020)\citenamefont {Qin},
  \citenamefont {Chung}, \citenamefont {Shi}, \citenamefont {Vitali},
  \citenamefont {Hubig}, \citenamefont {Schollwöck}, \citenamefont {White},\
  and\ \citenamefont {Zhang}}]{Qin2020}%
  \BibitemOpen
  \bibfield  {author} {\bibinfo {author} {\bibfnamefont {M.}~\bibnamefont
  {Qin}}, \bibinfo {author} {\bibfnamefont {C.-M.}\ \bibnamefont {Chung}},
  \bibinfo {author} {\bibfnamefont {H.}~\bibnamefont {Shi}}, \bibinfo {author}
  {\bibfnamefont {E.}~\bibnamefont {Vitali}}, \bibinfo {author} {\bibfnamefont
  {C.}~\bibnamefont {Hubig}}, \bibinfo {author} {\bibfnamefont
  {U.}~\bibnamefont {Schollwöck}}, \bibinfo {author} {\bibfnamefont {S.~R.}\
  \bibnamefont {White}},\ and\ \bibinfo {author} {\bibfnamefont
  {S.}~\bibnamefont {Zhang}},\ }\bibfield  {title} {\bibinfo {title} {{Absence
  of Superconductivity in the Pure Two-Dimensional Hubbard Model}},\ }\bibfield
   {journal} {\bibinfo  {journal} {Physical Review X}\ }\textbf {\bibinfo
  {volume} {10}},\ \href {https://doi.org/10.1103/physrevx.10.031016}
  {10.1103/physrevx.10.031016} (\bibinfo {year} {2020})\BibitemShut {NoStop}%
\bibitem [{\citenamefont {Xu}\ \emph {et~al.}(2024)\citenamefont {Xu},
  \citenamefont {Chung}, \citenamefont {Qin}, \citenamefont {Schollwöck},
  \citenamefont {White},\ and\ \citenamefont {Zhang}}]{Xu2024}%
  \BibitemOpen
  \bibfield  {author} {\bibinfo {author} {\bibfnamefont {H.}~\bibnamefont
  {Xu}}, \bibinfo {author} {\bibfnamefont {C.-M.}\ \bibnamefont {Chung}},
  \bibinfo {author} {\bibfnamefont {M.}~\bibnamefont {Qin}}, \bibinfo {author}
  {\bibfnamefont {U.}~\bibnamefont {Schollwöck}}, \bibinfo {author}
  {\bibfnamefont {S.~R.}\ \bibnamefont {White}},\ and\ \bibinfo {author}
  {\bibfnamefont {S.}~\bibnamefont {Zhang}},\ }\bibfield  {title} {\bibinfo
  {title} {{Coexistence of superconductivity with partially filled stripes in
  the Hubbard model}},\ }\bibfield  {journal} {\bibinfo  {journal} {Science}\
  }\textbf {\bibinfo {volume} {384}},\ \href
  {https://doi.org/10.1126/science.adh7691} {10.1126/science.adh7691} (\bibinfo
  {year} {2024})\BibitemShut {NoStop}%
\bibitem [{\citenamefont {Chakravarty}(2011)}]{Chakravarty2011}%
  \BibitemOpen
  \bibfield  {author} {\bibinfo {author} {\bibfnamefont {S.}~\bibnamefont
  {Chakravarty}},\ }\bibfield  {title} {\bibinfo {title} {{Quantum oscillations
  and key theoretical issues in high temperature superconductors from the
  perspective of density waves}},\ }\href
  {https://doi.org/10.1088/0034-4885/74/2/022501} {\bibfield  {journal}
  {\bibinfo  {journal} {Reports on Progress in Physics}\ }\textbf {\bibinfo
  {volume} {74}},\ \bibinfo {pages} {022501} (\bibinfo {year}
  {2011})}\BibitemShut {NoStop}%
\bibitem [{\citenamefont {{Jiang}}\ and\ \citenamefont
  {Devereaux}(2019)}]{Jiang2019}%
  \BibitemOpen
  \bibfield  {author} {\bibinfo {author} {\bibfnamefont {H.-C.}\ \bibnamefont
  {{Jiang}}}\ and\ \bibinfo {author} {\bibfnamefont {T.}~\bibnamefont
  {Devereaux}},\ }\bibfield  {title} {\bibinfo {title} {{Superconductivity in
  the doped Hubbard model and its interplay with next-nearest hopping t'}},\
  }\href {https://www.science.org/doi/epdf/10.1126/science.aal5304} {\bibfield
  {journal} {\bibinfo  {journal} {Science}\ }\textbf {\bibinfo {volume} {365}}
  (\bibinfo {year} {2019})}\BibitemShut {NoStop}%
\bibitem [{\citenamefont {Wang}\ \emph {et~al.}(2015)\citenamefont {Wang},
  \citenamefont {Agterberg},\ and\ \citenamefont {Chubukov}}]{Wang2015}%
  \BibitemOpen
  \bibfield  {author} {\bibinfo {author} {\bibfnamefont {Y.}~\bibnamefont
  {Wang}}, \bibinfo {author} {\bibfnamefont {D.~F.}\ \bibnamefont
  {Agterberg}},\ and\ \bibinfo {author} {\bibfnamefont {A.}~\bibnamefont
  {Chubukov}},\ }\bibfield  {title} {\bibinfo {title} {{Interplay between pair-
  and charge-density-wave orders in underdoped cuprates}},\ }\href
  {https://doi.org/10.1103/PhysRevB.91.115103} {\bibfield  {journal} {\bibinfo
  {journal} {Phys. Rev. B}\ }\textbf {\bibinfo {volume} {91}},\ \bibinfo
  {pages} {115103} (\bibinfo {year} {2015})}\BibitemShut {NoStop}%
\bibitem [{\citenamefont {Lao}\ and\ \citenamefont {Zhou}(2025)}]{Lao2025}%
  \BibitemOpen
  \bibfield  {author} {\bibinfo {author} {\bibfnamefont {J.}~\bibnamefont
  {Lao}}\ and\ \bibinfo {author} {\bibfnamefont {T.}~\bibnamefont {Zhou}},\
  }\bibfield  {title} {\bibinfo {title} {{Corner flat bands induced by $d$
  density wave and partial corner state modification due to competition with
  $d$-wave superconductivity}},\ }\href
  {https://doi.org/10.1103/PhysRevB.111.075169} {\bibfield  {journal} {\bibinfo
   {journal} {Phys. Rev. B}\ }\textbf {\bibinfo {volume} {111}},\ \bibinfo
  {pages} {075169} (\bibinfo {year} {2025})}\BibitemShut {NoStop}%
\bibitem [{\citenamefont {Wietek}\ \emph {et~al.}(2020)\citenamefont {Wietek},
  \citenamefont {He}, \citenamefont {White}, \citenamefont {Georges},\ and\
  \citenamefont {Stoudenmire}}]{Wietek2020}%
  \BibitemOpen
  \bibfield  {author} {\bibinfo {author} {\bibfnamefont {A.}~\bibnamefont
  {Wietek}}, \bibinfo {author} {\bibfnamefont {Y.-Y.}\ \bibnamefont {He}},
  \bibinfo {author} {\bibfnamefont {S.~R.}\ \bibnamefont {White}}, \bibinfo
  {author} {\bibfnamefont {A.}~\bibnamefont {Georges}},\ and\ \bibinfo {author}
  {\bibfnamefont {E.~M.}\ \bibnamefont {Stoudenmire}},\ }\href@noop {}
  {\bibinfo {title} {{Stripes, Antiferromagnetism, and the Pseudogap in the
  Doped Hubbard Model at Finite Temperature}}} (\bibinfo {year} {2020}),\
  \Eprint {https://arxiv.org/abs/2009.10736} {arXiv:2009.10736
  [cond-mat.str-el]} \BibitemShut {NoStop}%
\bibitem [{\citenamefont {Scholle}\ \emph {et~al.}(2026)\citenamefont
  {Scholle}, \citenamefont {Bonetti}, \citenamefont {Metzner},\ and\
  \citenamefont {Vilardi}}]{Scholle2026}%
  \BibitemOpen
  \bibfield  {author} {\bibinfo {author} {\bibfnamefont {R.}~\bibnamefont
  {Scholle}}, \bibinfo {author} {\bibfnamefont {M.}~\bibnamefont {Bonetti}},
  \bibinfo {author} {\bibfnamefont {W.}~\bibnamefont {Metzner}},\ and\ \bibinfo
  {author} {\bibfnamefont {D.}~\bibnamefont {Vilardi}},\ }\href
  {https://arxiv.org/abs/2602.20073} {\bibinfo {title} {{Coexisting magnetic,
  charge, and superconducting orders in the two-dimensional Hubbard model}}}
  (\bibinfo {year} {2026}),\ \Eprint {https://arxiv.org/abs/2602.20073}
  {arXiv:2602.20073} \BibitemShut {NoStop}%
\bibitem [{\citenamefont {Worm}\ \emph {et~al.}(2024)\citenamefont {Worm},
  \citenamefont {Reitner}, \citenamefont {Held},\ and\ \citenamefont
  {Toschi}}]{Reitner2024B}%
  \BibitemOpen
  \bibfield  {author} {\bibinfo {author} {\bibfnamefont {P.}~\bibnamefont
  {Worm}}, \bibinfo {author} {\bibfnamefont {M.}~\bibnamefont {Reitner}},
  \bibinfo {author} {\bibfnamefont {K.}~\bibnamefont {Held}},\ and\ \bibinfo
  {author} {\bibfnamefont {A.}~\bibnamefont {Toschi}},\ }\bibfield  {title}
  {\bibinfo {title} {Fermi and luttinger arcs: Two concepts, realized on one
  surface},\ }\href {https://doi.org/10.1103/PhysRevLett.133.166501} {\bibfield
   {journal} {\bibinfo  {journal} {Phys. Rev. Lett.}\ }\textbf {\bibinfo
  {volume} {133}},\ \bibinfo {pages} {166501} (\bibinfo {year}
  {2024})}\BibitemShut {NoStop}%
\bibitem [{\citenamefont {Adler}\ \emph
  {et~al.}(2024{\natexlab{a}})\citenamefont {Adler}, \citenamefont {Fus},
  \citenamefont {Malcolms}, \citenamefont {Vock}, \citenamefont {Held},
  \citenamefont {Katanin}, \citenamefont {Schäfer},\ and\ \citenamefont
  {Toschi}}]{Adler2024B}%
  \BibitemOpen
  \bibfield  {author} {\bibinfo {author} {\bibfnamefont {S.}~\bibnamefont
  {Adler}}, \bibinfo {author} {\bibfnamefont {D.~R.}\ \bibnamefont {Fus}},
  \bibinfo {author} {\bibfnamefont {M.~O.}\ \bibnamefont {Malcolms}}, \bibinfo
  {author} {\bibfnamefont {A.}~\bibnamefont {Vock}}, \bibinfo {author}
  {\bibfnamefont {K.}~\bibnamefont {Held}}, \bibinfo {author} {\bibfnamefont
  {A.~A.}\ \bibnamefont {Katanin}}, \bibinfo {author} {\bibfnamefont
  {T.}~\bibnamefont {Schäfer}},\ and\ \bibinfo {author} {\bibfnamefont
  {A.}~\bibnamefont {Toschi}},\ }\href {https://arxiv.org/abs/2409.04308}
  {\bibinfo {title} {Magnetic quantum criticality: dynamical mean-field
  perspective}} (\bibinfo {year} {2024}{\natexlab{a}}),\ \Eprint
  {https://arxiv.org/abs/2409.04308} {arXiv:2409.04308 [cond-mat.str-el]}
  \BibitemShut {NoStop}%
\bibitem [{\citenamefont {Rampon}\ \emph {et~al.}(2025)\citenamefont {Rampon},
  \citenamefont {\ifmmode~\check{S}\else \v{S}\fi{}imkovic},\ and\
  \citenamefont {Ferrero}}]{Rampon2025}%
  \BibitemOpen
  \bibfield  {author} {\bibinfo {author} {\bibfnamefont {L.}~\bibnamefont
  {Rampon}}, \bibinfo {author} {\bibfnamefont {F.}~\bibnamefont
  {\ifmmode~\check{S}\else \v{S}\fi{}imkovic}},\ and\ \bibinfo {author}
  {\bibfnamefont {M.}~\bibnamefont {Ferrero}},\ }\bibfield  {title} {\bibinfo
  {title} {Magnetic phase diagram of the three-dimensional doped hubbard
  model},\ }\href {https://doi.org/10.1103/PhysRevLett.134.066502} {\bibfield
  {journal} {\bibinfo  {journal} {Phys. Rev. Lett.}\ }\textbf {\bibinfo
  {volume} {134}},\ \bibinfo {pages} {066502} (\bibinfo {year}
  {2025})}\BibitemShut {NoStop}%
\bibitem [{\citenamefont {LeBlanc}\ \emph {et~al.}(2015)\citenamefont
  {LeBlanc}, \citenamefont {Antipov}, \citenamefont {Becca}, \citenamefont
  {Bulik}, \citenamefont {Chan}, \citenamefont {Chung}, \citenamefont {Deng},
  \citenamefont {Ferrero}, \citenamefont {Henderson}, \citenamefont
  {Jim\'enez-Hoyos}, \citenamefont {Kozik}, \citenamefont {Liu}, \citenamefont
  {Millis}, \citenamefont {Prokof'ev}, \citenamefont {Qin}, \citenamefont
  {Scuseria}, \citenamefont {Shi}, \citenamefont {Svistunov}, \citenamefont
  {Tocchio}, \citenamefont {Tupitsyn}, \citenamefont {White}, \citenamefont
  {Zhang}, \citenamefont {Zheng}, \citenamefont {Zhu},\ and\ \citenamefont
  {Gull}}]{Leblanc2015}%
  \BibitemOpen
  \bibfield  {author} {\bibinfo {author} {\bibfnamefont {J.~P.~F.}\
  \bibnamefont {LeBlanc}}, \bibinfo {author} {\bibfnamefont {A.~E.}\
  \bibnamefont {Antipov}}, \bibinfo {author} {\bibfnamefont {F.}~\bibnamefont
  {Becca}}, \bibinfo {author} {\bibfnamefont {I.~W.}\ \bibnamefont {Bulik}},
  \bibinfo {author} {\bibfnamefont {G.~K.-L.}\ \bibnamefont {Chan}}, \bibinfo
  {author} {\bibfnamefont {C.-M.}\ \bibnamefont {Chung}}, \bibinfo {author}
  {\bibfnamefont {Y.}~\bibnamefont {Deng}}, \bibinfo {author} {\bibfnamefont
  {M.}~\bibnamefont {Ferrero}}, \bibinfo {author} {\bibfnamefont {T.~M.}\
  \bibnamefont {Henderson}}, \bibinfo {author} {\bibfnamefont {C.~A.}\
  \bibnamefont {Jim\'enez-Hoyos}}, \bibinfo {author} {\bibfnamefont
  {E.}~\bibnamefont {Kozik}}, \bibinfo {author} {\bibfnamefont {X.-W.}\
  \bibnamefont {Liu}}, \bibinfo {author} {\bibfnamefont {A.~J.}\ \bibnamefont
  {Millis}}, \bibinfo {author} {\bibfnamefont {N.~V.}\ \bibnamefont
  {Prokof'ev}}, \bibinfo {author} {\bibfnamefont {M.}~\bibnamefont {Qin}},
  \bibinfo {author} {\bibfnamefont {G.~E.}\ \bibnamefont {Scuseria}}, \bibinfo
  {author} {\bibfnamefont {H.}~\bibnamefont {Shi}}, \bibinfo {author}
  {\bibfnamefont {B.~V.}\ \bibnamefont {Svistunov}}, \bibinfo {author}
  {\bibfnamefont {L.~F.}\ \bibnamefont {Tocchio}}, \bibinfo {author}
  {\bibfnamefont {I.~S.}\ \bibnamefont {Tupitsyn}}, \bibinfo {author}
  {\bibfnamefont {S.~R.}\ \bibnamefont {White}}, \bibinfo {author}
  {\bibfnamefont {S.}~\bibnamefont {Zhang}}, \bibinfo {author} {\bibfnamefont
  {B.-X.}\ \bibnamefont {Zheng}}, \bibinfo {author} {\bibfnamefont
  {Z.}~\bibnamefont {Zhu}},\ and\ \bibinfo {author} {\bibfnamefont
  {E.}~\bibnamefont {Gull}} (\bibinfo {collaboration} {Simons Collaboration on
  the Many-Electron Problem}),\ }\bibfield  {title} {\bibinfo {title}
  {{Solutions of the Two-Dimensional Hubbard Model: Benchmarks and Results from
  a Wide Range of Numerical Algorithms}},\ }\href
  {https://doi.org/10.1103/PhysRevX.5.041041} {\bibfield  {journal} {\bibinfo
  {journal} {Phys. Rev. X}\ }\textbf {\bibinfo {volume} {5}},\ \bibinfo {pages}
  {041041} (\bibinfo {year} {2015})}\BibitemShut {NoStop}%
\bibitem [{\citenamefont {Sch\"afer}\ \emph {et~al.}(2021)\citenamefont
  {Sch\"afer}, \citenamefont {Wentzell}, \citenamefont {\ifmmode~\check{S}\else
  \v{S}\fi{}imkovic}, \citenamefont {He}, \citenamefont {Hille}, \citenamefont
  {Klett}, \citenamefont {Eckhardt}, \citenamefont {Arzhang}, \citenamefont
  {Harkov}, \citenamefont {Le~R\'egent}, \citenamefont {Kirsch}, \citenamefont
  {Wang}, \citenamefont {Kim}, \citenamefont {Kozik}, \citenamefont {Stepanov},
  \citenamefont {Kauch}, \citenamefont {Andergassen}, \citenamefont {Hansmann},
  \citenamefont {Rohe}, \citenamefont {Vilk}, \citenamefont {LeBlanc},
  \citenamefont {Zhang}, \citenamefont {Tremblay}, \citenamefont {Ferrero},
  \citenamefont {Parcollet},\ and\ \citenamefont {Georges}}]{Schaefer2021}%
  \BibitemOpen
  \bibfield  {author} {\bibinfo {author} {\bibfnamefont {T.}~\bibnamefont
  {Sch\"afer}}, \bibinfo {author} {\bibfnamefont {N.}~\bibnamefont {Wentzell}},
  \bibinfo {author} {\bibfnamefont {F.}~\bibnamefont {\ifmmode~\check{S}\else
  \v{S}\fi{}imkovic}}, \bibinfo {author} {\bibfnamefont {Y.-Y.}\ \bibnamefont
  {He}}, \bibinfo {author} {\bibfnamefont {C.}~\bibnamefont {Hille}}, \bibinfo
  {author} {\bibfnamefont {M.}~\bibnamefont {Klett}}, \bibinfo {author}
  {\bibfnamefont {C.~J.}\ \bibnamefont {Eckhardt}}, \bibinfo {author}
  {\bibfnamefont {B.}~\bibnamefont {Arzhang}}, \bibinfo {author} {\bibfnamefont
  {V.}~\bibnamefont {Harkov}}, \bibinfo {author} {\bibfnamefont
  {F.}~\bibnamefont {Le~R\'egent}}, \bibinfo {author} {\bibfnamefont
  {A.}~\bibnamefont {Kirsch}}, \bibinfo {author} {\bibfnamefont
  {Y.}~\bibnamefont {Wang}}, \bibinfo {author} {\bibfnamefont {A.~J.}\
  \bibnamefont {Kim}}, \bibinfo {author} {\bibfnamefont {E.}~\bibnamefont
  {Kozik}}, \bibinfo {author} {\bibfnamefont {E.~A.}\ \bibnamefont {Stepanov}},
  \bibinfo {author} {\bibfnamefont {A.}~\bibnamefont {Kauch}}, \bibinfo
  {author} {\bibfnamefont {S.}~\bibnamefont {Andergassen}}, \bibinfo {author}
  {\bibfnamefont {P.}~\bibnamefont {Hansmann}}, \bibinfo {author}
  {\bibfnamefont {D.}~\bibnamefont {Rohe}}, \bibinfo {author} {\bibfnamefont
  {Y.~M.}\ \bibnamefont {Vilk}}, \bibinfo {author} {\bibfnamefont {J.~P.~F.}\
  \bibnamefont {LeBlanc}}, \bibinfo {author} {\bibfnamefont {S.}~\bibnamefont
  {Zhang}}, \bibinfo {author} {\bibfnamefont {A.-M.~S.}\ \bibnamefont
  {Tremblay}}, \bibinfo {author} {\bibfnamefont {M.}~\bibnamefont {Ferrero}},
  \bibinfo {author} {\bibfnamefont {O.}~\bibnamefont {Parcollet}},\ and\
  \bibinfo {author} {\bibfnamefont {A.}~\bibnamefont {Georges}},\ }\bibfield
  {title} {\bibinfo {title} {{Tracking the Footprints of Spin Fluctuations: A
  MultiMethod, MultiMessenger Study of the Two-Dimensional Hubbard Model}},\
  }\href {https://doi.org/10.1103/PhysRevX.11.011058} {\bibfield  {journal}
  {\bibinfo  {journal} {Phys. Rev. X}\ }\textbf {\bibinfo {volume} {11}},\
  \bibinfo {pages} {011058} (\bibinfo {year} {2021})}\BibitemShut {NoStop}%
\bibitem [{sup()}]{sup}%
  \BibitemOpen
  \href@noop {} {}\bibinfo {note} {See Supplemental Material at [URL will be
  inserted by publisher] for additional details on the model, derivations, and
  numerical results.}\BibitemShut {Stop}%
\bibitem [{\citenamefont {Zhang}(1990)}]{Zhang1990}%
  \BibitemOpen
  \bibfield  {author} {\bibinfo {author} {\bibfnamefont {S.}~\bibnamefont
  {Zhang}},\ }\bibfield  {title} {\bibinfo {title} {{Pseudospin symmetry and
  new collective modes of the Hubbard model}},\ }\href
  {https://doi.org/10.1103/PhysRevLett.65.120} {\bibfield  {journal} {\bibinfo
  {journal} {Phys. Rev. Lett.}\ }\textbf {\bibinfo {volume} {65}},\ \bibinfo
  {pages} {120} (\bibinfo {year} {1990})}\BibitemShut {NoStop}%
\bibitem [{\citenamefont {Keller}\ \emph {et~al.}(2001)\citenamefont {Keller},
  \citenamefont {Metzner},\ and\ \citenamefont {Schollw\"ock}}]{Keller2001}%
  \BibitemOpen
  \bibfield  {author} {\bibinfo {author} {\bibfnamefont {M.}~\bibnamefont
  {Keller}}, \bibinfo {author} {\bibfnamefont {W.}~\bibnamefont {Metzner}},\
  and\ \bibinfo {author} {\bibfnamefont {U.}~\bibnamefont {Schollw\"ock}},\
  }\bibfield  {title} {\bibinfo {title} {{Dynamical Mean-Field Theory for
  Pairing and Spin Gap in the Attractive Hubbard Model}},\ }\href
  {https://doi.org/10.1103/PhysRevLett.86.4612} {\bibfield  {journal} {\bibinfo
   {journal} {Phys. Rev. Lett.}\ }\textbf {\bibinfo {volume} {86}},\ \bibinfo
  {pages} {4612} (\bibinfo {year} {2001})}\BibitemShut {NoStop}%
\bibitem [{\citenamefont {Chakravarty}\ \emph {et~al.}(2001)\citenamefont
  {Chakravarty}, \citenamefont {Laughlin}, \citenamefont {Morr},\ and\
  \citenamefont {Nayak}}]{Chakravarty2001}%
  \BibitemOpen
  \bibfield  {author} {\bibinfo {author} {\bibfnamefont {S.}~\bibnamefont
  {Chakravarty}}, \bibinfo {author} {\bibfnamefont {R.~B.}\ \bibnamefont
  {Laughlin}}, \bibinfo {author} {\bibfnamefont {D.~K.}\ \bibnamefont {Morr}},\
  and\ \bibinfo {author} {\bibfnamefont {C.}~\bibnamefont {Nayak}},\ }\bibfield
   {title} {\bibinfo {title} {Hidden order in the cuprates},\ }\href
  {https://doi.org/10.1103/PhysRevB.63.094503} {\bibfield  {journal} {\bibinfo
  {journal} {Phys. Rev. B}\ }\textbf {\bibinfo {volume} {63}},\ \bibinfo
  {pages} {094503} (\bibinfo {year} {2001})}\BibitemShut {NoStop}%
\bibitem [{\citenamefont {Kotliar}\ \emph {et~al.}(2001)\citenamefont
  {Kotliar}, \citenamefont {Savrasov}, \citenamefont {P\'alsson},\ and\
  \citenamefont {Biroli}}]{Kotliar2001}%
  \BibitemOpen
  \bibfield  {author} {\bibinfo {author} {\bibfnamefont {G.}~\bibnamefont
  {Kotliar}}, \bibinfo {author} {\bibfnamefont {S.~Y.}\ \bibnamefont
  {Savrasov}}, \bibinfo {author} {\bibfnamefont {G.}~\bibnamefont
  {P\'alsson}},\ and\ \bibinfo {author} {\bibfnamefont {G.}~\bibnamefont
  {Biroli}},\ }\bibfield  {title} {\bibinfo {title} {{Cellular Dynamical Mean
  Field Approach to Strongly Correlated Systems}},\ }\href
  {https://doi.org/10.1103/PhysRevLett.87.186401} {\bibfield  {journal}
  {\bibinfo  {journal} {Phys. Rev. Lett.}\ }\textbf {\bibinfo {volume} {87}},\
  \bibinfo {pages} {186401} (\bibinfo {year} {2001})}\BibitemShut {NoStop}%
\bibitem [{\citenamefont {Meixner}\ \emph {et~al.}(2026)\citenamefont
  {Meixner}, \citenamefont {Reitner}, \citenamefont {Sch\"afer},\ and\
  \citenamefont {Toschi}}]{Meixner2026a}%
  \BibitemOpen
  \bibfield  {author} {\bibinfo {author} {\bibfnamefont {M.}~\bibnamefont
  {Meixner}}, \bibinfo {author} {\bibfnamefont {M.}~\bibnamefont {Reitner}},
  \bibinfo {author} {\bibfnamefont {T.}~\bibnamefont {Sch\"afer}},\ and\
  \bibinfo {author} {\bibfnamefont {A.}~\bibnamefont {Toschi}},\ }\href
  {https://arxiv.org/abs/2512.17716} {\bibinfo {title} {{Non-perturbative
  effects of short-range spatial correlations at the two-particle level}}}
  (\bibinfo {year} {2026}),\ \Eprint {https://arxiv.org/abs/2512.17716}
  {arXiv:2512.17716 [cond-mat.str-el]} \BibitemShut {NoStop}%
\bibitem [{\citenamefont {Hubbard}(1972)}]{Shib1972}%
  \BibitemOpen
  \bibfield  {author} {\bibinfo {author} {\bibfnamefont {H.}~\bibnamefont
  {Hubbard}},\ }\bibfield  {title} {\bibinfo {title} {{Thermodynamic Properties
  of the One-Dimensional Half-Filled-Band Hubbard Model. II}},\ }\href
  {https://doi.org/10.1143/PTP.48.2171} {\bibfield  {journal} {\bibinfo
  {journal} {Progress of Theoretical Physics}\ }\textbf {\bibinfo {volume}
  {48}},\ \bibinfo {pages} {2171} (\bibinfo {year} {1972})}\BibitemShut
  {NoStop}%
\bibitem [{\citenamefont {Singh}\ and\ \citenamefont
  {Scalettar}(1991)}]{Singh1991}%
  \BibitemOpen
  \bibfield  {author} {\bibinfo {author} {\bibfnamefont {R.~R.~P.}\
  \bibnamefont {Singh}}\ and\ \bibinfo {author} {\bibfnamefont {R.~T.}\
  \bibnamefont {Scalettar}},\ }\bibfield  {title} {\bibinfo {title} {{Exact
  demonstration of \ensuremath{\eta} pairing in the ground state of an
  attractive-U Hubbard model}},\ }\href
  {https://doi.org/10.1103/PhysRevLett.66.3203} {\bibfield  {journal} {\bibinfo
   {journal} {Phys. Rev. Lett.}\ }\textbf {\bibinfo {volume} {66}},\ \bibinfo
  {pages} {3203} (\bibinfo {year} {1991})}\BibitemShut {NoStop}%
\bibitem [{\citenamefont {E\ss{}l}\ \emph {et~al.}(2024)\citenamefont
  {E\ss{}l}, \citenamefont {Reitner}, \citenamefont {Sangiovanni},\ and\
  \citenamefont {Toschi}}]{Essl2024}%
  \BibitemOpen
  \bibfield  {author} {\bibinfo {author} {\bibfnamefont {H.}~\bibnamefont
  {E\ss{}l}}, \bibinfo {author} {\bibfnamefont {M.}~\bibnamefont {Reitner}},
  \bibinfo {author} {\bibfnamefont {G.}~\bibnamefont {Sangiovanni}},\ and\
  \bibinfo {author} {\bibfnamefont {A.}~\bibnamefont {Toschi}},\ }\bibfield
  {title} {\bibinfo {title} {{General Shiba mapping for on-site four-point
  correlation functions}},\ }\href
  {https://doi.org/10.1103/PhysRevResearch.6.033061} {\bibfield  {journal}
  {\bibinfo  {journal} {Phys. Rev. Res.}\ }\textbf {\bibinfo {volume} {6}},\
  \bibinfo {pages} {033061} (\bibinfo {year} {2024})}\BibitemShut {NoStop}%
\bibitem [{\citenamefont {Shen}\ and\ \citenamefont {Xie}(1996)}]{Shen1996}%
  \BibitemOpen
  \bibfield  {author} {\bibinfo {author} {\bibfnamefont {S.-Q.}\ \bibnamefont
  {Shen}}\ and\ \bibinfo {author} {\bibfnamefont {X.~C.}\ \bibnamefont {Xie}},\
  }\bibfield  {title} {\bibinfo {title} {{Pseudospin SU(2)-symmetry breaking,
  charge-density waves and superconductivity in the Hubbard model}},\ }\href
  {https://iopscience.iop.org/article/10.1088/0953-8984/8/26/012/pdf}
  {\bibfield  {journal} {\bibinfo  {journal} {J. Phys.: Condens. Matter}\
  }\textbf {\bibinfo {volume} {8}},\ \bibinfo {pages} {4805} (\bibinfo {year}
  {1996})}\BibitemShut {NoStop}%
\bibitem [{\citenamefont {Yang}\ and\ \citenamefont {Zhang}(1990)}]{Yang1990}%
  \BibitemOpen
  \bibfield  {author} {\bibinfo {author} {\bibfnamefont {C.}~\bibnamefont
  {Yang}}\ and\ \bibinfo {author} {\bibfnamefont {S.}~\bibnamefont {Zhang}},\
  }\bibfield  {title} {\bibinfo {title} {{SO4 symmetry in a Hubbard model}},\
  }\href {https://doi.org/10.1142/S0217984990000933} {\bibfield  {journal}
  {\bibinfo  {journal} {Modern Physics Letters B}\ }\textbf {\bibinfo {volume}
  {04}},\ \bibinfo {pages} {759} (\bibinfo {year} {1990})}\BibitemShut
  {NoStop}%
\bibitem [{\citenamefont {Pu}\ and\ \citenamefont {Shen}(1994)}]{Pu1994}%
  \BibitemOpen
  \bibfield  {author} {\bibinfo {author} {\bibfnamefont {F.-C.}\ \bibnamefont
  {Pu}}\ and\ \bibinfo {author} {\bibfnamefont {S.-Q.}\ \bibnamefont {Shen}},\
  }\bibfield  {title} {\bibinfo {title} {{Relation between pseudospin-rotation
  invariance and a supersolid}},\ }\href
  {https://doi.org/10.1103/PhysRevB.50.16086} {\bibfield  {journal} {\bibinfo
  {journal} {Phys. Rev. B}\ }\textbf {\bibinfo {volume} {50}},\ \bibinfo
  {pages} {16086} (\bibinfo {year} {1994})}\BibitemShut {NoStop}%
\bibitem [{\citenamefont {Scaletter}(2024)}]{Scaletter2024}%
  \BibitemOpen
  \bibfield  {author} {\bibinfo {author} {\bibfnamefont {R.}~\bibnamefont
  {Scaletter}},\ }\bibfield  {title} {\bibinfo {title} {{Understanding the
  Hubbard Model with Simple Calculations}},\ }in\ \href
  {https://www.cond-mat.de/events/correl24/manuscripts/scalettar.pdf} {\emph
  {\bibinfo {booktitle} {Correlations and Phase Transitions: Modeling and
  Simulation}}},\ Vol.~\bibinfo {volume} {14}\ (\bibinfo  {publisher}
  {Forschungszentrum Jülich},\ \bibinfo {year} {2024})\ Chap.~\bibinfo
  {chapter} {3}\BibitemShut {NoStop}%
\bibitem [{\citenamefont {Carmelo}(2026)}]{Carmelo2026}%
  \BibitemOpen
  \bibfield  {author} {\bibinfo {author} {\bibfnamefont {J.~M.~P.}\
  \bibnamefont {Carmelo}},\ }\bibfield  {title} {\bibinfo {title} {{Exact
  results for the Hubbard model on bipartite lattices in spatial dimensions
  $d>1$: Seven theorems from the full
  $[\mathrm{SU}(2)\ifmmode\times\else\texttimes\fi{}\mathrm{SU}(2)\ifmmode\times\else\texttimes\fi{}\mathrm{U}(1)]/{\mathbb{Z}}_{2}^{2}$
  symmetry}},\ }\href {https://doi.org/10.1103/p1sc-sx2d} {\bibfield  {journal}
  {\bibinfo  {journal} {Phys. Rev. B}\ }\textbf {\bibinfo {volume} {113}},\
  \bibinfo {pages} {155157} (\bibinfo {year} {2026})}\BibitemShut {NoStop}%
\bibitem [{Note1()}]{Note1}%
  \BibitemOpen
  \bibinfo {note} {This momentum space equation is related to the real-space
  Eq.~(\ref {Eq:real-space}) by Fourier transforms, where pp and ph sectors are
  subject to different Fourier conventions \cite {Pu1994,Rohringer2013a}(see
  Sec.~II of \cite {sup}) to implement the adequate operator
  ordering.}\BibitemShut {Stop}%
\bibitem [{Note2()}]{Note2}%
  \BibitemOpen
  \bibinfo {note} {Corresponding to the truncated unity expansion \cite
  {Lichtenstein2017}}\BibitemShut {NoStop}%
\bibitem [{\citenamefont {Affleck}\ and\ \citenamefont
  {Marston}(1988)}]{Affleck1988}%
  \BibitemOpen
  \bibfield  {author} {\bibinfo {author} {\bibfnamefont {I.}~\bibnamefont
  {Affleck}}\ and\ \bibinfo {author} {\bibfnamefont {J.~B.}\ \bibnamefont
  {Marston}},\ }\bibfield  {title} {\bibinfo {title} {{Large-n limit of the
  Heisenberg-Hubbard model: Implications for high-${T}_{c}$ superconductors}},\
  }\href {https://doi.org/10.1103/PhysRevB.37.3774} {\bibfield  {journal}
  {\bibinfo  {journal} {Phys. Rev. B}\ }\textbf {\bibinfo {volume} {37}},\
  \bibinfo {pages} {3774} (\bibinfo {year} {1988})}\BibitemShut {NoStop}%
\bibitem [{\citenamefont {Nayak}(2000{\natexlab{a}})}]{Nayak2000b}%
  \BibitemOpen
  \bibfield  {author} {\bibinfo {author} {\bibfnamefont {C.}~\bibnamefont
  {Nayak}},\ }\bibfield  {title} {\bibinfo {title} {{$O(4)$-invariant
  formulation of the nodal liquid}},\ }\href
  {https://doi.org/10.1103/PhysRevB.62.R6135} {\bibfield  {journal} {\bibinfo
  {journal} {Phys. Rev. B}\ }\textbf {\bibinfo {volume} {62}},\ \bibinfo
  {pages} {R6135} (\bibinfo {year} {2000}{\natexlab{a}})}\BibitemShut {NoStop}%
\bibitem [{\citenamefont {Chakravarty}(2002)}]{Chakravarty2002}%
  \BibitemOpen
  \bibfield  {author} {\bibinfo {author} {\bibfnamefont {S.}~\bibnamefont
  {Chakravarty}},\ }\bibfield  {title} {\bibinfo {title} {{Theory of the
  d-density wave from a vertex model and its implications}},\ }\href
  {https://doi.org/10.1103/PhysRevB.66.224505} {\bibfield  {journal} {\bibinfo
  {journal} {Phys. Rev. B}\ }\textbf {\bibinfo {volume} {66}},\ \bibinfo
  {pages} {224505} (\bibinfo {year} {2002})}\BibitemShut {NoStop}%
\bibitem [{\citenamefont {Leeb}\ and\ \citenamefont {Knolle}(2026)}]{Leeb2026}%
  \BibitemOpen
  \bibfield  {author} {\bibinfo {author} {\bibfnamefont {V.}~\bibnamefont
  {Leeb}}\ and\ \bibinfo {author} {\bibfnamefont {J.}~\bibnamefont {Knolle}},\
  }\href {https://arxiv.org/pdf/2601.07418} {\bibinfo {title} {{Collinear
  p-wave magnetism and hidden orbital ferrimagnetism}}} (\bibinfo {year}
  {2026})\BibitemShut {NoStop}%
\bibitem [{Note3()}]{Note3}%
  \BibitemOpen
  \bibinfo {note} {We note that an alike relation of $d$-DW and pairing
  instability has been discussed in the different context of the $t-J$ model
  \cite {Kotliar1988,Cappelluti1999} and, recently, the $\sigma _z$-Hubbard
  model \cite {Zhu2025}}\BibitemShut {NoStop}%
\bibitem [{\citenamefont {Anderson}(1961)}]{Anderson1961}%
  \BibitemOpen
  \bibfield  {author} {\bibinfo {author} {\bibfnamefont {P.}~\bibnamefont
  {Anderson}},\ }\bibfield  {title} {\bibinfo {title} {Localized magnetic
  states in metals},\ }\href@noop {} {\bibfield  {journal} {\bibinfo  {journal}
  {Phys. Rev.}\ }\textbf {\bibinfo {volume} {124}},\ \bibinfo {pages} {41}
  (\bibinfo {year} {1961})}\BibitemShut {NoStop}%
\bibitem [{\citenamefont {van Loon}\ and\ \citenamefont
  {Strand}(2024)}]{vanLoon2024-2}%
  \BibitemOpen
  \bibfield  {author} {\bibinfo {author} {\bibfnamefont {E.~G. C.~P.}\
  \bibnamefont {van Loon}}\ and\ \bibinfo {author} {\bibfnamefont {H.~U.~R.}\
  \bibnamefont {Strand}},\ }\bibfield  {title} {\bibinfo {title} {{Dual
  Bethe-Salpeter equation for the multiorbital lattice susceptibility within
  dynamical mean-field theory}},\ }\href
  {https://doi.org/10.1103/PhysRevB.109.155157} {\bibfield  {journal} {\bibinfo
   {journal} {Phys. Rev. B}\ }\textbf {\bibinfo {volume} {109}},\ \bibinfo
  {pages} {155157} (\bibinfo {year} {2024})}\BibitemShut {NoStop}%
\bibitem [{\citenamefont {Jarrell}\ \emph {et~al.}(2001)\citenamefont
  {Jarrell}, \citenamefont {Maier}, \citenamefont {Huscroft},\ and\
  \citenamefont {Moukouri}}]{Jarrell2001}%
  \BibitemOpen
  \bibfield  {author} {\bibinfo {author} {\bibfnamefont {M.}~\bibnamefont
  {Jarrell}}, \bibinfo {author} {\bibfnamefont {T.}~\bibnamefont {Maier}},
  \bibinfo {author} {\bibfnamefont {C.}~\bibnamefont {Huscroft}},\ and\
  \bibinfo {author} {\bibfnamefont {S.}~\bibnamefont {Moukouri}},\ }\bibfield
  {title} {\bibinfo {title} {{Quantum Monte Carlo algorithm for nonlocal
  corrections to the dynamical mean-field theory}},\ }\href@noop {} {\bibfield
  {journal} {\bibinfo  {journal} {Phys. Rev. B}\ }\textbf {\bibinfo {volume}
  {64}},\ \bibinfo {pages} {195130} (\bibinfo {year} {2001})}\BibitemShut
  {NoStop}%
\bibitem [{\citenamefont {Georges}\ and\ \citenamefont
  {Krauth}(1992)}]{Georges1992}%
  \BibitemOpen
  \bibfield  {author} {\bibinfo {author} {\bibfnamefont {A.}~\bibnamefont
  {Georges}}\ and\ \bibinfo {author} {\bibfnamefont {W.}~\bibnamefont
  {Krauth}},\ }\bibfield  {title} {\bibinfo {title} {{Numerical solution of the
  $d=\infty$ Hubbard model: Evidence for a Mott transition}},\ }\href
  {https://doi.org/10.1103/PhysRevLett.69.1240} {\bibfield  {journal} {\bibinfo
   {journal} {Phys. Rev. Lett.}\ }\textbf {\bibinfo {volume} {69}},\ \bibinfo
  {pages} {1240} (\bibinfo {year} {1992})}\BibitemShut {NoStop}%
\bibitem [{\citenamefont {Kotliar}\ \emph {et~al.}(2000)\citenamefont
  {Kotliar}, \citenamefont {Lange},\ and\ \citenamefont
  {Rozenberg}}]{Kotliar2000}%
  \BibitemOpen
  \bibfield  {author} {\bibinfo {author} {\bibfnamefont {G.}~\bibnamefont
  {Kotliar}}, \bibinfo {author} {\bibfnamefont {E.}~\bibnamefont {Lange}},\
  and\ \bibinfo {author} {\bibfnamefont {M.~J.}\ \bibnamefont {Rozenberg}},\
  }\bibfield  {title} {\bibinfo {title} {{Landau Theory of the Finite
  Temperature Mott Transition}},\ }\href
  {https://doi.org/10.1103/PhysRevLett.84.5180} {\bibfield  {journal} {\bibinfo
   {journal} {Phys. Rev. Lett.}\ }\textbf {\bibinfo {volume} {84}},\ \bibinfo
  {pages} {5180} (\bibinfo {year} {2000})}\BibitemShut {NoStop}%
\bibitem [{Note4()}]{Note4}%
  \BibitemOpen
  \bibinfo {note} {The $2\times 2$ CDMFT cell is sufficient to capture the
  $d-$wave instability, consequently, the BSE-superlattice vector is set to
  0.}\BibitemShut {Stop}%
\bibitem [{\citenamefont {Park}\ \emph
  {et~al.}(2008{\natexlab{a}})\citenamefont {Park}, \citenamefont {Haule},\
  and\ \citenamefont {Kotliar}}]{Park2008}%
  \BibitemOpen
  \bibfield  {author} {\bibinfo {author} {\bibfnamefont {H.}~\bibnamefont
  {Park}}, \bibinfo {author} {\bibfnamefont {K.}~\bibnamefont {Haule}},\ and\
  \bibinfo {author} {\bibfnamefont {G.}~\bibnamefont {Kotliar}},\ }\bibfield
  {title} {\bibinfo {title} {{Cluster Dynamical Mean Field Theory of the Mott
  Transition}},\ }\href {https://doi.org/10.1103/PhysRevLett.101.186403}
  {\bibfield  {journal} {\bibinfo  {journal} {Phys. Rev. Lett.}\ }\textbf
  {\bibinfo {volume} {101}},\ \bibinfo {pages} {186403} (\bibinfo {year}
  {2008}{\natexlab{a}})}\BibitemShut {NoStop}%
\bibitem [{\citenamefont {Fratino}\ \emph {et~al.}(2016)\citenamefont
  {Fratino}, \citenamefont {S\'emon}, \citenamefont {Sordi},\ and\
  \citenamefont {Tremblay}}]{Fratino2016}%
  \BibitemOpen
  \bibfield  {author} {\bibinfo {author} {\bibfnamefont {L.}~\bibnamefont
  {Fratino}}, \bibinfo {author} {\bibfnamefont {P.}~\bibnamefont {S\'emon}},
  \bibinfo {author} {\bibfnamefont {G.}~\bibnamefont {Sordi}},\ and\ \bibinfo
  {author} {\bibfnamefont {A.~M.~S.}\ \bibnamefont {Tremblay}},\ }\bibfield
  {title} {\bibinfo {title} {{An organizing principle for two-dimensional
  strongly correlated superconductivity}},\ }\href
  {https://doi.org/https://doi.org/10.1038/srep22715} {\bibfield  {journal}
  {\bibinfo  {journal} {Scientific Reports}\ }\textbf {\bibinfo {volume} {6}},\
  \bibinfo {pages} {22715} (\bibinfo {year} {2016})}\BibitemShut {NoStop}%
\bibitem [{\citenamefont {Varma}(1997)}]{Varma1997}%
  \BibitemOpen
  \bibfield  {author} {\bibinfo {author} {\bibfnamefont {C.~M.}\ \bibnamefont
  {Varma}},\ }\bibfield  {title} {\bibinfo {title} {{Non-Fermi-liquid states
  and pairing instability of a general model of copper oxide metals}},\ }\href
  {https://doi.org/10.1103/PhysRevB.55.14554} {\bibfield  {journal} {\bibinfo
  {journal} {Phys. Rev. B}\ }\textbf {\bibinfo {volume} {55}},\ \bibinfo
  {pages} {14554} (\bibinfo {year} {1997})}\BibitemShut {NoStop}%
\bibitem [{\citenamefont {Varma}(1999)}]{Varma1999}%
  \BibitemOpen
  \bibfield  {author} {\bibinfo {author} {\bibfnamefont {C.~M.}\ \bibnamefont
  {Varma}},\ }\bibfield  {title} {\bibinfo {title} {{Pseudogap Phase and the
  Quantum-Critical Point in Copper-Oxide Metals}},\ }\href
  {https://doi.org/10.1103/PhysRevLett.83.3538} {\bibfield  {journal} {\bibinfo
   {journal} {Phys. Rev. Lett.}\ }\textbf {\bibinfo {volume} {83}},\ \bibinfo
  {pages} {3538} (\bibinfo {year} {1999})}\BibitemShut {NoStop}%
\bibitem [{\citenamefont {Goswami}(2013)}]{Goswami2013}%
  \BibitemOpen
  \bibfield  {author} {\bibinfo {author} {\bibfnamefont {P.}~\bibnamefont
  {Goswami}},\ }\bibfield  {title} {\bibinfo {title} {A theoretical approach to
  pseudogap and superconducting transitions in hole-doped cuprates},\ }\href
  {https://doi.org/https://doi.org/10.1155/2013/210384} {\bibfield  {journal}
  {\bibinfo  {journal} {International Scholarly Research Notices}\ }\textbf
  {\bibinfo {volume} {2013}},\ \bibinfo {pages} {210384} (\bibinfo {year}
  {2013})},\ \Eprint
  {https://arxiv.org/abs/https://onlinelibrary.wiley.com/doi/pdf/10.1155/2013/210384}
  {https://onlinelibrary.wiley.com/doi/pdf/10.1155/2013/210384} \BibitemShut
  {NoStop}%
\bibitem [{\citenamefont {Nayak}(2000{\natexlab{b}})}]{Nayak2000}%
  \BibitemOpen
  \bibfield  {author} {\bibinfo {author} {\bibfnamefont {C.}~\bibnamefont
  {Nayak}},\ }\bibfield  {title} {\bibinfo {title} {{Density-wave states of
  nonzero angular momentum}},\ }\href
  {https://doi.org/10.1103/PhysRevB.62.4880} {\bibfield  {journal} {\bibinfo
  {journal} {Phys. Rev. B}\ }\textbf {\bibinfo {volume} {62}},\ \bibinfo
  {pages} {4880} (\bibinfo {year} {2000}{\natexlab{b}})}\BibitemShut {NoStop}%
\bibitem [{\citenamefont {Mermin}\ and\ \citenamefont
  {Wagner}(1966)}]{Mermin1966}%
  \BibitemOpen
  \bibfield  {author} {\bibinfo {author} {\bibfnamefont {N.~D.}\ \bibnamefont
  {Mermin}}\ and\ \bibinfo {author} {\bibfnamefont {H.}~\bibnamefont
  {Wagner}},\ }\bibfield  {title} {\bibinfo {title} {{Absence of Ferromagnetism
  or Antiferromagnetism in One- or Two-Dimensional Isotropic Heisenberg
  Models}},\ }\href {https://doi.org/10.1103/PhysRevLett.17.1307} {\bibfield
  {journal} {\bibinfo  {journal} {Phys. Rev. Lett.}\ }\textbf {\bibinfo
  {volume} {17}},\ \bibinfo {pages} {1307} (\bibinfo {year}
  {1966})}\BibitemShut {NoStop}%
\bibitem [{\citenamefont {Hohenberg}(1967)}]{Hohenberg1967}%
  \BibitemOpen
  \bibfield  {author} {\bibinfo {author} {\bibfnamefont {P.~C.}\ \bibnamefont
  {Hohenberg}},\ }\bibfield  {title} {\bibinfo {title} {{Existence of
  Long-Range Order in One and Two Dimensions}},\ }\href
  {https://doi.org/10.1103/PhysRev.158.383} {\bibfield  {journal} {\bibinfo
  {journal} {Phys. Rev.}\ }\textbf {\bibinfo {volume} {158}},\ \bibinfo {pages}
  {383} (\bibinfo {year} {1967})}\BibitemShut {NoStop}%
\bibitem [{\citenamefont {Dar\'e}\ \emph {et~al.}(1996)\citenamefont {Dar\'e},
  \citenamefont {Vilk},\ and\ \citenamefont {Tremblay}}]{Dare1996}%
  \BibitemOpen
  \bibfield  {author} {\bibinfo {author} {\bibfnamefont {A.-M.}\ \bibnamefont
  {Dar\'e}}, \bibinfo {author} {\bibfnamefont {Y.~M.}\ \bibnamefont {Vilk}},\
  and\ \bibinfo {author} {\bibfnamefont {A.~M.~S.}\ \bibnamefont {Tremblay}},\
  }\bibfield  {title} {\bibinfo {title} {Crossover from two- to
  three-dimensional critical behavior for nearly antiferromagnetic itinerant
  electrons},\ }\href {https://doi.org/10.1103/PhysRevB.53.14236} {\bibfield
  {journal} {\bibinfo  {journal} {Phys. Rev. B}\ }\textbf {\bibinfo {volume}
  {53}},\ \bibinfo {pages} {14236} (\bibinfo {year} {1996})}\BibitemShut
  {NoStop}%
\bibitem [{\citenamefont {Vilk}\ and\ \citenamefont
  {Tremblay}(1997)}]{Vilk1997}%
  \BibitemOpen
  \bibfield  {author} {\bibinfo {author} {\bibfnamefont {Y.~M.}\ \bibnamefont
  {Vilk}}\ and\ \bibinfo {author} {\bibfnamefont {A.-M.~S.}\ \bibnamefont
  {Tremblay}},\ }\bibfield  {title} {\bibinfo {title} {{Non-Perturbative
  Many-Body Approach to the Hubbard Model and Single-Particle Pseudogap}},\
  }\href {https://doi.org/10.1051/jp1:1997135} {\bibfield  {journal} {\bibinfo
  {journal} {J. Phys. I France}\ }\textbf {\bibinfo {volume} {7}},\ \bibinfo
  {pages} {1309} (\bibinfo {year} {1997})}\BibitemShut {NoStop}%
\bibitem [{\citenamefont {Sch\"afer}\ \emph {et~al.}(2015)\citenamefont
  {Sch\"afer}, \citenamefont {Geles}, \citenamefont {Rost}, \citenamefont
  {Rohringer}, \citenamefont {Arrigoni}, \citenamefont {Held}, \citenamefont
  {Bl\"umer}, \citenamefont {Aichhorn},\ and\ \citenamefont
  {Toschi}}]{Schaefer2015b}%
  \BibitemOpen
  \bibfield  {author} {\bibinfo {author} {\bibfnamefont {T.}~\bibnamefont
  {Sch\"afer}}, \bibinfo {author} {\bibfnamefont {F.}~\bibnamefont {Geles}},
  \bibinfo {author} {\bibfnamefont {D.}~\bibnamefont {Rost}}, \bibinfo {author}
  {\bibfnamefont {G.}~\bibnamefont {Rohringer}}, \bibinfo {author}
  {\bibfnamefont {E.}~\bibnamefont {Arrigoni}}, \bibinfo {author}
  {\bibfnamefont {K.}~\bibnamefont {Held}}, \bibinfo {author} {\bibfnamefont
  {N.}~\bibnamefont {Bl\"umer}}, \bibinfo {author} {\bibfnamefont
  {M.}~\bibnamefont {Aichhorn}},\ and\ \bibinfo {author} {\bibfnamefont
  {A.}~\bibnamefont {Toschi}},\ }\bibfield  {title} {\bibinfo {title} {{Fate of
  the false Mott-Hubbard transition in two dimensions}},\ }\href
  {https://doi.org/10.1103/PhysRevB.91.125109} {\bibfield  {journal} {\bibinfo
  {journal} {Phys. Rev. B}\ }\textbf {\bibinfo {volume} {91}},\ \bibinfo
  {pages} {125109} (\bibinfo {year} {2015})}\BibitemShut {NoStop}%
\bibitem [{\citenamefont {Mu\ss{}hoff}\ \emph {et~al.}(2021)\citenamefont
  {Mu\ss{}hoff}, \citenamefont {Kiani},\ and\ \citenamefont
  {Pavarini}}]{Musshoff2021}%
  \BibitemOpen
  \bibfield  {author} {\bibinfo {author} {\bibfnamefont {J.}~\bibnamefont
  {Mu\ss{}hoff}}, \bibinfo {author} {\bibfnamefont {A.}~\bibnamefont {Kiani}},\
  and\ \bibinfo {author} {\bibfnamefont {E.}~\bibnamefont {Pavarini}},\
  }\bibfield  {title} {\bibinfo {title} {{Magnetic response trends in cuprates
  and the $t\ensuremath{-}{t}^{\ensuremath{'}}$ Hubbard model}},\ }\href
  {https://doi.org/10.1103/PhysRevB.103.075136} {\bibfield  {journal} {\bibinfo
   {journal} {Phys. Rev. B}\ }\textbf {\bibinfo {volume} {103}},\ \bibinfo
  {pages} {075136} (\bibinfo {year} {2021})}\BibitemShut {NoStop}%
\bibitem [{\citenamefont {Krien}\ \emph {et~al.}(2019)\citenamefont {Krien},
  \citenamefont {Valli},\ and\ \citenamefont {Capone}}]{Krien2019c}%
  \BibitemOpen
  \bibfield  {author} {\bibinfo {author} {\bibfnamefont {F.}~\bibnamefont
  {Krien}}, \bibinfo {author} {\bibfnamefont {A.}~\bibnamefont {Valli}},\ and\
  \bibinfo {author} {\bibfnamefont {M.}~\bibnamefont {Capone}},\ }\bibfield
  {title} {\bibinfo {title} {{Single-boson exchange decomposition of the vertex
  function}},\ }\href {https://doi.org/10.1103/PhysRevB.100.155149} {\bibfield
  {journal} {\bibinfo  {journal} {Phys. Rev. B}\ }\textbf {\bibinfo {volume}
  {100}},\ \bibinfo {pages} {155149} (\bibinfo {year} {2019})}\BibitemShut
  {NoStop}%
\bibitem [{\citenamefont {Bonetti}\ \emph {et~al.}(2022)\citenamefont
  {Bonetti}, \citenamefont {Toschi}, \citenamefont {Hille}, \citenamefont
  {Andergassen},\ and\ \citenamefont {Vilardi}}]{Bonetti2022}%
  \BibitemOpen
  \bibfield  {author} {\bibinfo {author} {\bibfnamefont {P.~M.}\ \bibnamefont
  {Bonetti}}, \bibinfo {author} {\bibfnamefont {A.}~\bibnamefont {Toschi}},
  \bibinfo {author} {\bibfnamefont {C.}~\bibnamefont {Hille}}, \bibinfo
  {author} {\bibfnamefont {S.}~\bibnamefont {Andergassen}},\ and\ \bibinfo
  {author} {\bibfnamefont {D.}~\bibnamefont {Vilardi}},\ }\bibfield  {title}
  {\bibinfo {title} {Single-boson exchange representation of the functional
  renormalization group for strongly interacting many-electron systems},\
  }\href {https://doi.org/10.1103/PhysRevResearch.4.013034} {\bibfield
  {journal} {\bibinfo  {journal} {Phys. Rev. Res.}\ }\textbf {\bibinfo {volume}
  {4}},\ \bibinfo {pages} {013034} (\bibinfo {year} {2022})}\BibitemShut
  {NoStop}%
\bibitem [{\citenamefont {Adler}\ \emph
  {et~al.}(2024{\natexlab{b}})\citenamefont {Adler}, \citenamefont {Krien},
  \citenamefont {Chalupa-Gantner}, \citenamefont {Sangiovanni},\ and\
  \citenamefont {Toschi}}]{Adler2024}%
  \BibitemOpen
  \bibfield  {author} {\bibinfo {author} {\bibfnamefont {S.}~\bibnamefont
  {Adler}}, \bibinfo {author} {\bibfnamefont {F.}~\bibnamefont {Krien}},
  \bibinfo {author} {\bibfnamefont {P.}~\bibnamefont {Chalupa-Gantner}},
  \bibinfo {author} {\bibfnamefont {G.}~\bibnamefont {Sangiovanni}},\ and\
  \bibinfo {author} {\bibfnamefont {A.}~\bibnamefont {Toschi}},\ }\bibfield
  {title} {\bibinfo {title} {{Non-perturbative intertwining between spin and
  charge correlations: A ``smoking gun'' single-boson-exchange result}},\
  }\href {https://doi.org/10.21468/SciPostPhys.16.2.054} {\bibfield  {journal}
  {\bibinfo  {journal} {SciPost Phys.}\ }\textbf {\bibinfo {volume} {16}},\
  \bibinfo {pages} {054} (\bibinfo {year} {2024}{\natexlab{b}})}\BibitemShut
  {NoStop}%
\bibitem [{\citenamefont {Gievers}(2025)}]{Gievers2025}%
  \BibitemOpen
  \bibfield  {author} {\bibinfo {author} {\bibfnamefont {M.}~\bibnamefont
  {Gievers}},\ }\emph {\bibinfo {title} {Functional approaches to Fermi
  polarons in cold atomic gases and solid-state systems}},\ \href
  {https://edoc.ub.uni-muenchen.de/34995/11/Gievers_Marcel.pdf} {Ph.D.
  thesis},\ \bibinfo  {school} {Ludwig-Maximilians Universit\"at München}
  (\bibinfo {year} {2025})\BibitemShut {NoStop}%
\bibitem [{Note5()}]{Note5}%
  \BibitemOpen
  \bibinfo {note} {This corresponds to applying the SBE decomposition to the
  two-particle irreducible vertex $\Gamma _{\protect \mathrm {pp}}$ \cite
  {Krien2020b,Meixner2026a}.}\BibitemShut {Stop}%
\bibitem [{\citenamefont {Meixner}\ \emph {et~al.}(2025)\citenamefont
  {Meixner}, \citenamefont {Krämer}, \citenamefont {Wentzell}, \citenamefont
  {Bonetti}, \citenamefont {Andergassen}, \citenamefont {Toschi},\ and\
  \citenamefont {Schäfer}}]{Meixner2025}%
  \BibitemOpen
  \bibfield  {author} {\bibinfo {author} {\bibfnamefont {M.}~\bibnamefont
  {Meixner}}, \bibinfo {author} {\bibfnamefont {M.}~\bibnamefont {Krämer}},
  \bibinfo {author} {\bibfnamefont {N.}~\bibnamefont {Wentzell}}, \bibinfo
  {author} {\bibfnamefont {P.~M.}\ \bibnamefont {Bonetti}}, \bibinfo {author}
  {\bibfnamefont {S.}~\bibnamefont {Andergassen}}, \bibinfo {author}
  {\bibfnamefont {A.}~\bibnamefont {Toschi}},\ and\ \bibinfo {author}
  {\bibfnamefont {T.}~\bibnamefont {Schäfer}},\ }\bibfield  {title} {\bibinfo
  {title} {Disentangling real space fluctuations: The diagnostics of
  metal-insulator transitions beyond single-particle spectral functions},\
  }\bibfield  {journal} {\bibinfo  {journal} {Physical Review Research}\
  }\textbf {\bibinfo {volume} {7}},\ \href {https://doi.org/10.1103/1nt5-swsk}
  {10.1103/1nt5-swsk} (\bibinfo {year} {2025})\BibitemShut {NoStop}%
\bibitem [{\citenamefont {Kyung}\ \emph {et~al.}(2003)\citenamefont {Kyung},
  \citenamefont {Landry},\ and\ \citenamefont {Tremblay}}]{Kyung2003}%
  \BibitemOpen
  \bibfield  {author} {\bibinfo {author} {\bibfnamefont {B.}~\bibnamefont
  {Kyung}}, \bibinfo {author} {\bibfnamefont {J.-S.}\ \bibnamefont {Landry}},\
  and\ \bibinfo {author} {\bibfnamefont {A.-M.~S.}\ \bibnamefont {Tremblay}},\
  }\bibfield  {title} {\bibinfo {title} {Antiferromagnetic fluctuations and
  d-wave superconductivity in electron-doped high-temperature
  superconductors},\ }\href {https://doi.org/10.1103/PhysRevB.68.174502}
  {\bibfield  {journal} {\bibinfo  {journal} {Phys. Rev. B}\ }\textbf {\bibinfo
  {volume} {68}},\ \bibinfo {pages} {174502} (\bibinfo {year}
  {2003})}\BibitemShut {NoStop}%
\bibitem [{\citenamefont {Kancharla}\ \emph {et~al.}(2008)\citenamefont
  {Kancharla}, \citenamefont {Kyung}, \citenamefont {S\'en\'echal},
  \citenamefont {Civelli}, \citenamefont {Capone}, \citenamefont {Kotliar},\
  and\ \citenamefont {Tremblay}}]{Kancharla2008}%
  \BibitemOpen
  \bibfield  {author} {\bibinfo {author} {\bibfnamefont {S.~S.}\ \bibnamefont
  {Kancharla}}, \bibinfo {author} {\bibfnamefont {B.}~\bibnamefont {Kyung}},
  \bibinfo {author} {\bibfnamefont {D.}~\bibnamefont {S\'en\'echal}}, \bibinfo
  {author} {\bibfnamefont {M.}~\bibnamefont {Civelli}}, \bibinfo {author}
  {\bibfnamefont {M.}~\bibnamefont {Capone}}, \bibinfo {author} {\bibfnamefont
  {G.}~\bibnamefont {Kotliar}},\ and\ \bibinfo {author} {\bibfnamefont
  {A.-M.~S.}\ \bibnamefont {Tremblay}},\ }\bibfield  {title} {\bibinfo {title}
  {{Anomalous superconductivity and its competition with antiferromagnetism in
  doped Mott insulators}},\ }\href {https://doi.org/10.1103/PhysRevB.77.184516}
  {\bibfield  {journal} {\bibinfo  {journal} {Phys. Rev. B}\ }\textbf {\bibinfo
  {volume} {77}},\ \bibinfo {pages} {184516} (\bibinfo {year}
  {2008})}\BibitemShut {NoStop}%
\bibitem [{\citenamefont {Kotliar}(1988)}]{Kotliar1988}%
  \BibitemOpen
  \bibfield  {author} {\bibinfo {author} {\bibfnamefont {G.}~\bibnamefont
  {Kotliar}},\ }\bibfield  {title} {\bibinfo {title} {{Resonating valence bonds
  and d-wave superconductivity}},\ }\href
  {https://doi.org/10.1103/PhysRevB.37.3664} {\bibfield  {journal} {\bibinfo
  {journal} {Phys. Rev. B}\ }\textbf {\bibinfo {volume} {37}},\ \bibinfo
  {pages} {3664} (\bibinfo {year} {1988})}\BibitemShut {NoStop}%
\bibitem [{\citenamefont {Georges}\ and\ \citenamefont
  {Kotliar}(1992)}]{Georges1992a}%
  \BibitemOpen
  \bibfield  {author} {\bibinfo {author} {\bibfnamefont {A.}~\bibnamefont
  {Georges}}\ and\ \bibinfo {author} {\bibfnamefont {G.}~\bibnamefont
  {Kotliar}},\ }\bibfield  {title} {\bibinfo {title} {{Hubbard model in
  infinite dimensions}},\ }\href {https://doi.org/10.1103/PhysRevB.45.6479}
  {\bibfield  {journal} {\bibinfo  {journal} {Phys. Rev. B}\ }\textbf {\bibinfo
  {volume} {45}},\ \bibinfo {pages} {6479} (\bibinfo {year}
  {1992})}\BibitemShut {NoStop}%
\bibitem [{\citenamefont {Parcollet}\ \emph {et~al.}(2004)\citenamefont
  {Parcollet}, \citenamefont {Biroli},\ and\ \citenamefont
  {Kotliar}}]{Parcollet2004}%
  \BibitemOpen
  \bibfield  {author} {\bibinfo {author} {\bibfnamefont {O.}~\bibnamefont
  {Parcollet}}, \bibinfo {author} {\bibfnamefont {G.}~\bibnamefont {Biroli}},\
  and\ \bibinfo {author} {\bibfnamefont {G.}~\bibnamefont {Kotliar}},\
  }\bibfield  {title} {\bibinfo {title} {{Cluster Dynamical Mean Field Analysis
  of the Mott Transition}},\ }\href
  {https://doi.org/10.1103/PhysRevLett.92.226402} {\bibfield  {journal}
  {\bibinfo  {journal} {Phys. Rev. Lett.}\ }\textbf {\bibinfo {volume} {92}},\
  \bibinfo {pages} {226402} (\bibinfo {year} {2004})}\BibitemShut {NoStop}%
\bibitem [{\citenamefont {Reitner}\ \emph {et~al.}(2020)\citenamefont
  {Reitner}, \citenamefont {Chalupa}, \citenamefont {Del~Re}, \citenamefont
  {Springer}, \citenamefont {Ciuchi}, \citenamefont {Sangiovanni},\ and\
  \citenamefont {Toschi}}]{Reitner2020}%
  \BibitemOpen
  \bibfield  {author} {\bibinfo {author} {\bibfnamefont {M.}~\bibnamefont
  {Reitner}}, \bibinfo {author} {\bibfnamefont {P.}~\bibnamefont {Chalupa}},
  \bibinfo {author} {\bibfnamefont {L.}~\bibnamefont {Del~Re}}, \bibinfo
  {author} {\bibfnamefont {D.}~\bibnamefont {Springer}}, \bibinfo {author}
  {\bibfnamefont {S.}~\bibnamefont {Ciuchi}}, \bibinfo {author} {\bibfnamefont
  {G.}~\bibnamefont {Sangiovanni}},\ and\ \bibinfo {author} {\bibfnamefont
  {A.}~\bibnamefont {Toschi}},\ }\bibfield  {title} {\bibinfo {title}
  {{Attractive Effect of a Strong Electronic Repulsion: The Physics of Vertex
  Divergences}},\ }\bibfield  {journal} {\bibinfo  {journal} {Physical Review
  Letters}\ }\textbf {\bibinfo {volume} {125}},\ \href
  {https://doi.org/10.1103/physrevlett.125.196403}
  {10.1103/physrevlett.125.196403} (\bibinfo {year} {2020})\BibitemShut
  {NoStop}%
\bibitem [{\citenamefont {Sordi}\ \emph
  {et~al.}(2012{\natexlab{a}})\citenamefont {Sordi}, \citenamefont {S\'emon},
  \citenamefont {Haule},\ and\ \citenamefont {Tremblay}}]{Sordi2012}%
  \BibitemOpen
  \bibfield  {author} {\bibinfo {author} {\bibfnamefont {G.}~\bibnamefont
  {Sordi}}, \bibinfo {author} {\bibfnamefont {P.}~\bibnamefont {S\'emon}},
  \bibinfo {author} {\bibfnamefont {K.}~\bibnamefont {Haule}},\ and\ \bibinfo
  {author} {\bibfnamefont {A.-M.~S.}\ \bibnamefont {Tremblay}},\ }\bibfield
  {title} {\bibinfo {title} {{Strong Coupling Superconductivity, Pseudogap, and
  Mott Transition}},\ }\href {https://doi.org/10.1103/PhysRevLett.108.216401}
  {\bibfield  {journal} {\bibinfo  {journal} {Phys. Rev. Lett.}\ }\textbf
  {\bibinfo {volume} {108}},\ \bibinfo {pages} {216401} (\bibinfo {year}
  {2012}{\natexlab{a}})}\BibitemShut {NoStop}%
\bibitem [{\citenamefont {Sordi}\ \emph
  {et~al.}(2012{\natexlab{b}})\citenamefont {Sordi}, \citenamefont {S{\'e}mon},
  \citenamefont {Haule},\ and\ \citenamefont {Tremblay}}]{Sordi2012b}%
  \BibitemOpen
  \bibfield  {author} {\bibinfo {author} {\bibfnamefont {G.}~\bibnamefont
  {Sordi}}, \bibinfo {author} {\bibfnamefont {P.}~\bibnamefont {S{\'e}mon}},
  \bibinfo {author} {\bibfnamefont {K.}~\bibnamefont {Haule}},\ and\ \bibinfo
  {author} {\bibfnamefont {A.-M.}\ \bibnamefont {Tremblay}},\ }\bibfield
  {title} {\bibinfo {title} {{Pseudogap temperature as a Widom line in doped
  Mott insulators}},\ }\href {https://www.nature.com/articles/srep00547}
  {\bibfield  {journal} {\bibinfo  {journal} {Scientific reports}\ }\textbf
  {\bibinfo {volume} {2}},\ \bibinfo {pages} {1} (\bibinfo {year}
  {2012}{\natexlab{b}})}\BibitemShut {NoStop}%
\bibitem [{\citenamefont {Patricolo}\ \emph {et~al.}(2025)\citenamefont
  {Patricolo}, \citenamefont {Gievers}, \citenamefont {Fraboulet},
  \citenamefont {Al-Eryani}, \citenamefont {Heinzelmann}, \citenamefont
  {Bonetti}, \citenamefont {Toschi}, \citenamefont {Vilardi},\ and\
  \citenamefont {Andergassen}}]{Patricolo2025}%
  \BibitemOpen
  \bibfield  {author} {\bibinfo {author} {\bibfnamefont {M.}~\bibnamefont
  {Patricolo}}, \bibinfo {author} {\bibfnamefont {M.}~\bibnamefont {Gievers}},
  \bibinfo {author} {\bibfnamefont {K.}~\bibnamefont {Fraboulet}}, \bibinfo
  {author} {\bibfnamefont {A.}~\bibnamefont {Al-Eryani}}, \bibinfo {author}
  {\bibfnamefont {S.}~\bibnamefont {Heinzelmann}}, \bibinfo {author}
  {\bibfnamefont {P.~M.}\ \bibnamefont {Bonetti}}, \bibinfo {author}
  {\bibfnamefont {A.}~\bibnamefont {Toschi}}, \bibinfo {author} {\bibfnamefont
  {D.}~\bibnamefont {Vilardi}},\ and\ \bibinfo {author} {\bibfnamefont
  {S.}~\bibnamefont {Andergassen}},\ }\bibfield  {title} {\bibinfo {title}
  {{Single-boson exchange formulation of the Schwinger-Dyson equation and its
  application to the functional renormalization group}},\ }\href
  {https://doi.org/10.21468/SciPostPhys.18.3.078} {\bibfield  {journal}
  {\bibinfo  {journal} {SciPost Phys.}\ }\textbf {\bibinfo {volume} {18}},\
  \bibinfo {pages} {078} (\bibinfo {year} {2025})}\BibitemShut {NoStop}%
\bibitem [{\citenamefont {Fraboulet}\ \emph {et~al.}(2025)\citenamefont
  {Fraboulet}, \citenamefont {Al-Eryani}, \citenamefont {Heinzelmann},
  \citenamefont {Kauch},\ and\ \citenamefont {Andergassen}}]{Fraboulet2025}%
  \BibitemOpen
  \bibfield  {author} {\bibinfo {author} {\bibfnamefont {K.}~\bibnamefont
  {Fraboulet}}, \bibinfo {author} {\bibfnamefont {A.}~\bibnamefont
  {Al-Eryani}}, \bibinfo {author} {\bibfnamefont {S.}~\bibnamefont
  {Heinzelmann}}, \bibinfo {author} {\bibfnamefont {A.}~\bibnamefont {Kauch}},\
  and\ \bibinfo {author} {\bibfnamefont {S.}~\bibnamefont {Andergassen}},\
  }\href {https://arxiv.org/pdf/2512.11190} {\bibinfo {title} {{Multiloop
  functional renormalization group from single bosons}}} (\bibinfo {year}
  {2025}),\ \Eprint {https://arxiv.org/abs/2512.11190} {arXiv:2512.11190}
  \BibitemShut {NoStop}%
\bibitem [{\citenamefont {Rohringer}(2013)}]{Rohringer2013a}%
  \BibitemOpen
  \bibfield  {author} {\bibinfo {author} {\bibfnamefont {G.}~\bibnamefont
  {Rohringer}},\ }\emph {\bibinfo {title} {New routes towards a theoretical
  treatment of nonlocal electronic correlations}},\ \href
  {http://digital.obvsg.at/download/pdf/1631831} {Ph.D. thesis},\ \bibinfo
  {school} {Vienna University of Technology} (\bibinfo {year}
  {2013})\BibitemShut {NoStop}%
\bibitem [{\citenamefont {Lichtenstein}\ \emph {et~al.}(2017)\citenamefont
  {Lichtenstein}, \citenamefont {{Sánchez de la Peña}}, \citenamefont {Rohe},
  \citenamefont {{Di Napoli}}, \citenamefont {Honerkamp},\ and\ \citenamefont
  {Maier}}]{Lichtenstein2017}%
  \BibitemOpen
  \bibfield  {author} {\bibinfo {author} {\bibfnamefont {J.}~\bibnamefont
  {Lichtenstein}}, \bibinfo {author} {\bibfnamefont {D.}~\bibnamefont
  {{Sánchez de la Peña}}}, \bibinfo {author} {\bibfnamefont {D.}~\bibnamefont
  {Rohe}}, \bibinfo {author} {\bibfnamefont {E.}~\bibnamefont {{Di Napoli}}},
  \bibinfo {author} {\bibfnamefont {C.}~\bibnamefont {Honerkamp}},\ and\
  \bibinfo {author} {\bibfnamefont {S.}~\bibnamefont {Maier}},\ }\bibfield
  {title} {\bibinfo {title} {High-performance functional renormalization group
  calculations for interacting fermions},\ }\href
  {https://doi.org/https://doi.org/10.1016/j.cpc.2016.12.013} {\bibfield
  {journal} {\bibinfo  {journal} {Computer Physics Communications}\ }\textbf
  {\bibinfo {volume} {213}},\ \bibinfo {pages} {100} (\bibinfo {year}
  {2017})}\BibitemShut {NoStop}%
\bibitem [{\citenamefont {Cappelluti}\ and\ \citenamefont
  {Zeyher}(1999)}]{Cappelluti1999}%
  \BibitemOpen
  \bibfield  {author} {\bibinfo {author} {\bibfnamefont {E.}~\bibnamefont
  {Cappelluti}}\ and\ \bibinfo {author} {\bibfnamefont {R.}~\bibnamefont
  {Zeyher}},\ }\bibfield  {title} {\bibinfo {title} {Interplay between
  superconductivity and flux phase in the $t\ensuremath{-}j$ model},\ }\href
  {https://doi.org/10.1103/PhysRevB.59.6475} {\bibfield  {journal} {\bibinfo
  {journal} {Phys. Rev. B}\ }\textbf {\bibinfo {volume} {59}},\ \bibinfo
  {pages} {6475} (\bibinfo {year} {1999})}\BibitemShut {NoStop}%
\bibitem [{\citenamefont {Zhu}\ \emph {et~al.}(2025)\citenamefont {Zhu},
  \citenamefont {Sun}, \citenamefont {Gong}, \citenamefont {Huang},
  \citenamefont {Feng}, \citenamefont {Scalettar},\ and\ \citenamefont
  {Guo}}]{Zhu2025}%
  \BibitemOpen
  \bibfield  {author} {\bibinfo {author} {\bibfnamefont {X.}~\bibnamefont
  {Zhu}}, \bibinfo {author} {\bibfnamefont {J.}~\bibnamefont {Sun}}, \bibinfo
  {author} {\bibfnamefont {S.-S.}\ \bibnamefont {Gong}}, \bibinfo {author}
  {\bibfnamefont {W.}~\bibnamefont {Huang}}, \bibinfo {author} {\bibfnamefont
  {S.}~\bibnamefont {Feng}}, \bibinfo {author} {\bibfnamefont {R.~T.}\
  \bibnamefont {Scalettar}},\ and\ \bibinfo {author} {\bibfnamefont
  {H.}~\bibnamefont {Guo}},\ }\bibfield  {title} {\bibinfo {title} {{Rigorous
  demonstration of pair-density-wave superconductivity in the
  ${\ensuremath{\sigma}}_{z}$-Hubbard model}},\ }\href
  {https://doi.org/10.1103/PhysRevB.111.045158} {\bibfield  {journal} {\bibinfo
   {journal} {Phys. Rev. B}\ }\textbf {\bibinfo {volume} {111}},\ \bibinfo
  {pages} {045158} (\bibinfo {year} {2025})}\BibitemShut {NoStop}%
\bibitem [{\citenamefont {Krien}\ \emph {et~al.}(2021)\citenamefont {Krien},
  \citenamefont {Kauch},\ and\ \citenamefont {Held}}]{Krien2020b}%
  \BibitemOpen
  \bibfield  {author} {\bibinfo {author} {\bibfnamefont {F.}~\bibnamefont
  {Krien}}, \bibinfo {author} {\bibfnamefont {A.}~\bibnamefont {Kauch}},\ and\
  \bibinfo {author} {\bibfnamefont {K.}~\bibnamefont {Held}},\ }\bibfield
  {title} {\bibinfo {title} {{Tiling with triangles: parquet and
  $GW\ensuremath{\gamma}$ methods unified}},\ }\href
  {https://doi.org/10.1103/PhysRevResearch.3.013149} {\bibfield  {journal}
  {\bibinfo  {journal} {Phys. Rev. Research}\ }\textbf {\bibinfo {volume}
  {3}},\ \bibinfo {pages} {013149} (\bibinfo {year} {2021})}\BibitemShut
  {NoStop}%
\bibitem [{\citenamefont {Zhang}\ and\ \citenamefont
  {Imada}(2007)}]{Zhang2007}%
  \BibitemOpen
  \bibfield  {author} {\bibinfo {author} {\bibfnamefont {Y.~Z.}\ \bibnamefont
  {Zhang}}\ and\ \bibinfo {author} {\bibfnamefont {M.}~\bibnamefont {Imada}},\
  }\bibfield  {title} {\bibinfo {title} {Pseudogap and mott transition studied
  by cellular dynamical mean-field theory},\ }\href
  {https://doi.org/10.1103/PhysRevB.76.045108} {\bibfield  {journal} {\bibinfo
  {journal} {Phys. Rev. B}\ }\textbf {\bibinfo {volume} {76}},\ \bibinfo
  {pages} {045108} (\bibinfo {year} {2007})}\BibitemShut {NoStop}%
\bibitem [{\citenamefont {Ferrero}\ \emph {et~al.}(2009)\citenamefont
  {Ferrero}, \citenamefont {Cornaglia}, \citenamefont {De~Leo}, \citenamefont
  {Parcollet}, \citenamefont {Kotliar},\ and\ \citenamefont
  {Georges}}]{Ferrero2009}%
  \BibitemOpen
  \bibfield  {author} {\bibinfo {author} {\bibfnamefont {M.}~\bibnamefont
  {Ferrero}}, \bibinfo {author} {\bibfnamefont {P.~S.}\ \bibnamefont
  {Cornaglia}}, \bibinfo {author} {\bibfnamefont {L.}~\bibnamefont {De~Leo}},
  \bibinfo {author} {\bibfnamefont {O.}~\bibnamefont {Parcollet}}, \bibinfo
  {author} {\bibfnamefont {G.}~\bibnamefont {Kotliar}},\ and\ \bibinfo {author}
  {\bibfnamefont {A.}~\bibnamefont {Georges}},\ }\bibfield  {title} {\bibinfo
  {title} {{Pseudogap opening and formation of Fermi arcs as an
  orbital-selective Mott transition in momentum space}},\ }\href
  {https://doi.org/10.1103/PhysRevB.80.064501} {\bibfield  {journal} {\bibinfo
  {journal} {Phys. Rev. B}\ }\textbf {\bibinfo {volume} {80}},\ \bibinfo
  {pages} {064501} (\bibinfo {year} {2009})}\BibitemShut {NoStop}%
\bibitem [{\citenamefont {Bragan\ifmmode~\mbox{\c{c}}\else \c{c}\fi{}a}\ \emph
  {et~al.}(2018)\citenamefont {Bragan\ifmmode~\mbox{\c{c}}\else \c{c}\fi{}a},
  \citenamefont {Sakai}, \citenamefont {Aguiar},\ and\ \citenamefont
  {Civelli}}]{Braganca2018}%
  \BibitemOpen
  \bibfield  {author} {\bibinfo {author} {\bibfnamefont {H.}~\bibnamefont
  {Bragan\ifmmode~\mbox{\c{c}}\else \c{c}\fi{}a}}, \bibinfo {author}
  {\bibfnamefont {S.}~\bibnamefont {Sakai}}, \bibinfo {author} {\bibfnamefont
  {M.~C.~O.}\ \bibnamefont {Aguiar}},\ and\ \bibinfo {author} {\bibfnamefont
  {M.}~\bibnamefont {Civelli}},\ }\bibfield  {title} {\bibinfo {title}
  {{Correlation-Driven Lifshitz Transition at the Emergence of the Pseudogap
  Phase in the Two-Dimensional Hubbard Model}},\ }\href
  {https://doi.org/10.1103/PhysRevLett.120.067002} {\bibfield  {journal}
  {\bibinfo  {journal} {Phys. Rev. Lett.}\ }\textbf {\bibinfo {volume} {120}},\
  \bibinfo {pages} {067002} (\bibinfo {year} {2018})}\BibitemShut {NoStop}%
\bibitem [{\citenamefont {Moutenet}\ \emph {et~al.}(2018)\citenamefont
  {Moutenet}, \citenamefont {Wu},\ and\ \citenamefont
  {Ferrero}}]{Moutenet2018}%
  \BibitemOpen
  \bibfield  {author} {\bibinfo {author} {\bibfnamefont {A.}~\bibnamefont
  {Moutenet}}, \bibinfo {author} {\bibfnamefont {W.}~\bibnamefont {Wu}},\ and\
  \bibinfo {author} {\bibfnamefont {M.}~\bibnamefont {Ferrero}},\ }\bibfield
  {title} {\bibinfo {title} {{Determinant Monte Carlo algorithms for dynamical
  quantities in fermionic systems}},\ }\href
  {https://doi.org/10.1103/PhysRevB.97.085117} {\bibfield  {journal} {\bibinfo
  {journal} {Phys. Rev. B}\ }\textbf {\bibinfo {volume} {97}},\ \bibinfo
  {pages} {085117} (\bibinfo {year} {2018})}\BibitemShut {NoStop}%
\bibitem [{\citenamefont {Ortiz}\ \emph {et~al.}(2022)\citenamefont {Ortiz},
  \citenamefont {Puphal}, \citenamefont {Klett}, \citenamefont {Hotz},
  \citenamefont {Kremer}, \citenamefont {Trepka}, \citenamefont {Hemmida},
  \citenamefont {von Nidda}, \citenamefont {Isobe}, \citenamefont {Khasanov},
  \citenamefont {Luetkens}, \citenamefont {Hansmann}, \citenamefont {Keimer},
  \citenamefont {Sch\"afer},\ and\ \citenamefont {Hepting}}]{Ortiz2022}%
  \BibitemOpen
  \bibfield  {author} {\bibinfo {author} {\bibfnamefont {R.~A.}\ \bibnamefont
  {Ortiz}}, \bibinfo {author} {\bibfnamefont {P.}~\bibnamefont {Puphal}},
  \bibinfo {author} {\bibfnamefont {M.}~\bibnamefont {Klett}}, \bibinfo
  {author} {\bibfnamefont {F.}~\bibnamefont {Hotz}}, \bibinfo {author}
  {\bibfnamefont {R.~K.}\ \bibnamefont {Kremer}}, \bibinfo {author}
  {\bibfnamefont {H.}~\bibnamefont {Trepka}}, \bibinfo {author} {\bibfnamefont
  {M.}~\bibnamefont {Hemmida}}, \bibinfo {author} {\bibfnamefont {H.-A.~K.}\
  \bibnamefont {von Nidda}}, \bibinfo {author} {\bibfnamefont {M.}~\bibnamefont
  {Isobe}}, \bibinfo {author} {\bibfnamefont {R.}~\bibnamefont {Khasanov}},
  \bibinfo {author} {\bibfnamefont {H.}~\bibnamefont {Luetkens}}, \bibinfo
  {author} {\bibfnamefont {P.}~\bibnamefont {Hansmann}}, \bibinfo {author}
  {\bibfnamefont {B.}~\bibnamefont {Keimer}}, \bibinfo {author} {\bibfnamefont
  {T.}~\bibnamefont {Sch\"afer}},\ and\ \bibinfo {author} {\bibfnamefont
  {M.}~\bibnamefont {Hepting}},\ }\bibfield  {title} {\bibinfo {title}
  {{Magnetic correlations in infinite-layer nickelates: An experimental and
  theoretical multimethod study}},\ }\href
  {https://doi.org/10.1103/PhysRevResearch.4.023093} {\bibfield  {journal}
  {\bibinfo  {journal} {Phys. Rev. Res.}\ }\textbf {\bibinfo {volume} {4}},\
  \bibinfo {pages} {023093} (\bibinfo {year} {2022})}\BibitemShut {NoStop}%
\bibitem [{\citenamefont {Meixner}\ \emph {et~al.}(2024)\citenamefont
  {Meixner}, \citenamefont {Menke}, \citenamefont {Klett}, \citenamefont
  {Heinzelmann}, \citenamefont {Andergassen}, \citenamefont {Hansmann},\ and\
  \citenamefont {Schäfer}}]{Meixner2024}%
  \BibitemOpen
  \bibfield  {author} {\bibinfo {author} {\bibfnamefont {M.}~\bibnamefont
  {Meixner}}, \bibinfo {author} {\bibfnamefont {H.}~\bibnamefont {Menke}},
  \bibinfo {author} {\bibfnamefont {M.}~\bibnamefont {Klett}}, \bibinfo
  {author} {\bibfnamefont {S.}~\bibnamefont {Heinzelmann}}, \bibinfo {author}
  {\bibfnamefont {S.}~\bibnamefont {Andergassen}}, \bibinfo {author}
  {\bibfnamefont {P.}~\bibnamefont {Hansmann}},\ and\ \bibinfo {author}
  {\bibfnamefont {T.}~\bibnamefont {Schäfer}},\ }\bibfield  {title} {\bibinfo
  {title} {{Mott transition and pseudogap of the square-lattice Hubbard model:
  Results from center-focused cellular dynamical mean-field theory}},\ }\href
  {https://doi.org/10.21468/SciPostPhys.16.2.059} {\bibfield  {journal}
  {\bibinfo  {journal} {SciPost Phys.}\ }\textbf {\bibinfo {volume} {16}},\
  \bibinfo {pages} {059} (\bibinfo {year} {2024})}\BibitemShut {NoStop}%
\bibitem [{\citenamefont {Lichtenstein}\ and\ \citenamefont
  {Katsnelson}(2000)}]{Lichtenstein2000}%
  \BibitemOpen
  \bibfield  {author} {\bibinfo {author} {\bibfnamefont {A.~I.}\ \bibnamefont
  {Lichtenstein}}\ and\ \bibinfo {author} {\bibfnamefont {M.~I.}\ \bibnamefont
  {Katsnelson}},\ }\bibfield  {title} {\bibinfo {title} {{Antiferromagnetism
  and \textit{d}-wave superconductivity in cuprates: A cluster dynamical
  mean-field theory}},\ }\href {https://doi.org/10.1103/PhysRevB.62.R9283}
  {\bibfield  {journal} {\bibinfo  {journal} {Phys. Rev. B}\ }\textbf {\bibinfo
  {volume} {62}},\ \bibinfo {pages} {R9283} (\bibinfo {year}
  {2000})}\BibitemShut {NoStop}%
\bibitem [{\citenamefont {Haule}\ and\ \citenamefont
  {Kotliar}(2007)}]{Haule2007b}%
  \BibitemOpen
  \bibfield  {author} {\bibinfo {author} {\bibfnamefont {K.}~\bibnamefont
  {Haule}}\ and\ \bibinfo {author} {\bibfnamefont {G.}~\bibnamefont
  {Kotliar}},\ }\bibfield  {title} {\bibinfo {title} {{Strongly correlated
  superconductivity: A plaquette dynamical mean-field theory study}},\ }\href
  {https://doi.org/10.1103/PhysRevB.76.104509} {\bibfield  {journal} {\bibinfo
  {journal} {Phys. Rev. B}\ }\textbf {\bibinfo {volume} {76}},\ \bibinfo
  {pages} {104509} (\bibinfo {year} {2007})}\BibitemShut {NoStop}%
\bibitem [{\citenamefont {Tremblay}(2013)}]{Tremblay2013}%
  \BibitemOpen
  \bibfield  {author} {\bibinfo {author} {\bibfnamefont {A.-M.}\ \bibnamefont
  {Tremblay}},\ }\bibfield  {title} {\bibinfo {title} {{Strongly Correlated
  Superconductivity}},\ }in\ \href
  {https://www.cond-mat.de/events/correl13/manuscripts/tremblay.pdf} {\emph
  {\bibinfo {booktitle} {Emergent Phenomena in Correlated Matter: Modeling and
  Simulation}}},\ Vol.~\bibinfo {volume} {3}\ (\bibinfo  {publisher}
  {Forschungszentrum Jülich},\ \bibinfo {year} {2013})\ Chap.~\bibinfo
  {chapter} {10}\BibitemShut {NoStop}%
\bibitem [{\citenamefont {Fratino}\ \emph {et~al.}(2017)\citenamefont
  {Fratino}, \citenamefont {S\'emon}, \citenamefont {Charlebois}, \citenamefont
  {Sordi},\ and\ \citenamefont {Tremblay}}]{Fratino2017}%
  \BibitemOpen
  \bibfield  {author} {\bibinfo {author} {\bibfnamefont {L.}~\bibnamefont
  {Fratino}}, \bibinfo {author} {\bibfnamefont {P.}~\bibnamefont {S\'emon}},
  \bibinfo {author} {\bibfnamefont {M.}~\bibnamefont {Charlebois}}, \bibinfo
  {author} {\bibfnamefont {G.}~\bibnamefont {Sordi}},\ and\ \bibinfo {author}
  {\bibfnamefont {A.-M.~S.}\ \bibnamefont {Tremblay}},\ }\bibfield  {title}
  {\bibinfo {title} {{Signatures of the Mott transition in the
  antiferromagnetic state of the two-dimensional Hubbard model}},\ }\href
  {https://doi.org/10.1103/PhysRevB.95.235109} {\bibfield  {journal} {\bibinfo
  {journal} {Phys. Rev. B}\ }\textbf {\bibinfo {volume} {95}},\ \bibinfo
  {pages} {235109} (\bibinfo {year} {2017})}\BibitemShut {NoStop}%
\bibitem [{\citenamefont {Harland}\ \emph {et~al.}(2019)\citenamefont
  {Harland}, \citenamefont {Brener}, \citenamefont {Lichtenstein},\ and\
  \citenamefont {Katsnelson}}]{Harland2019}%
  \BibitemOpen
  \bibfield  {author} {\bibinfo {author} {\bibfnamefont {M.}~\bibnamefont
  {Harland}}, \bibinfo {author} {\bibfnamefont {S.}~\bibnamefont {Brener}},
  \bibinfo {author} {\bibfnamefont {A.~I.}\ \bibnamefont {Lichtenstein}},\ and\
  \bibinfo {author} {\bibfnamefont {M.~I.}\ \bibnamefont {Katsnelson}},\
  }\bibfield  {title} {\bibinfo {title} {Josephson lattice model for phase
  fluctuations of local pairs in copper oxide superconductors},\ }\href
  {https://doi.org/10.1103/PhysRevB.100.024510} {\bibfield  {journal} {\bibinfo
   {journal} {Phys. Rev. B}\ }\textbf {\bibinfo {volume} {100}},\ \bibinfo
  {pages} {024510} (\bibinfo {year} {2019})}\BibitemShut {NoStop}%
\bibitem [{\citenamefont {Walsh}\ \emph {et~al.}(2023)\citenamefont {Walsh},
  \citenamefont {Charlebois}, \citenamefont {S\'emon}, \citenamefont
  {Tremblay},\ and\ \citenamefont {Sordi}}]{Walsh2023}%
  \BibitemOpen
  \bibfield  {author} {\bibinfo {author} {\bibfnamefont {C.}~\bibnamefont
  {Walsh}}, \bibinfo {author} {\bibfnamefont {M.}~\bibnamefont {Charlebois}},
  \bibinfo {author} {\bibfnamefont {P.}~\bibnamefont {S\'emon}}, \bibinfo
  {author} {\bibfnamefont {A.-M.~S.}\ \bibnamefont {Tremblay}},\ and\ \bibinfo
  {author} {\bibfnamefont {G.}~\bibnamefont {Sordi}},\ }\bibfield  {title}
  {\bibinfo {title} {{Superconductivity in the two-dimensional Hubbard model
  with cellular dynamical mean-field theory: A quantum impurity model
  analysis}},\ }\href {https://doi.org/10.1103/PhysRevB.108.075163} {\bibfield
  {journal} {\bibinfo  {journal} {Phys. Rev. B}\ }\textbf {\bibinfo {volume}
  {108}},\ \bibinfo {pages} {075163} (\bibinfo {year} {2023})}\BibitemShut
  {NoStop}%
\bibitem [{\citenamefont {Sakai}(2023)}]{Sakai2023}%
  \BibitemOpen
  \bibfield  {author} {\bibinfo {author} {\bibfnamefont {S.}~\bibnamefont
  {Sakai}},\ }\bibfield  {title} {\bibinfo {title} {{Nonperturbative
  Calculations for Spectroscopic Properties of Cuprate High-Temperature
  Superconductors}},\ }\href {https://doi.org/10.7566/JPSJ.92.092001}
  {\bibfield  {journal} {\bibinfo  {journal} {Journal of the Physical Society
  of Japan}\ }\textbf {\bibinfo {volume} {92}},\ \bibinfo {pages} {092001}
  (\bibinfo {year} {2023})}\BibitemShut {NoStop}%
\bibitem [{\citenamefont {Verret}\ \emph {et~al.}(2019)\citenamefont {Verret},
  \citenamefont {Roy}, \citenamefont {Foley}, \citenamefont {Charlebois},
  \citenamefont {S\'en\'echal},\ and\ \citenamefont {Tremblay}}]{Verret2019}%
  \BibitemOpen
  \bibfield  {author} {\bibinfo {author} {\bibfnamefont {S.}~\bibnamefont
  {Verret}}, \bibinfo {author} {\bibfnamefont {J.}~\bibnamefont {Roy}},
  \bibinfo {author} {\bibfnamefont {A.}~\bibnamefont {Foley}}, \bibinfo
  {author} {\bibfnamefont {M.}~\bibnamefont {Charlebois}}, \bibinfo {author}
  {\bibfnamefont {D.}~\bibnamefont {S\'en\'echal}},\ and\ \bibinfo {author}
  {\bibfnamefont {A.-M.~S.}\ \bibnamefont {Tremblay}},\ }\bibfield  {title}
  {\bibinfo {title} {{Intrinsic cluster-shaped density waves in cellular
  dynamical mean-field theory}},\ }\href
  {https://doi.org/10.1103/PhysRevB.100.224520} {\bibfield  {journal} {\bibinfo
   {journal} {Phys. Rev. B}\ }\textbf {\bibinfo {volume} {100}},\ \bibinfo
  {pages} {224520} (\bibinfo {year} {2019})}\BibitemShut {NoStop}%
\bibitem [{\citenamefont {Metzner}\ and\ \citenamefont
  {Vollhardt}(1989)}]{Metzner1989}%
  \BibitemOpen
  \bibfield  {author} {\bibinfo {author} {\bibfnamefont {W.}~\bibnamefont
  {Metzner}}\ and\ \bibinfo {author} {\bibfnamefont {D.}~\bibnamefont
  {Vollhardt}},\ }\bibfield  {title} {\bibinfo {title} {{Correlated Lattice
  Fermions in $d=\infty$ Dimensions}},\ }\href
  {https://doi.org/10.1103/PhysRevLett.62.324} {\bibfield  {journal} {\bibinfo
  {journal} {Phys. Rev. Lett.}\ }\textbf {\bibinfo {volume} {62}},\ \bibinfo
  {pages} {324} (\bibinfo {year} {1989})}\BibitemShut {NoStop}%
\bibitem [{\citenamefont {Lichtenstein}\ \emph {et~al.}(2001)\citenamefont
  {Lichtenstein}, \citenamefont {Katsnelson},\ and\ \citenamefont
  {Kotliar}}]{Lichtenstein2001}%
  \BibitemOpen
  \bibfield  {author} {\bibinfo {author} {\bibfnamefont {A.~I.}\ \bibnamefont
  {Lichtenstein}}, \bibinfo {author} {\bibfnamefont {M.~I.}\ \bibnamefont
  {Katsnelson}},\ and\ \bibinfo {author} {\bibfnamefont {G.}~\bibnamefont
  {Kotliar}},\ }\bibfield  {title} {\bibinfo {title} {Finite-temperature
  magnetism of transition metals: An {\sl ab initio} dynamical mean-field
  theory},\ }\href@noop {} {\bibfield  {journal} {\bibinfo  {journal} {Phys.
  Rev. Lett.}\ }\textbf {\bibinfo {volume} {87}},\ \bibinfo {pages} {067205}
  (\bibinfo {year} {2001})}\BibitemShut {NoStop}%
\bibitem [{\citenamefont {Bolech}\ \emph {et~al.}(2003)\citenamefont {Bolech},
  \citenamefont {Kancharla},\ and\ \citenamefont {Kotliar}}]{Bolech03}%
  \BibitemOpen
  \bibfield  {author} {\bibinfo {author} {\bibfnamefont {C.~J.}\ \bibnamefont
  {Bolech}}, \bibinfo {author} {\bibfnamefont {S.~S.}\ \bibnamefont
  {Kancharla}},\ and\ \bibinfo {author} {\bibfnamefont {G.}~\bibnamefont
  {Kotliar}},\ }\bibfield  {title} {\bibinfo {title} {Cellular dynamical
  mean-field theory for the one-dimensional extended hubbard model},\ }\href
  {https://doi.org/10.1103/PhysRevB.67.075110} {\bibfield  {journal} {\bibinfo
  {journal} {Phys. Rev. B}\ }\textbf {\bibinfo {volume} {67}},\ \bibinfo
  {pages} {075110} (\bibinfo {year} {2003})}\BibitemShut {NoStop}%
\bibitem [{\citenamefont {Maier}\ \emph {et~al.}(2005)\citenamefont {Maier},
  \citenamefont {Jarrell}, \citenamefont {Pruschke},\ and\ \citenamefont
  {Hettler}}]{Maier2005}%
  \BibitemOpen
  \bibfield  {author} {\bibinfo {author} {\bibfnamefont {T.~A.}\ \bibnamefont
  {Maier}}, \bibinfo {author} {\bibfnamefont {M.}~\bibnamefont {Jarrell}},
  \bibinfo {author} {\bibfnamefont {T.}~\bibnamefont {Pruschke}},\ and\
  \bibinfo {author} {\bibfnamefont {M.}~\bibnamefont {Hettler}},\ }\bibfield
  {title} {\bibinfo {title} {{Quantum Cluster Theories}},\ }\href
  {https://doi.org/10.1103/RevModPhys.77.1027} {\bibfield  {journal} {\bibinfo
  {journal} {Rev. Mod. Phys.}\ }\textbf {\bibinfo {volume} {77}},\ \bibinfo
  {pages} {1027} (\bibinfo {year} {2005})}\BibitemShut {NoStop}%
\bibitem [{\citenamefont {Georges}\ \emph {et~al.}(1996)\citenamefont
  {Georges}, \citenamefont {Kotliar}, \citenamefont {Krauth},\ and\
  \citenamefont {Rozenberg}}]{Georges1996}%
  \BibitemOpen
  \bibfield  {author} {\bibinfo {author} {\bibfnamefont {A.}~\bibnamefont
  {Georges}}, \bibinfo {author} {\bibfnamefont {G.}~\bibnamefont {Kotliar}},
  \bibinfo {author} {\bibfnamefont {W.}~\bibnamefont {Krauth}},\ and\ \bibinfo
  {author} {\bibfnamefont {M.~J.}\ \bibnamefont {Rozenberg}},\ }\bibfield
  {title} {\bibinfo {title} {{Dynamical mean-field theory of strongly
  correlated fermion systems and the limit of infinite dimensions}},\ }\href
  {https://doi.org/10.1103/RevModPhys.68.13} {\bibfield  {journal} {\bibinfo
  {journal} {Rev. Mod. Phys.}\ }\textbf {\bibinfo {volume} {68}},\ \bibinfo
  {pages} {13} (\bibinfo {year} {1996})}\BibitemShut {NoStop}%
\bibitem [{\citenamefont {Potthoff}(2018)}]{Potthoff2018}%
  \BibitemOpen
  \bibfield  {author} {\bibinfo {author} {\bibfnamefont {M.}~\bibnamefont
  {Potthoff}},\ }\bibfield  {title} {\bibinfo {title} {Cluster extensions of
  dynamical mean-field theory},\ }in\ \href {http://hdl.handle.net/2128/19720}
  {\emph {\bibinfo {booktitle} {{DMFT}:{F}rom {I}nfinite {D}imensions to {R}eal
  {M}aterials}}},\ Vol.~\bibinfo {volume} {8}\ (\bibinfo  {publisher}
  {Forschungszentrum Jülich},\ \bibinfo {year} {2018})\ Chap.~\bibinfo
  {chapter} {5}\BibitemShut {NoStop}%
\bibitem [{\citenamefont {Rubtsov}\ \emph {et~al.}(2005)\citenamefont
  {Rubtsov}, \citenamefont {Savkin},\ and\ \citenamefont
  {Lichtenstein}}]{Rubtsov2005}%
  \BibitemOpen
  \bibfield  {author} {\bibinfo {author} {\bibfnamefont {A.~N.}\ \bibnamefont
  {Rubtsov}}, \bibinfo {author} {\bibfnamefont {V.~V.}\ \bibnamefont
  {Savkin}},\ and\ \bibinfo {author} {\bibfnamefont {A.~I.}\ \bibnamefont
  {Lichtenstein}},\ }\bibfield  {title} {\bibinfo {title} {{Continuous-time
  quantum Monte Carlo method for fermions}},\ }\href
  {https://doi.org/10.1103/PhysRevB.72.035122} {\bibfield  {journal} {\bibinfo
  {journal} {Phys. Rev. B}\ }\textbf {\bibinfo {volume} {72}},\ \bibinfo
  {pages} {035122} (\bibinfo {year} {2005})}\BibitemShut {NoStop}%
\bibitem [{\citenamefont {Gull}\ \emph {et~al.}(2008)\citenamefont {Gull},
  \citenamefont {Werner}, \citenamefont {Parcollet},\ and\ \citenamefont
  {Troyer}}]{Gull2008a}%
  \BibitemOpen
  \bibfield  {author} {\bibinfo {author} {\bibfnamefont {E.}~\bibnamefont
  {Gull}}, \bibinfo {author} {\bibfnamefont {P.}~\bibnamefont {Werner}},
  \bibinfo {author} {\bibfnamefont {O.}~\bibnamefont {Parcollet}},\ and\
  \bibinfo {author} {\bibfnamefont {M.}~\bibnamefont {Troyer}},\ }\bibfield
  {title} {\bibinfo {title} {{Continuous-time auxiliary-field Monte Carlo for
  quantum impurity models}},\ }\href
  {http://stacks.iop.org/0295-5075/82/i=5/a=57003} {\bibfield  {journal}
  {\bibinfo  {journal} {EPL (Europhysics Letters)}\ }\textbf {\bibinfo {volume}
  {82}},\ \bibinfo {pages} {57003} (\bibinfo {year} {2008})}\BibitemShut
  {NoStop}%
\bibitem [{\citenamefont {Parcollet}\ \emph {et~al.}(2015)\citenamefont
  {Parcollet}, \citenamefont {Ferrero}, \citenamefont {Ayral}, \citenamefont
  {Hafermann}, \citenamefont {Krivenko}, \citenamefont {Messio},\ and\
  \citenamefont {Seth}}]{TRIQS}%
  \BibitemOpen
  \bibfield  {author} {\bibinfo {author} {\bibfnamefont {O.}~\bibnamefont
  {Parcollet}}, \bibinfo {author} {\bibfnamefont {M.}~\bibnamefont {Ferrero}},
  \bibinfo {author} {\bibfnamefont {T.}~\bibnamefont {Ayral}}, \bibinfo
  {author} {\bibfnamefont {H.}~\bibnamefont {Hafermann}}, \bibinfo {author}
  {\bibfnamefont {I.}~\bibnamefont {Krivenko}}, \bibinfo {author}
  {\bibfnamefont {L.}~\bibnamefont {Messio}},\ and\ \bibinfo {author}
  {\bibfnamefont {P.}~\bibnamefont {Seth}},\ }\bibfield  {title} {\bibinfo
  {title} {Triqs: A toolbox for research on interacting quantum systems},\
  }\href {https://doi.org/https://doi.org/10.1016/j.cpc.2015.04.023} {\bibfield
   {journal} {\bibinfo  {journal} {Computer Physics Communications}\ }\textbf
  {\bibinfo {volume} {196}},\ \bibinfo {pages} {398 } (\bibinfo {year}
  {2015})}\BibitemShut {NoStop}%
\bibitem [{\citenamefont {Hettler}\ \emph {et~al.}(2000)\citenamefont
  {Hettler}, \citenamefont {Mukherjee}, \citenamefont {Jarrell},\ and\
  \citenamefont {Krishnamurthy}}]{Hettler2000}%
  \BibitemOpen
  \bibfield  {author} {\bibinfo {author} {\bibfnamefont {M.~H.}\ \bibnamefont
  {Hettler}}, \bibinfo {author} {\bibfnamefont {M.}~\bibnamefont {Mukherjee}},
  \bibinfo {author} {\bibfnamefont {M.}~\bibnamefont {Jarrell}},\ and\ \bibinfo
  {author} {\bibfnamefont {H.~R.}\ \bibnamefont {Krishnamurthy}},\ }\bibfield
  {title} {\bibinfo {title} {Dynamical cluster approximation: Nonlocal dynamics
  of correlated electron systems},\ }\href
  {https://doi.org/10.1103/PhysRevB.61.12739} {\bibfield  {journal} {\bibinfo
  {journal} {Phys. Rev. B}\ }\textbf {\bibinfo {volume} {61}},\ \bibinfo
  {pages} {12739} (\bibinfo {year} {2000})}\BibitemShut {NoStop}%
\bibitem [{\citenamefont {Potthoff}(2003)}]{Potthoff2003}%
  \BibitemOpen
  \bibfield  {author} {\bibinfo {author} {\bibfnamefont {M.}~\bibnamefont
  {Potthoff}},\ }\bibfield  {title} {\bibinfo {title} {Self-energy-functional
  approach to systems of correlated electrons},\ }\href
  {https://doi.org/10.1140/epjb/e2003-00121-8} {\bibfield  {journal} {\bibinfo
  {journal} {Eur. Phys. J. B}\ }\textbf {\bibinfo {volume} {32}},\ \bibinfo
  {pages} {429} (\bibinfo {year} {2003})}\BibitemShut {NoStop}%
\bibitem [{\citenamefont {David~S\'en\'echal}(2004)}]{Senechal2004Book}%
  \BibitemOpen
  \bibfield  {author} {\bibinfo {author} {\bibfnamefont {C.~B.}\ \bibnamefont
  {David~S\'en\'echal}, \bibfnamefont {Andr\'e-Marie~Tremblay}},\ }\href@noop
  {} {\emph {\bibinfo {title} {Theoretical Methods for Strongly Correlated
  Electrons}}}\ (\bibinfo  {publisher} {Springer, New York},\ \bibinfo {year}
  {2004})\BibitemShut {NoStop}%
\bibitem [{\citenamefont {Bickers}(2004)}]{Bickers04}%
  \BibitemOpen
  \bibfield  {author} {\bibinfo {author} {\bibfnamefont {N.~E.}\ \bibnamefont
  {Bickers}},\ }\bibfield  {title} {\bibinfo {title} {Self-consistent many-body
  theory for condensed matter systems},\ }in\ \href
  {https://link.springer.com/chapter/10.1007/0-387-21717-7_6} {\emph {\bibinfo
  {booktitle} {Theoretical Methods for Strongly Correlated Electrons. CRM
  Series in Mathematical Physics}}},\ \bibinfo {editor} {edited by\ \bibinfo
  {editor} {\bibfnamefont {D.}~\bibnamefont {S\'en\'echal}}, \bibinfo {editor}
  {\bibfnamefont {A.-M.}\ \bibnamefont {Tremblay}},\ and\ \bibinfo {editor}
  {\bibfnamefont {C.}~\bibnamefont {Bourbonnais}}}\ (\bibinfo  {publisher}
  {Springer},\ \bibinfo {year} {2004})\BibitemShut {NoStop}%
\bibitem [{\citenamefont {Rohringer}\ \emph {et~al.}(2012)\citenamefont
  {Rohringer}, \citenamefont {Valli},\ and\ \citenamefont
  {Toschi}}]{Rohringer2012}%
  \BibitemOpen
  \bibfield  {author} {\bibinfo {author} {\bibfnamefont {G.}~\bibnamefont
  {Rohringer}}, \bibinfo {author} {\bibfnamefont {A.}~\bibnamefont {Valli}},\
  and\ \bibinfo {author} {\bibfnamefont {A.}~\bibnamefont {Toschi}},\
  }\bibfield  {title} {\bibinfo {title} {Local electronic correlation at the
  two-particle level},\ }\href {https://doi.org/10.1103/PhysRevB.86.125114}
  {\bibfield  {journal} {\bibinfo  {journal} {Phys. Rev. B}\ }\textbf {\bibinfo
  {volume} {86}},\ \bibinfo {pages} {125114} (\bibinfo {year}
  {2012})}\BibitemShut {NoStop}%
\bibitem [{\citenamefont {Jarrell}(1992)}]{Jarrell1992}%
  \BibitemOpen
  \bibfield  {author} {\bibinfo {author} {\bibfnamefont {M.}~\bibnamefont
  {Jarrell}},\ }\bibfield  {title} {\bibinfo {title} {{Hubbard model in
  infinite dimensions: A quantum Monte Carlo study}},\ }\href
  {https://doi.org/10.1103/PhysRevLett.69.168} {\bibfield  {journal} {\bibinfo
  {journal} {Phys. Rev. Lett.}\ }\textbf {\bibinfo {volume} {69}},\ \bibinfo
  {pages} {168} (\bibinfo {year} {1992})}\BibitemShut {NoStop}%
\bibitem [{\citenamefont {Craven}(1969)}]{Craven1969}%
  \BibitemOpen
  \bibfield  {author} {\bibinfo {author} {\bibfnamefont {B.~D.}\ \bibnamefont
  {Craven}},\ }\bibfield  {title} {\bibinfo {title} {Complex symmetric
  matrices},\ }\href {https://doi.org/10.1017/S1446788700007588} {\bibfield
  {journal} {\bibinfo  {journal} {Journal of the Australian Mathematical
  Society}\ }\textbf {\bibinfo {volume} {10}},\ \bibinfo {pages} {341–354}
  (\bibinfo {year} {1969})}\BibitemShut {NoStop}%
\bibitem [{\citenamefont {Reitner}\ \emph {et~al.}(2024)\citenamefont
  {Reitner}, \citenamefont {Crippa}, \citenamefont {Fus}, \citenamefont
  {Budich}, \citenamefont {Toschi},\ and\ \citenamefont
  {Sangiovanni}}]{Reitner2024}%
  \BibitemOpen
  \bibfield  {author} {\bibinfo {author} {\bibfnamefont {M.}~\bibnamefont
  {Reitner}}, \bibinfo {author} {\bibfnamefont {L.}~\bibnamefont {Crippa}},
  \bibinfo {author} {\bibfnamefont {D.~R.}\ \bibnamefont {Fus}}, \bibinfo
  {author} {\bibfnamefont {J.~C.}\ \bibnamefont {Budich}}, \bibinfo {author}
  {\bibfnamefont {A.}~\bibnamefont {Toschi}},\ and\ \bibinfo {author}
  {\bibfnamefont {G.}~\bibnamefont {Sangiovanni}},\ }\bibfield  {title}
  {\bibinfo {title} {{Protection of correlation-induced phase instabilities by
  exceptional susceptibilities}},\ }\href
  {https://doi.org/10.1103/PhysRevResearch.6.L022031} {\bibfield  {journal}
  {\bibinfo  {journal} {Phys. Rev. Res.}\ }\textbf {\bibinfo {volume} {6}},\
  \bibinfo {pages} {L022031} (\bibinfo {year} {2024})}\BibitemShut {NoStop}%
\bibitem [{\citenamefont {van Loon}\ \emph {et~al.}(2020)\citenamefont {van
  Loon}, \citenamefont {Krien},\ and\ \citenamefont {Katanin}}]{vanLoon2020}%
  \BibitemOpen
  \bibfield  {author} {\bibinfo {author} {\bibfnamefont {E.~G. C.~P.}\
  \bibnamefont {van Loon}}, \bibinfo {author} {\bibfnamefont {F.}~\bibnamefont
  {Krien}},\ and\ \bibinfo {author} {\bibfnamefont {A.~A.}\ \bibnamefont
  {Katanin}},\ }\bibfield  {title} {\bibinfo {title} {Bethe-salpeter equation
  at the critical end point of the mott transition},\ }\href
  {https://doi.org/10.1103/PhysRevLett.125.136402} {\bibfield  {journal}
  {\bibinfo  {journal} {Phys. Rev. Lett.}\ }\textbf {\bibinfo {volume} {125}},\
  \bibinfo {pages} {136402} (\bibinfo {year} {2020})}\BibitemShut {NoStop}%
\bibitem [{\citenamefont {Nagaoka}(1966)}]{Nagaoka1966}%
  \BibitemOpen
  \bibfield  {author} {\bibinfo {author} {\bibfnamefont {Y.}~\bibnamefont
  {Nagaoka}},\ }\bibfield  {title} {\bibinfo {title} {Ferromagnetism in a
  narrow, almost half-filled $s$ band},\ }\href
  {https://doi.org/10.1103/PhysRev.147.392} {\bibfield  {journal} {\bibinfo
  {journal} {Phys. Rev.}\ }\textbf {\bibinfo {volume} {147}},\ \bibinfo {pages}
  {392} (\bibinfo {year} {1966})}\BibitemShut {NoStop}%
\bibitem [{\citenamefont {Obermeier}\ \emph {et~al.}(1997)\citenamefont
  {Obermeier}, \citenamefont {Pruschke},\ and\ \citenamefont
  {Keller}}]{Obermeier1997}%
  \BibitemOpen
  \bibfield  {author} {\bibinfo {author} {\bibfnamefont {T.}~\bibnamefont
  {Obermeier}}, \bibinfo {author} {\bibfnamefont {T.}~\bibnamefont
  {Pruschke}},\ and\ \bibinfo {author} {\bibfnamefont {J.}~\bibnamefont
  {Keller}},\ }\bibfield  {title} {\bibinfo {title} {Ferromagnetism in the
  large-$u$ hubbard model},\ }\href {https://doi.org/10.1103/PhysRevB.56.R8479}
  {\bibfield  {journal} {\bibinfo  {journal} {Phys. Rev. B}\ }\textbf {\bibinfo
  {volume} {56}},\ \bibinfo {pages} {R8479(R)} (\bibinfo {year}
  {1997})}\BibitemShut {NoStop}%
\bibitem [{\citenamefont {Park}\ \emph
  {et~al.}(2008{\natexlab{b}})\citenamefont {Park}, \citenamefont {Haule},
  \citenamefont {Marianetti},\ and\ \citenamefont {Kotliar}}]{Park2008a}%
  \BibitemOpen
  \bibfield  {author} {\bibinfo {author} {\bibfnamefont {H.}~\bibnamefont
  {Park}}, \bibinfo {author} {\bibfnamefont {K.}~\bibnamefont {Haule}},
  \bibinfo {author} {\bibfnamefont {C.~A.}\ \bibnamefont {Marianetti}},\ and\
  \bibinfo {author} {\bibfnamefont {G.}~\bibnamefont {Kotliar}},\ }\bibfield
  {title} {\bibinfo {title} {{Dynamical mean-field theory study of Nagaoka
  ferromagnetism}},\ }\href {https://doi.org/10.1103/PhysRevB.77.035107}
  {\bibfield  {journal} {\bibinfo  {journal} {Phys. Rev. B}\ }\textbf {\bibinfo
  {volume} {77}},\ \bibinfo {pages} {035107} (\bibinfo {year}
  {2008}{\natexlab{b}})}\BibitemShut {NoStop}%
\bibitem [{\citenamefont {Kamogawa}\ \emph {et~al.}(2019)\citenamefont
  {Kamogawa}, \citenamefont {Nasu},\ and\ \citenamefont {Koga}}]{Kamogawa2019}%
  \BibitemOpen
  \bibfield  {author} {\bibinfo {author} {\bibfnamefont {Y.}~\bibnamefont
  {Kamogawa}}, \bibinfo {author} {\bibfnamefont {J.}~\bibnamefont {Nasu}},\
  and\ \bibinfo {author} {\bibfnamefont {A.}~\bibnamefont {Koga}},\ }\bibfield
  {title} {\bibinfo {title} {{Ferromagnetic instability for the single-band
  Hubbard model in the strong-coupling regime}},\ }\href
  {https://doi.org/10.1103/PhysRevB.99.235107} {\bibfield  {journal} {\bibinfo
  {journal} {Phys. Rev. B}\ }\textbf {\bibinfo {volume} {99}},\ \bibinfo
  {pages} {235107} (\bibinfo {year} {2019})}\BibitemShut {NoStop}%
\bibitem [{\citenamefont {Li}\ \emph {et~al.}(2024)\citenamefont {Li},
  \citenamefont {Liu},\ and\ \citenamefont {Bryant}}]{Li2024}%
  \BibitemOpen
  \bibfield  {author} {\bibinfo {author} {\bibfnamefont {Y.}~\bibnamefont
  {Li}}, \bibinfo {author} {\bibfnamefont {K.}~\bibnamefont {Liu}},\ and\
  \bibinfo {author} {\bibfnamefont {G.~W.}\ \bibnamefont {Bryant}},\ }\bibfield
   {title} {\bibinfo {title} {Nagaoka ferromagnetism in
  $3\ifmmode\times\else\texttimes\fi{}3$ arrays and beyond},\ }\href
  {https://doi.org/10.1103/PhysRevB.110.245141} {\bibfield  {journal} {\bibinfo
   {journal} {Phys. Rev. B}\ }\textbf {\bibinfo {volume} {110}},\ \bibinfo
  {pages} {245141} (\bibinfo {year} {2024})}\BibitemShut {NoStop}%
\end{thebibliography}%


\begin{thebibliography}{24}%
\makeatletter
\providecommand \@ifxundefined [1]{%
 \@ifx{#1\undefined}
}%
\providecommand \@ifnum [1]{%
 \ifnum #1\expandafter \@firstoftwo
 \else \expandafter \@secondoftwo
 \fi
}%
\providecommand \@ifx [1]{%
 \ifx #1\expandafter \@firstoftwo
 \else \expandafter \@secondoftwo
 \fi
}%
\providecommand \natexlab [1]{#1}%
\providecommand \enquote  [1]{``#1''}%
\providecommand \bibnamefont  [1]{#1}%
\providecommand \bibfnamefont [1]{#1}%
\providecommand \citenamefont [1]{#1}%
\providecommand \href@noop [0]{\@secondoftwo}%
\providecommand \href [0]{\begingroup \@sanitize@url \@href}%
\providecommand \@href[1]{\@@startlink{#1}\@@href}%
\providecommand \@@href[1]{\endgroup#1\@@endlink}%
\providecommand \@sanitize@url [0]{\catcode `\\12\catcode `\$12\catcode
  `\&12\catcode `\#12\catcode `\^12\catcode `\_12\catcode `\%12\relax}%
\providecommand \@@startlink[1]{}%
\providecommand \@@endlink[0]{}%
\providecommand \url  [0]{\begingroup\@sanitize@url \@url }%
\providecommand \@url [1]{\endgroup\@href {#1}{\urlprefix }}%
\providecommand \urlprefix  [0]{URL }%
\providecommand \Eprint [0]{\href }%
\providecommand \doibase [0]{https://doi.org/}%
\providecommand \selectlanguage [0]{\@gobble}%
\providecommand \bibinfo  [0]{\@secondoftwo}%
\providecommand \bibfield  [0]{\@secondoftwo}%
\providecommand \translation [1]{[#1]}%
\providecommand \BibitemOpen [0]{}%
\providecommand \bibitemStop [0]{}%
\providecommand \bibitemNoStop [0]{.\EOS\space}%
\providecommand \EOS [0]{\spacefactor3000\relax}%
\providecommand \BibitemShut  [1]{\csname bibitem#1\endcsname}%
\let\auto@bib@innerbib\@empty
%</preamble>
\bibitem [{\citenamefont {Rohringer}(2013)}]{Rohringer2013a}%
  \BibitemOpen
  \bibfield  {author} {\bibinfo {author} {\bibfnamefont {G.}~\bibnamefont
  {Rohringer}},\ }\emph {\bibinfo {title} {New routes towards a theoretical
  treatment of nonlocal electronic correlations}},\ \href
  {http://digital.obvsg.at/download/pdf/1631831} {Ph.D. thesis},\ \bibinfo
  {school} {Vienna University of Technology} (\bibinfo {year}
  {2013})\BibitemShut {NoStop}%
\bibitem [{\citenamefont {Shiba}(1972)}]{Shiba1972}%
  \BibitemOpen
  \bibfield  {author} {\bibinfo {author} {\bibfnamefont {H.}~\bibnamefont
  {Shiba}},\ }\bibfield  {title} {\bibinfo {title} {Thermodynamic properties of
  the one-dimensional half-filled-band hubbard model. ii},\ }\href@noop {}
  {\bibfield  {journal} {\bibinfo  {journal} {Prog. Theor. Phys.}\ }\textbf
  {\bibinfo {volume} {48}},\ \bibinfo {pages} {2171} (\bibinfo {year}
  {1972})}\BibitemShut {NoStop}%
\bibitem [{\citenamefont {Carmelo}(2026)}]{Carmelo2026}%
  \BibitemOpen
  \bibfield  {author} {\bibinfo {author} {\bibfnamefont {J.~M.~P.}\
  \bibnamefont {Carmelo}},\ }\bibfield  {title} {\bibinfo {title} {{Exact
  results for the Hubbard model on bipartite lattices in spatial dimensions
  $d>1$: Seven theorems from the full
  $[\mathrm{SU}(2)\ifmmode\times\else\texttimes\fi{}\mathrm{SU}(2)\ifmmode\times\else\texttimes\fi{}\mathrm{U}(1)]/{\mathbb{Z}}_{2}^{2}$
  symmetry}},\ }\href {https://doi.org/10.1103/p1sc-sx2d} {\bibfield  {journal}
  {\bibinfo  {journal} {Phys. Rev. B}\ }\textbf {\bibinfo {volume} {113}},\
  \bibinfo {pages} {155157} (\bibinfo {year} {2026})}\BibitemShut {NoStop}%
\bibitem [{\citenamefont {Yang}\ and\ \citenamefont {Zhang}(1990)}]{Yang1990}%
  \BibitemOpen
  \bibfield  {author} {\bibinfo {author} {\bibfnamefont {C.}~\bibnamefont
  {Yang}}\ and\ \bibinfo {author} {\bibfnamefont {S.}~\bibnamefont {Zhang}},\
  }\bibfield  {title} {\bibinfo {title} {{SO4 symmetry in a Hubbard model}},\
  }\href {https://doi.org/10.1142/S0217984990000933} {\bibfield  {journal}
  {\bibinfo  {journal} {Modern Physics Letters B}\ }\textbf {\bibinfo {volume}
  {04}},\ \bibinfo {pages} {759} (\bibinfo {year} {1990})}\BibitemShut
  {NoStop}%
\bibitem [{\citenamefont {E\ss{}l}\ \emph {et~al.}(2024)\citenamefont
  {E\ss{}l}, \citenamefont {Reitner}, \citenamefont {Sangiovanni},\ and\
  \citenamefont {Toschi}}]{Essl2024}%
  \BibitemOpen
  \bibfield  {author} {\bibinfo {author} {\bibfnamefont {H.}~\bibnamefont
  {E\ss{}l}}, \bibinfo {author} {\bibfnamefont {M.}~\bibnamefont {Reitner}},
  \bibinfo {author} {\bibfnamefont {G.}~\bibnamefont {Sangiovanni}},\ and\
  \bibinfo {author} {\bibfnamefont {A.}~\bibnamefont {Toschi}},\ }\bibfield
  {title} {\bibinfo {title} {{General Shiba mapping for on-site four-point
  correlation functions}},\ }\href
  {https://doi.org/10.1103/PhysRevResearch.6.033061} {\bibfield  {journal}
  {\bibinfo  {journal} {Phys. Rev. Res.}\ }\textbf {\bibinfo {volume} {6}},\
  \bibinfo {pages} {033061} (\bibinfo {year} {2024})}\BibitemShut {NoStop}%
\bibitem [{Note1()}]{Note1}%
  \BibitemOpen
  \bibinfo {note} {Note that complex conjugation can not be straightforwardly
  applied to CDMFT lattice quantities, as they also carry the super-lattice
  momentum vector $\protect \tilde {\protect \ensuremath {\protect \mathbf
  {q}}}$.}\BibitemShut {Stop}%
\bibitem [{\citenamefont {Reitner}\ \emph {et~al.}(2025)\citenamefont
  {Reitner}, \citenamefont {Del~Re}, \citenamefont {Capone},\ and\
  \citenamefont {Toschi}}]{Reitner2025}%
  \BibitemOpen
  \bibfield  {author} {\bibinfo {author} {\bibfnamefont {M.}~\bibnamefont
  {Reitner}}, \bibinfo {author} {\bibfnamefont {L.}~\bibnamefont {Del~Re}},
  \bibinfo {author} {\bibfnamefont {M.}~\bibnamefont {Capone}},\ and\ \bibinfo
  {author} {\bibfnamefont {A.}~\bibnamefont {Toschi}},\ }\bibfield  {title}
  {\bibinfo {title} {{Nonperturbative feats in the physics of correlated
  antiferromagnets}},\ }\href {https://doi.org/10.1103/ympr-9m73} {\bibfield
  {journal} {\bibinfo  {journal} {Phys. Rev. Res.}\ }\textbf {\bibinfo {volume}
  {7}},\ \bibinfo {pages} {033264} (\bibinfo {year} {2025})}\BibitemShut
  {NoStop}%
\bibitem [{\citenamefont {Eckhardt}\ \emph {et~al.}(2018)\citenamefont
  {Eckhardt}, \citenamefont {Schober}, \citenamefont {Ehrlich},\ and\
  \citenamefont {Honerkamp}}]{Eckhardt18}%
  \BibitemOpen
  \bibfield  {author} {\bibinfo {author} {\bibfnamefont {C.~J.}\ \bibnamefont
  {Eckhardt}}, \bibinfo {author} {\bibfnamefont {G.~A.~H.}\ \bibnamefont
  {Schober}}, \bibinfo {author} {\bibfnamefont {J.}~\bibnamefont {Ehrlich}},\
  and\ \bibinfo {author} {\bibfnamefont {C.}~\bibnamefont {Honerkamp}},\
  }\bibfield  {title} {\bibinfo {title} {{Truncated-unity parquet equations:
  Application to the repulsive Hubbard model}},\ }\href
  {https://doi.org/10.1103/PhysRevB.98.075143} {\bibfield  {journal} {\bibinfo
  {journal} {Phys. Rev. B}\ }\textbf {\bibinfo {volume} {98}},\ \bibinfo
  {pages} {075143} (\bibinfo {year} {2018})}\BibitemShut {NoStop}%
\bibitem [{\citenamefont {Georges}\ and\ \citenamefont
  {Krauth}(1992)}]{Georges1992}%
  \BibitemOpen
  \bibfield  {author} {\bibinfo {author} {\bibfnamefont {A.}~\bibnamefont
  {Georges}}\ and\ \bibinfo {author} {\bibfnamefont {W.}~\bibnamefont
  {Krauth}},\ }\bibfield  {title} {\bibinfo {title} {{Numerical solution of the
  $d=\infty$ Hubbard model: Evidence for a Mott transition}},\ }\href
  {https://doi.org/10.1103/PhysRevLett.69.1240} {\bibfield  {journal} {\bibinfo
   {journal} {Phys. Rev. Lett.}\ }\textbf {\bibinfo {volume} {69}},\ \bibinfo
  {pages} {1240} (\bibinfo {year} {1992})}\BibitemShut {NoStop}%
\bibitem [{\citenamefont {Jarrell}\ \emph {et~al.}(2001)\citenamefont
  {Jarrell}, \citenamefont {Maier}, \citenamefont {Huscroft},\ and\
  \citenamefont {Moukouri}}]{Jarrell2001}%
  \BibitemOpen
  \bibfield  {author} {\bibinfo {author} {\bibfnamefont {M.}~\bibnamefont
  {Jarrell}}, \bibinfo {author} {\bibfnamefont {T.}~\bibnamefont {Maier}},
  \bibinfo {author} {\bibfnamefont {C.}~\bibnamefont {Huscroft}},\ and\
  \bibinfo {author} {\bibfnamefont {S.}~\bibnamefont {Moukouri}},\ }\bibfield
  {title} {\bibinfo {title} {{Quantum Monte Carlo algorithm for nonlocal
  corrections to the dynamical mean-field theory}},\ }\href@noop {} {\bibfield
  {journal} {\bibinfo  {journal} {Phys. Rev. B}\ }\textbf {\bibinfo {volume}
  {64}},\ \bibinfo {pages} {195130} (\bibinfo {year} {2001})}\BibitemShut
  {NoStop}%
\bibitem [{\citenamefont {Kotliar}\ \emph {et~al.}(2000)\citenamefont
  {Kotliar}, \citenamefont {Lange},\ and\ \citenamefont
  {Rozenberg}}]{Kotliar2000}%
  \BibitemOpen
  \bibfield  {author} {\bibinfo {author} {\bibfnamefont {G.}~\bibnamefont
  {Kotliar}}, \bibinfo {author} {\bibfnamefont {E.}~\bibnamefont {Lange}},\
  and\ \bibinfo {author} {\bibfnamefont {M.~J.}\ \bibnamefont {Rozenberg}},\
  }\bibfield  {title} {\bibinfo {title} {{Landau Theory of the Finite
  Temperature Mott Transition}},\ }\href
  {https://doi.org/10.1103/PhysRevLett.84.5180} {\bibfield  {journal} {\bibinfo
   {journal} {Phys. Rev. Lett.}\ }\textbf {\bibinfo {volume} {84}},\ \bibinfo
  {pages} {5180} (\bibinfo {year} {2000})}\BibitemShut {NoStop}%
\bibitem [{\citenamefont {van Loon}\ and\ \citenamefont
  {Strand}(2024)}]{vanLoon2024-2}%
  \BibitemOpen
  \bibfield  {author} {\bibinfo {author} {\bibfnamefont {E.~G. C.~P.}\
  \bibnamefont {van Loon}}\ and\ \bibinfo {author} {\bibfnamefont {H.~U.~R.}\
  \bibnamefont {Strand}},\ }\bibfield  {title} {\bibinfo {title} {{Dual
  Bethe-Salpeter equation for the multiorbital lattice susceptibility within
  dynamical mean-field theory}},\ }\href
  {https://doi.org/10.1103/PhysRevB.109.155157} {\bibfield  {journal} {\bibinfo
   {journal} {Phys. Rev. B}\ }\textbf {\bibinfo {volume} {109}},\ \bibinfo
  {pages} {155157} (\bibinfo {year} {2024})}\BibitemShut {NoStop}%
\bibitem [{\citenamefont {Maier}\ \emph {et~al.}(2005)\citenamefont {Maier},
  \citenamefont {Jarrell}, \citenamefont {Pruschke},\ and\ \citenamefont
  {Hettler}}]{Maier2005}%
  \BibitemOpen
  \bibfield  {author} {\bibinfo {author} {\bibfnamefont {T.~A.}\ \bibnamefont
  {Maier}}, \bibinfo {author} {\bibfnamefont {M.}~\bibnamefont {Jarrell}},
  \bibinfo {author} {\bibfnamefont {T.}~\bibnamefont {Pruschke}},\ and\
  \bibinfo {author} {\bibfnamefont {M.}~\bibnamefont {Hettler}},\ }\bibfield
  {title} {\bibinfo {title} {{Quantum Cluster Theories}},\ }\href
  {https://doi.org/10.1103/RevModPhys.77.1027} {\bibfield  {journal} {\bibinfo
  {journal} {Rev. Mod. Phys.}\ }\textbf {\bibinfo {volume} {77}},\ \bibinfo
  {pages} {1027} (\bibinfo {year} {2005})}\BibitemShut {NoStop}%
\bibitem [{\citenamefont {Meixner}\ \emph {et~al.}(2026)\citenamefont
  {Meixner}, \citenamefont {Reitner}, \citenamefont {Sch\"afer},\ and\
  \citenamefont {Toschi}}]{Meixner2026a}%
  \BibitemOpen
  \bibfield  {author} {\bibinfo {author} {\bibfnamefont {M.}~\bibnamefont
  {Meixner}}, \bibinfo {author} {\bibfnamefont {M.}~\bibnamefont {Reitner}},
  \bibinfo {author} {\bibfnamefont {T.}~\bibnamefont {Sch\"afer}},\ and\
  \bibinfo {author} {\bibfnamefont {A.}~\bibnamefont {Toschi}},\ }\href
  {https://arxiv.org/abs/2512.17716} {\bibinfo {title} {{Non-perturbative
  effects of short-range spatial correlations at the two-particle level}}}
  (\bibinfo {year} {2026}),\ \Eprint {https://arxiv.org/abs/2512.17716}
  {arXiv:2512.17716 [cond-mat.str-el]} \BibitemShut {NoStop}%
\bibitem [{\citenamefont {Georges}\ \emph {et~al.}(1996)\citenamefont
  {Georges}, \citenamefont {Kotliar}, \citenamefont {Krauth},\ and\
  \citenamefont {Rozenberg}}]{georges:1996}%
  \BibitemOpen
  \bibfield  {author} {\bibinfo {author} {\bibfnamefont {A.}~\bibnamefont
  {Georges}}, \bibinfo {author} {\bibfnamefont {G.}~\bibnamefont {Kotliar}},
  \bibinfo {author} {\bibfnamefont {W.}~\bibnamefont {Krauth}},\ and\ \bibinfo
  {author} {\bibfnamefont {M.~J.}\ \bibnamefont {Rozenberg}},\ }\bibfield
  {title} {\bibinfo {title} {Dynamical mean-field theory of strongly correlated
  fermion systems and the limit of infinite dimensions},\ }\href
  {https://doi.org/10.1103/RevModPhys.68.13} {\bibfield  {journal} {\bibinfo
  {journal} {Rev. Mod. Phys.}\ }\textbf {\bibinfo {volume} {68}},\ \bibinfo
  {pages} {13} (\bibinfo {year} {1996})}\BibitemShut {NoStop}%
\bibitem [{\citenamefont {Hettler}\ \emph {et~al.}(2000)\citenamefont
  {Hettler}, \citenamefont {Mukherjee}, \citenamefont {Jarrell},\ and\
  \citenamefont {Krishnamurthy}}]{Hettler2000}%
  \BibitemOpen
  \bibfield  {author} {\bibinfo {author} {\bibfnamefont {M.~H.}\ \bibnamefont
  {Hettler}}, \bibinfo {author} {\bibfnamefont {M.}~\bibnamefont {Mukherjee}},
  \bibinfo {author} {\bibfnamefont {M.}~\bibnamefont {Jarrell}},\ and\ \bibinfo
  {author} {\bibfnamefont {H.~R.}\ \bibnamefont {Krishnamurthy}},\ }\bibfield
  {title} {\bibinfo {title} {Dynamical cluster approximation: Nonlocal dynamics
  of correlated electron systems},\ }\href
  {https://doi.org/10.1103/PhysRevB.61.12739} {\bibfield  {journal} {\bibinfo
  {journal} {Phys. Rev. B}\ }\textbf {\bibinfo {volume} {61}},\ \bibinfo
  {pages} {12739} (\bibinfo {year} {2000})}\BibitemShut {NoStop}%
\bibitem [{\citenamefont {Kotliar}\ \emph {et~al.}(2001)\citenamefont
  {Kotliar}, \citenamefont {Savrasov}, \citenamefont {P\'alsson},\ and\
  \citenamefont {Biroli}}]{Kotliar2001}%
  \BibitemOpen
  \bibfield  {author} {\bibinfo {author} {\bibfnamefont {G.}~\bibnamefont
  {Kotliar}}, \bibinfo {author} {\bibfnamefont {S.~Y.}\ \bibnamefont
  {Savrasov}}, \bibinfo {author} {\bibfnamefont {G.}~\bibnamefont
  {P\'alsson}},\ and\ \bibinfo {author} {\bibfnamefont {G.}~\bibnamefont
  {Biroli}},\ }\bibfield  {title} {\bibinfo {title} {{Cellular Dynamical Mean
  Field Approach to Strongly Correlated Systems}},\ }\href
  {https://doi.org/10.1103/PhysRevLett.87.186401} {\bibfield  {journal}
  {\bibinfo  {journal} {Phys. Rev. Lett.}\ }\textbf {\bibinfo {volume} {87}},\
  \bibinfo {pages} {186401} (\bibinfo {year} {2001})}\BibitemShut {NoStop}%
\bibitem [{\citenamefont {Mu\ss{}hoff}\ \emph {et~al.}(2021)\citenamefont
  {Mu\ss{}hoff}, \citenamefont {Kiani},\ and\ \citenamefont
  {Pavarini}}]{Musshoff2021}%
  \BibitemOpen
  \bibfield  {author} {\bibinfo {author} {\bibfnamefont {J.}~\bibnamefont
  {Mu\ss{}hoff}}, \bibinfo {author} {\bibfnamefont {A.}~\bibnamefont {Kiani}},\
  and\ \bibinfo {author} {\bibfnamefont {E.}~\bibnamefont {Pavarini}},\
  }\bibfield  {title} {\bibinfo {title} {{Magnetic response trends in cuprates
  and the $t\ensuremath{-}{t}^{\ensuremath{'}}$ Hubbard model}},\ }\href
  {https://doi.org/10.1103/PhysRevB.103.075136} {\bibfield  {journal} {\bibinfo
   {journal} {Phys. Rev. B}\ }\textbf {\bibinfo {volume} {103}},\ \bibinfo
  {pages} {075136} (\bibinfo {year} {2021})}\BibitemShut {NoStop}%
\bibitem [{\citenamefont {Walsh}\ \emph {et~al.}(2023)\citenamefont {Walsh},
  \citenamefont {Charlebois}, \citenamefont {S\'emon}, \citenamefont
  {Tremblay},\ and\ \citenamefont {Sordi}}]{Walsh2023}%
  \BibitemOpen
  \bibfield  {author} {\bibinfo {author} {\bibfnamefont {C.}~\bibnamefont
  {Walsh}}, \bibinfo {author} {\bibfnamefont {M.}~\bibnamefont {Charlebois}},
  \bibinfo {author} {\bibfnamefont {P.}~\bibnamefont {S\'emon}}, \bibinfo
  {author} {\bibfnamefont {A.-M.~S.}\ \bibnamefont {Tremblay}},\ and\ \bibinfo
  {author} {\bibfnamefont {G.}~\bibnamefont {Sordi}},\ }\bibfield  {title}
  {\bibinfo {title} {{Superconductivity in the two-dimensional Hubbard model
  with cellular dynamical mean-field theory: A quantum impurity model
  analysis}},\ }\href {https://doi.org/10.1103/PhysRevB.108.075163} {\bibfield
  {journal} {\bibinfo  {journal} {Phys. Rev. B}\ }\textbf {\bibinfo {volume}
  {108}},\ \bibinfo {pages} {075163} (\bibinfo {year} {2023})}\BibitemShut
  {NoStop}%
\bibitem [{\citenamefont {Harland}\ \emph {et~al.}(2019)\citenamefont
  {Harland}, \citenamefont {Brener}, \citenamefont {Lichtenstein},\ and\
  \citenamefont {Katsnelson}}]{Harland2019}%
  \BibitemOpen
  \bibfield  {author} {\bibinfo {author} {\bibfnamefont {M.}~\bibnamefont
  {Harland}}, \bibinfo {author} {\bibfnamefont {S.}~\bibnamefont {Brener}},
  \bibinfo {author} {\bibfnamefont {A.~I.}\ \bibnamefont {Lichtenstein}},\ and\
  \bibinfo {author} {\bibfnamefont {M.~I.}\ \bibnamefont {Katsnelson}},\
  }\bibfield  {title} {\bibinfo {title} {Josephson lattice model for phase
  fluctuations of local pairs in copper oxide superconductors},\ }\href
  {https://doi.org/10.1103/PhysRevB.100.024510} {\bibfield  {journal} {\bibinfo
   {journal} {Phys. Rev. B}\ }\textbf {\bibinfo {volume} {100}},\ \bibinfo
  {pages} {024510} (\bibinfo {year} {2019})}\BibitemShut {NoStop}%
\bibitem [{\citenamefont {Fratino}\ \emph {et~al.}(2016)\citenamefont
  {Fratino}, \citenamefont {S\'emon}, \citenamefont {Sordi},\ and\
  \citenamefont {Tremblay}}]{Fratino2016}%
  \BibitemOpen
  \bibfield  {author} {\bibinfo {author} {\bibfnamefont {L.}~\bibnamefont
  {Fratino}}, \bibinfo {author} {\bibfnamefont {P.}~\bibnamefont {S\'emon}},
  \bibinfo {author} {\bibfnamefont {G.}~\bibnamefont {Sordi}},\ and\ \bibinfo
  {author} {\bibfnamefont {A.~M.~S.}\ \bibnamefont {Tremblay}},\ }\bibfield
  {title} {\bibinfo {title} {{An organizing principle for two-dimensional
  strongly correlated superconductivity}},\ }\href
  {https://doi.org/https://doi.org/10.1038/srep22715} {\bibfield  {journal}
  {\bibinfo  {journal} {Scientific Reports}\ }\textbf {\bibinfo {volume} {6}},\
  \bibinfo {pages} {22715} (\bibinfo {year} {2016})}\BibitemShut {NoStop}%
\bibitem [{\citenamefont {Krien}\ \emph {et~al.}(2019)\citenamefont {Krien},
  \citenamefont {Valli},\ and\ \citenamefont {Capone}}]{Krien2019c}%
  \BibitemOpen
  \bibfield  {author} {\bibinfo {author} {\bibfnamefont {F.}~\bibnamefont
  {Krien}}, \bibinfo {author} {\bibfnamefont {A.}~\bibnamefont {Valli}},\ and\
  \bibinfo {author} {\bibfnamefont {M.}~\bibnamefont {Capone}},\ }\bibfield
  {title} {\bibinfo {title} {{Single-boson exchange decomposition of the vertex
  function}},\ }\href {https://doi.org/10.1103/PhysRevB.100.155149} {\bibfield
  {journal} {\bibinfo  {journal} {Phys. Rev. B}\ }\textbf {\bibinfo {volume}
  {100}},\ \bibinfo {pages} {155149} (\bibinfo {year} {2019})}\BibitemShut
  {NoStop}%
\bibitem [{\citenamefont {Krien}\ \emph {et~al.}(2021)\citenamefont {Krien},
  \citenamefont {Kauch},\ and\ \citenamefont {Held}}]{Krien2020b}%
  \BibitemOpen
  \bibfield  {author} {\bibinfo {author} {\bibfnamefont {F.}~\bibnamefont
  {Krien}}, \bibinfo {author} {\bibfnamefont {A.}~\bibnamefont {Kauch}},\ and\
  \bibinfo {author} {\bibfnamefont {K.}~\bibnamefont {Held}},\ }\bibfield
  {title} {\bibinfo {title} {{Tiling with triangles: parquet and
  $GW\ensuremath{\gamma}$ methods unified}},\ }\href
  {https://doi.org/10.1103/PhysRevResearch.3.013149} {\bibfield  {journal}
  {\bibinfo  {journal} {Phys. Rev. Research}\ }\textbf {\bibinfo {volume}
  {3}},\ \bibinfo {pages} {013149} (\bibinfo {year} {2021})}\BibitemShut
  {NoStop}%
\bibitem [{Note2()}]{Note2}%
  \BibitemOpen
  \bibinfo {note} {We use the right- and left-hand side eigenvectors for the
  respective sides of the scalar product.}\BibitemShut {Stop}%
\end{thebibliography}%
\clearpage % Ensures the Supplemental Material starts on a new page
\onecolumngrid % Switch to single-column format for supplemental material
\setcounter{page}{1}
\setcounter{secnumdepth}{1}

\appendix
\begin{center}
    \textbf{\large End Matter}\\[0.5em]
\end{center}
\vspace{0.5cm}

\thispagestyle{empty}
\twocolumngrid
 
\noindent
\section{Cellular dynamical mean-field theory at the two-particle level}
\label{sec:cdmft}
A widely used, non-perturbative method to study the impact of real-space correlations of 2D systems and the intertwining of thermodynamic instabilities such as the metal-insulator transitions (MIT) \cite{Parcollet2004,Zhang2007,Park2008}, pseudogap regimes  \cite{Ferrero2009,Braganca2018,Moutenet2018,Musshoff2021,Ortiz2022,Meixner2024}, strongly correlated superconductivity \cite{Lichtenstein2000,Haule2007b,Sordi2012b,Tremblay2013,Fratino2017,Harland2019,Walsh2023,Sakai2023}, and charge order \cite{Verret2019,Meixner2024} is CDMFT \cite{Metzner1989,Georges1992a,Kotliar2001,Lichtenstein2001}. There, the original lattice model is understood as a superlattice of $N_c$ sites \cite{Bolech03,Maier2005}, resulting in an enlarged unit cell and a basis with $N_c$ atoms. A mapping from the lattice to an impurity problem is then undertaken by integrating out the real space dependencies in the single electron Green function between the lattice cells. For one specific cell, the other superlattice cells are mimicked by an energy-dependent, non-interacting bath. This results in an Anderson impurity model \cite{Georges1996}, the solution of which only involves computing a cell-local \cite{Potthoff2018} self-energy, obtained from an interaction expansion continuous time quantum Monte Carlo solver \cite{Rubtsov2005,Gull2008a} provided by TRIQS \cite{TRIQS}. Mapping the result back to the lattice yields a cycle which can be solved self-consistently. 
While the formulation of the mapping on the level of the single electron Green function makes electronic spectra in principle accessible, an evaluation of thermodynamic instabilities on the two-particle level is more involved. Here, the conserving quantity \cite{Hettler2000,Potthoff2003,Senechal2004Book} is the two-particle irreducible vertex $\Gamma$ \cite{Bickers04,Rohringer2012}, which allows to relate the CDMFT susceptibilities of the impurity $\chi^{imp}$ and lattice $\chi^{\vtq}$, $\vtq\in$RBZ via a Bethe-Salpeter treatment in the respective channel \cite{Jarrell1992,Musshoff2021,Meixner2026a}, e.g.,~particle-particle \cite{Rohringer2013a,Meixner2026a} 
\begin{equation}
\begin{split}
\label{Eq:BSE_pp}
\chi_{\mathrm{pp},\uparrow\downarrow}^{\vtq}=&\left( T^2\Gamma_{\mathrm{pp},\uparrow\downarrow}+\left[\chi^{\vtq}_{0,\mathrm{pp},\uparrow\downarrow}\right]^{-1}\right)^{-1}\\
=&\Big( \left[\chi^{imp}_{\pp}\right]^{-1}+\underbrace{\left[\chi^{\vtq}_{0,\pp\up\down}\right]^{-1}-\left[\chi^{imp}_{0,\pp\up\down}\right]^{-1}}_{t^2_\vtq\equiv t_\text{eff}^2}\Big)^{-1},
\end{split}
\end{equation} with the disconnected contribution \begin{equation}
\label{eq:def_bubble}
\chi^{\vtq}_{0,\mathrm{pp},\uparrow\downarrow}(\omega)=\frac{\delta_{\nu\nu'}}{TN_{\vtk}}\sum_{\vtk}G^{\l\i}_{\vtk,\up}(\nu)G^{\j\h}_{\vtq-\vtk,\down}(\omega-\nu).
\end{equation} While these quantities are dependent on frequency and real space indices, we employ a multi-index notation of left- and right hand side variables to obtain matrix objects \cite{Musshoff2021,Meixner2026a}.
In Eq.~(\ref{eq:def_bubble}), we averaged over $N_\vtk$ points in the reduced Brillouin zone (RBZ) of the superlattice. The embedding correction $t^2_\vtq$, accounting for particle hopping between neighboring superlattice cells, parametrized by $\vtk,\vtq\in\mathrm{RBZ}$. 
Note that approximating the lattice by truncating the length of the correlations to one superlattice cell allows to treat all fluctuations on equal footing, preserving the degeneracies indicated previously.

In general, CDMFT fulfills the identities derived in the main text, both on the impurity and for its lattice approximation. For the latter, a flip of the superlattice vector $\vtq_\ph=-\vtq_\pp$ is necessary, see Sec.~VI of \cite{sup}, however, for the $d-SC$ example given in the main text, the instability has $\vtq_\ph=-\vtq_\pp=\vnull$, hence we omit this index in the main text.

\section{Consequences for thermodynamic phase transitions}\label{sec:PT}
As already discussed in the main text the simultaneous phase transition in $d$-DW and $d-$SC at particle-hole symmetry [\cref{fig:CDMFT_ph_symmetry}], together with the continuous deviation from this degeneracy when introducing $t^\prime$ [\cref{fig:CDMFT_tp}], naturally raises the question why the MIT is not accompanied by a superconducting transition since, according to \cref{fig:table}, $\chi_\text{ch}^{s,\vq=\vnull}=\chi_\text{pp}^{s,\vq=\vpi}$.
This behavior can be understood by investigating the symmetries of the eigenvectors of the generalized susceptibilities. 
For this we write the generalized susceptibility in its spectral form as
\begin{align}
\label{Eq:sus_spectral}
    \chi^{kk^\prime q}=\sum_\alpha v^{k^\prime,q}_\alpha\lambda^q_\alpha v^{k,q}_\alpha,
\end{align}
which can be done since the generalized susceptibilities in the ch, sp and pp channel are symmetric matrices (see Sec.~III of \cite{sup}). Therefore, they can be diagonalized by an orthonormal transformation \cite{Craven1969} when excluding the special case of exceptional points \cite{Reitner2024}.
From now on we will consider the generalized susceptibilities as matrices in the fermionic arguments $k$ and $k^\prime$ with $q$ being an parameter. 

Under $SU(2)_P$-symmetry the generalized susceptibilities in charge and spin channel commute with \mbox{$J^{kk^\prime q}=\delta_{k\overline{k^\prime}}$} (see Sec.~VI of \cite{sup}), therefore the eigenvectors of the charge and spin channels satisfy the relation \mbox{$v_{\text{ch/sp},\pm}^q(k)=\pm v_{\text{ch/sp},\pm}^q(\overline{k})$}, i.e. the eigenvectors have a defined $\overline{k}$-parity.
Hence, we can rewrite \cref{Eq:momentumspace} as
\begin{align}
\label{eq:Projectors_pp}
\begin{split}
    \chi_\pp^{-q-\Pi}=&\sum_\alpha v^{k^\prime,q}_{\text{ch},\alpha}\lambda^q_{\text{ch},\alpha} \left(v^{k,q}_{\text{ch},\alpha} +v^{\overline{k},q}_{\text{ch},\alpha}\right) \\
    &+ \sum_\alpha v^{k^\prime,q}_{\text{sp},\alpha}\lambda_{\text{sp},\alpha}^q \left(v^{k,q}_{\text{sp},\alpha} -v^{\overline{k},q}_{\text{sp},\alpha}\right)\\
    &=\sum_{k_1,k_2}P_+^{kk_1q}\chi_\text{ch}^{k_1k_2q}P_+^{k2k^\prime q} + P_-^{kk_1q}\chi_\text{sp}^{k_1k_2q}P_-^{k2k^\prime q},
\end{split}
\end{align}
where we introduced the projectors $P_\pm^{kk^\prime q}=\frac{1}{2}(\delta_{kk^\prime}\pm J^{kk^\prime q})$ onto the symmetric and antisymmetric subspaces. For a simplified notation we will drop the indices $k$ and $k^\prime$ in the following.
As a result, the set of eigenvectors of $\chi_\pp^{-q-\Pi}$ are the union of the symmetric/antisymmetric subspace of the charge/spin-channel eigenvectors, i.e., $\{v_\pp^{-q-\Pi}\}=\{v_{\text{ch},+}^q\}\cup \{v_{\text{sp},-}^q\}$ with the set of eigenvalues given by $\{\lambda_\pp^{-q-\Pi}\}=\{\lambda_{+,\text{ch}}^q\} \cup \{\lambda_{-,\text{sp}}^q\}$.

Further, the physical susceptibility can be written as
\begin{align}
\label{Eq:weights_phys_sus}
    \chi^{r,q}=\sum_\alpha W^{r,q}_\alpha\lambda_\alpha^{q}\text{ with }W^{r,q}_\alpha=\left(\frac{1}{\sqrt{N}\beta}\sum_{k}f_r(k)v^q_\alpha(k)\right)^2,
\end{align}
where $\lambda_\alpha^q,v_\alpha^q$ denote the eigenvalues and eigenvectors of the generalized susceptibility, while $W_\alpha^{r,q}$ represent the weight's contributing to the physical response function. The weight of an eigenvalue can vanish in two different scenarios: First, when the eigenvector has a antisymmetric frequency structure, i.e., $v(\vk,\nu)=-v(\vk,-\nu)$, or, second, when the $\overline{k}$-parity of the eigenvector and the form factor is different in momentum space, i.e. $v(\vk,\nu)=\pm v(\overline{\vk},\nu)$, but $f(\vk)=\mp f(\overline{\vk})$.

Since we are interested in phase transitions that are associated with a specific form factor, it is reasonable to assume that the momentum-overlap between the form factor and the eigenvector that drives the transition does not vanish, i.e.,  $\sum_{\vk}f(\vk)v(\vk,\nu)\neq0$. 
Consequently, we shall exclude the case of mismatched momentum $\overline{k}$-parity from the following discussion. 
Based on this, we can infer that for an even form factor [$f(\vk)=f(\bar\vk)$], a vector $v\in V_-$ has zero weight due to its antisymmetric frequency structure. Conversely, a vector $v\in V_+$ possesses finite weight. For odd form factors, the situation is reversed: the weight vanishes for $v\in V_+$ and is finite for $v\in V_-$.

Equipped with these symmetry arguments we can understand the difference between the $d$-DW transition and the MIT. Starting with the $d$-DW, we see in \cref{fig:CDMFT_ph_symmetry} that $\chi_\text{ch}^{d,\vq=\vpi}$ diverges at $T_c$, which is realized through a diverging eigenvalue $\lambda_\text{ch}^{\vq=\vpi}\rightarrow\infty$ with $W_\text{ch}^{d,\vq=\vpi}\neq 0$. Since the phase transition has $d$-wave symmetry, the $\vk$-structure of the eigenvector that belongs to the diverging eigenvalue must also be ``even", i.e., $v_\text{ch}^{\vq=\vpi}(\vk,\nu)=v_\text{ch}^{\vq=\vpi}(-\vk,\nu)$. Consequently, in order to generate a non-vanishing weight $W_\text{ch}^{d,\vq=\vpi}\neq 0$, the frequency structure needs to be even, i.e., $v_\text{ch}^{\vq=\vpi}(\vk,\nu)=v_\text{ch}^{\vq=\vpi}(\vk,-\nu)$. 

As $v_\text{ch}^{\vq=\vpi}$ is in the symmetric subspace of the generalized susceptibility $\chi_\text{ch}^{\vq=\vpi}$, due to Eq.~(\ref{eq:Projectors_pp}), $\chi_\text{pp}^{d,\vq=\vnull}$ must also diverge, see \cref{fig:CDMFT_ph_symmetry}. The corresponding eigenvector-eigenvalue pair is also part of the eigensystem of the generalized susceptibility $\chi_\text{pp}^{\vq=\vnull}$. 
The fact, that both generalized susceptibilities $\chi_\text{ch}^{\vq=\vpi}$ and $\chi_\text{pp}^{\vq=\vnull}$ possess a diverging eigenvalue has profound implications when particle-hole symmetry is broken. Both physical susceptibilities diverge also out of particle-hole symmetry, albeit at gradually different parameter points, which suggests a competition between these two orders. As shown in \cref{fig:CDMFT_tp} the superconductivity prevails.

The situation for the MIT is qualitatively different. While an eigenvalue of the generalized susceptibility $\chi_\text{ch}^{\vq=\vnull}$ diverges at the critical endpoint of the MIT, the associated physical susceptibility $\chi_\text{ch}^{s,\vq=\vnull}$ remains finite. This is because the weight $W_\text{ch}^{s,\vq=\vnull}$ of the diverging eigenvalue vanishes at particle-hole symmetry \cite{Reitner2020,vanLoon2020,Meixner2026a}.  
Since the phase transition has dominantely $s$-wave character, the frequency structure of the corresponding eigenvector must be even, i.e., $v_\text{ch}^{\vq=\vnull}(\vk,\nu)=v_\text{ch}^{\vq=\vnull}(-\vk-\vpi,\nu)$, the vanishing weight is due to an odd frequency structure, i.e., $v_\text{ch}^{\vq=\vnull}(\vk,\nu)=v_\text{ch}^{\vq=\vnull}(\vk,-\nu)$, as found in \cite{Reitner2020,vanLoon2020,Meixner2026a}. 
Consequently, $v_\text{ch}^{\vq=\vnull}(\vk,\nu)$ is in the antisymmetric subspace of $\chi_\text{ch}^{\vq=\vnull}$ and therefore not present in the eigensystem of the generalized susceptibility $\chi_\text{pp}^{\vq=\vpi}$.
Further, we argue that $\chi_\text{sp}^{\vq=\vnull}$ has no diverging eigenvalue in its antisymmetric subspace. To justify this we need to consider two situations. The first involves a diverging eigenvalue in $\chi_\text{sp}^{\vq=\vnull}$ at weak coupling that is driven by a van Hove singularity. This eigenvalue resides in the symmetric subspace of the generalized susceptibility, as it directly results in the divergence of the physical susceptibility. 

The second scenario considers Nagaoka ferromagnetism. Nagaoka's theorem \cite{Nagaoka1966} states that the ground state of the Hubbard model for $U\rightarrow\infty$ with exactly one hole is fully spin polarized if the lattice is sufficiently connected, which is the case for bipartite lattices. For this phase transition it is plausible that a diverging eigenvalue in the antisymmetric subspace of $\chi_\text{sp}^{\vq=\vnull}$ is involved. This suspicion arises because the transition occurs away from particle-hole symmetry as soon as the system is doped \cite{Obermeier1997,Park2008a,Kamogawa2019}, a behavior compatible with a divergent eigenvalue that possesses vanishing weight at particle-hole symmetry and is therefore part of the antisymmetric subspace. However, we exclude this case for our analysis, as this ferromagnetism requires infinitely large interaction strengths in the thermodynamic limit \cite{Li2024,Park2008a,Obermeier1997,Kamogawa2019}.

$\chi_\text{pp}^{\vq=\vpi}$, hence, has no diverging eigenvalue at particle-hole symmetry. Lifting the particle-hole symmetry causes $\chi_\text{ch}^{s,\vq=\vnull}$ to diverge which is associated to a phase separation \cite{Reitner2020}, while $\chi_\text{pp}^{s,\vq=\vpi}$ remains finite. This abruptly breaks the degeneracy between the two susceptibilities. The investigation of these two cases reveals a qualitative distinction between phase transitions characterized by a vanishing weight at particle-hole symmetry, like the phase separation, and those maintaining a non-zero weight like the $d$-DW. 

Moreover, the known fact that a genuine $s$-wave superconducting phase transition ($\chi_\text{pp}^{s,\vq=0}$) is not appearing in the repulsive Hubbard model matches the fact that the antiferromagnetic phase transition, that is related to $\chi_\text{sp}^{s,\vq=\vpi}$, is driven by an eigenvector with symmetric frequency structure and therefore non-zero weight. 
Consequently, this eigenvector is in the symmetric subspace of the generalized susceptibility $\chi_\text{sp}^{\vq=\vpi}$, and is, therefore, not present in $\chi_\text{pp}^{\vq=\vnull}$.
\end{document}

\title{Supplemental Material for ``On Degeneracies of Density, Magnetic, and Pairing Responses: How Competing Orders Echo Underlying Symmetries in the Hubbard Model''}

\author{Michael Meixner}
\thanks{Both authors contributed equally.}
\MPI
\author{Herbert E{\ss}l\orcidlink{0009-0005-9883-8104}}
\thanks{Both authors contributed equally.}
\TUVienna
\author{Matthias Reitner\orcidlink{0000-0002-2529-0847}}
\TUVienna
\author{Alessandro Toschi\orcidlink{0000-0001-5669-3377}}
\TUVienna
\author{Thomas Schäfer\orcidlink{0000-0002-7550-4807}}
\Trieste
\MPI

\date{\today}

\maketitle
\section{Symmetries of the Hubbard Hamiltonian}
The Hubbard Hamiltonian on the square lattice reads
\begin{align}
\label{Eq:Hamiltonian}
    H=-\sum_{\i\j,\sigma}t_{\i\j}c_{\i,\sigma}^\dagger c_{\j,\sigma} -\delta\mu\sum_{\i,\sigma}n_{\i,\sigma}-B_z\sum_{\i} \left(n_{\i,\up}-n_{\i\down}\right)+U\sum_\i (n_{\i,\uparrow}-1/2)(n_{\i,\downarrow}-1/2),
\end{align}
with annihilation(creation) operator for an electron with spin $\sigma$ and at site $i$ $c^{(\dagger)}_{i,\sigma}$, density operator $n_{i,\sigma}=c^\dagger_{i,\sigma}c_{i,\sigma}$, hopping matrix $t_{ij}$, chemical potential $\delta\mu$ ($\delta\mu=0$ corresponds to half-filling), external magnetic field $B_z$ and onsite interaction $U$. 

Following Rohringer \cite{Rohringer2013a}, we define spinors
\begin{equation}
    \Vec{c}_i=\begin{pmatrix} c_{i,\uparrow} \\ c_{i,\downarrow} \end{pmatrix}, \qquad \Vec{c}^{\,\dagger}_i=\begin{pmatrix} c^\dagger_{i,\uparrow} \\ c^\dagger_{i,\downarrow} \end{pmatrix},
\end{equation} and we can express the spin operators by means of Pauli matrices as $S_{\alpha}^i=\frac{1}{2}\Vec{c}_i^\dagger \sigma_\alpha \Vec{c}_i$:
\begin{equation}
\label{Eq:local_S}
    \begin{split}
        S_{x}^i&=\frac{1}{2}\left(c^\dagger_{i,\uparrow} c_{i,\downarrow}+c^\dagger_{i,\downarrow} c_{i,\uparrow}\right),\\
        S_{y}^i&=\frac{\i}{2}\left(c^\dagger_{i,\downarrow} c_{i,\uparrow}-c^\dagger_{i,\uparrow} c_{i,\downarrow}\right),\\
        S_{z}^i&=\frac{1}{2}\left(c^\dagger_{i,\uparrow} c_{i,\uparrow}-c^\dagger_{i,\downarrow} c_{i,\downarrow}\right).\\
    \end{split}
\end{equation}
Further we can define the pseudo-spinors as
\begin{equation}
    \Vec{c}_{P,i}=\begin{pmatrix} c_{P,i,\uparrow} \\ c_{P,i,\downarrow} \end{pmatrix}=\begin{pmatrix} c_{i,\uparrow} \\ \e^{-\i\mathbf{k_P}\vR_1}c_{i,\downarrow}^\dagger \end{pmatrix}, \qquad \Vec{c}\,^{\dagger}_{P,i}=\begin{pmatrix} c^\dagger_{P,i,\uparrow} \\ c^\dagger_{P,i,\downarrow} \end{pmatrix}=\begin{pmatrix} c^\dagger_{i,\uparrow} \\ \e^{\i\mathbf{k_P}\vR_1}c_{i,\downarrow} \end{pmatrix},
\end{equation}
where the spinors and pseudo-spinors are related through the Shiba transformation \cite{Shiba1972}, which is a partial particle-hole transformation where only the down-spin is transformed while the up-spin is left unchanged, i.e.,
\begin{align}
\label{Eq:Shiba}
\begin{split}
c^{\dagger}_{i,\down}&\rightarrow\e^{\i\mathbf{k_P}\vR_i}c_{i,\down},\\
c_{i,\down}&\rightarrow\e^{-\i\mathbf{k_P}\vR_i}c^\dagger_{i,\down},
\end{split}
\end{align}
where $\mathbf{k_P}$ has to be chosen such that $e^{i \vR\mathbf{k_P}}=\pm 1$ for any lattice vectors of the sublattices A or B \cite{Carmelo2026}. Specifically, for the hyper-cubic lattice $\mathbf{k_P}=\vpi=(\pi,\pi,\dots)$ holds.
We define the pseudo-spin operator $S_{P,\alpha}^{i}=\frac{1}{2}\Vec{c}_{P,i}^\dagger \sigma_\alpha \Vec{c}_{P,i}$ analogous to the spin operator:
\begin{equation}
\label{Eq:local_S_P}
    \begin{split}
        S_{P,x}^i&=\frac{1}{2}\left(c^\dagger_{i,\uparrow} c^\dagger_{i,\downarrow}\e^{-\i\vpi\vR_i}+c_{i,\downarrow} c_{i,\uparrow}\e^{\i\vpi\vR_i}\right),\\
        S_{P,y}^i&=\frac{\i}{2}\left(c_{i,\downarrow} c_{i,\uparrow}\e^{-\i\vpi\vR_i}-c^\dagger_{i,\uparrow} c^\dagger_{i,\downarrow}\e^{\i\vpi\vR_i}\right),\\
        S_{P,z}^i&=\frac{1}{2}\left(c^\dagger_{i,\uparrow} c_{i,\uparrow}+c^\dagger_{i,\downarrow} c_{i,\downarrow}-1\right),\\
    \end{split}
\end{equation}
Note that by setting the lattice space to 1 the the phase factor in the Shiba transformation is just a sign, i.e., $\e^{\i\vpi\vR_i}=(-1)^i$.

A rotation in the (pseudo-)spin can be conducted by applying the rotation operator $\mathcal{D}(\varphi,\vn)$ on the (pseudo-)spinor \cite{Rohringer2013a}
\begin{align}
\label{Eq:spinor_rot}
    \vec{c}_{(P)}\,(\varphi,\vn)=\mathcal{D}(\varphi,\vn)\vec{c}_{(P)} \text{ with }\mathcal{D}(\varphi,\vn)=\e^{-i\varphi\vn\cdot\boldsymbol{\sigma}}
    =\left[ \mathbf{1}\cos\left(\frac{\varphi}{2}\right)-i\vn \cdot\boldsymbol{\sigma}\sin\left(\frac{\varphi}{2}\right)\right],
\end{align}
where $\mathcal{D}(\varphi,\vn)$ rotates the (pseudo-)spinor an angle $\varphi$ around an axis in the direction of the unit vector $\vn$. The relation for $\vec{c}^\dagger_{(P)}$ in given by the adjoint equation.

A Hamiltonian obeys $SU(2)_{(P)}$-symmetry if it commutes with all total (pseudo-)spin components, i.e. $\left[\sum_i S_{(P),\alpha}^i,H\right]=0$ for $\alpha=x,y,z$. Further we call the Hamiltonian particle-hole(ph)-symmetric if it is $SU(2)$- \emph{and} $SU(2)_P$-symmetric, the related symmetry is $SO(4)=[SU(2)\cross SU(2)]/\mathrm{Z}_2$ \cite{Yang1990,Carmelo2026}. We restrict our derivation to the particle number conserving (non-superconducting) state.

The Hamiltonian in \cref{Eq:Hamiltonian} is ph-symmetric if $\delta\mu= B_z=0$ and if $t_{\i\j}$ is only non-zeros between an even and an odd site. In general our derivations work if the model has a bipartite lattice structure such as hyper-cubic, honeycomb, body-centered cubic, face-centered cubic and diamond lattices with only nearest neighbor hopping if the vector $\vk_\textbf{P}$ is adapted accordingly. Further, it is $SU(2)_P$-symmetric if $\delta\mu=t^\prime=0$ (or more general $\delta\mu=0$ and a bipartite lattice) and $SU(2)$-symmetric if $B_z=0$.

\section{Definition of one- and two-particle quantities}
We define the one-particle Green function as
\begin{align}
\label{Eq:Def_G}
\begin{split}
    G^k_\sigma&=\frac{1}{\beta}\int_0^\beta \mathrm{d}\tau_1 \mathrm{d}\tau_2 \mathrm{e}^{-\i\nu\tau_1} \mathrm{e}^{\i\nu\tau_2} \frac{1}{N}\sum_{12} \mathrm{e}^{-\i\vk\vR_1} \mathrm{e}^{\i\vk\vR_2} G^{12}_\sigma\\
    G^{12}_\sigma&=-\langle T_{\tau} c_{1,\sigma} (\tau_1)c^\dagger_{2,\sigma}(\tau_2)\rangle,
\end{split}
\end{align}
with the imaginary time ordering operator $T_{\tau}$, inverse temperature $\beta$, fermionic Matsubara frequency $\nu$, imaginary times $\tau_i$, lattice vector $\vR_i$, momentum $\vk$, spin $\sigma$ and the number of lattice sites $N$. Further we denote the four momenta as $k=(\nu,\vk)$ and the space-time argument as $i=(\tau_i,\vR_i)$.
With this we can define the ph and pp bubble as
\begin{align}
\begin{split}
\label{Eq:Def_chi0_ph}
    \chi^{kk^\prime q}_{0,\ph,\sigma\sigma^\prime} =&\frac{1}{\beta}\int_0^\beta \mathrm{d}\tau_1 \mathrm{d}\tau_2 \mathrm{d}\tau_3 \mathrm{d}\tau_4  \mathrm{e}^{-\i\nu\tau_1} \mathrm{e}^{\i(\omega+\nu)\tau_2}\mathrm{e}^{-\i(\omega+\nu')\tau_3} \mathrm{e}^{\i\nu'\tau_4}\\
&\frac{1}{N^2}\sum_{1234} \mathrm{e}^{-\i\vk\vR_1} \mathrm{e}^{\i(\vk+\vq)\vR_2}\mathrm{e}^{-\i(\vk^\prime+\vq)\vR_3}\mathrm{e}^{\i\vk^\prime\vR_4} \chi_{0,\ph,\sigma\sigma'}^{12 |34},\\
\chi_{0,\ph,\sigma\sigma'}^{12 |34}=&- G_\sigma^{41}G_{\sigma}^{23}\delta_{\sigma\sigma^\prime},
\end{split}
\end{align}
\begin{align}
\begin{split}
\label{Eq:Def_chi0_pp}
    \chi^{kk^\prime q}_{0,\pp,\sigma\sigma^\prime} =&\frac{1}{\beta}\int_0^\beta \mathrm{d}\tau_1 \mathrm{d}\tau_2 \mathrm{d}\tau_3 \mathrm{d}\tau_4  \mathrm{e}^{-\i\nu\tau_1} \mathrm{e}^{\i(\omega-\nu^\prime)\tau_2}\mathrm{e}^{-\i(\omega-\nu)\tau_3} \mathrm{e}^{\i\nu'\tau_4}\\
    &\frac{1}{N^2}\sum_{1234} \mathrm{e}^{-\i\vk\vR_1} \mathrm{e}^{\i(\vq-\vk^\prime)\vR_2}\mathrm{e}^{-\i(\vq-\vk)\vR_3}\mathrm{e}^{\i\vk^\prime\vR_4} \chi_{0,\ph,\sigma\sigma'}^{13 |24},\\
    \chi_{0,\pp,\sigma\sigma'}^{13 |24}=& G_\sigma^{41}G_{\sigma^\prime}^{23},
\end{split}
\end{align}
where $\omega$ is a bosonic Matsubara frequency. Further, we define the generalized susceptibility in the longitudinal ph-channel as
\begin{equation}
\begin{split}
\label{Eq:Def_ph}
\chi^{kk^\prime q}_{\mathrm{ph},\sigma\sigma'}=&\frac{1}{\beta}\int_0^\beta \mathrm{d}\tau_1 \mathrm{d}\tau_2 \mathrm{d}\tau_3 \mathrm{d}\tau_4  \mathrm{e}^{-\i\nu\tau_1} \mathrm{e}^{\i(\omega+\nu)\tau_2}\mathrm{e}^{-\i(\omega+\nu')\tau_3} \mathrm{e}^{\i\nu'\tau_4}\\
&\frac{1}{N^2}\sum_{1234} \mathrm{e}^{-\i\vk\vR_1} \mathrm{e}^{\i(\vk+\vq)\vR_2}\mathrm{e}^{-\i(\vk^\prime+\vq)\vR_3}\mathrm{e}^{\i\vk^\prime\vR_4} \chi_{\ph,\sigma\sigma'}^{12 |34}\\ \chi_{\ph,\sigma\sigma'}^{12 |34}=&\langle T_{\tau}c_{\mathrm{1},\sigma}^\dagger (\tau_1)c_{\mathrm{2},\sigma}(\tau_2)c^\dagger_{\mathrm{3},\sigma'}(\tau_3)c_{4,\sigma^\prime}(\tau_4)\rangle -G^{\mathrm{21}}_{\sigma}G^{43}_{\sigma'},
\end{split}
\end{equation}
and in the transversal ph-channel as
\begin{equation}
\begin{split}
\label{Eq:Def_ph_transverse}
\chi^{kk^\prime q}_{\mathrm{ph},\overline{\sigma\sigma'}}=&\frac{1}{\beta}\int_0^\beta \mathrm{d}\tau_1 \mathrm{d}\tau_2 \mathrm{d}\tau_3 \mathrm{d}\tau_4  \mathrm{e}^{-\i\nu\tau_1} \mathrm{e}^{\i(\omega+\nu)\tau_2}\mathrm{e}^{-\i(\omega+\nu')\tau_3} \mathrm{e}^{\i\nu'\tau_4}\\
&\frac{1}{N^2}\sum_{1234} \mathrm{e}^{-\i\vk\vR_1} \mathrm{e}^{\i(\vk+\vq)\vR_2}\mathrm{e}^{-\i(\vk^\prime+\vq)\vR_3}\mathrm{e}^{\i\vk^\prime\vR_4}\chi_{\ph,\overline{\sigma\sigma'}}^{12 |34},\\
\chi_{\ph,\overline{\sigma\sigma'}}^{12 |34}=&\langle T_{\tau}c_{\mathrm{1},\sigma}^\dagger (\tau_1)c_{\mathrm{2},\sigma^\prime}(\tau_2)c^\dagger_{\mathrm{3},\sigma'}(\tau_3)c_{4,\sigma}(\tau_4)\rangle.
\end{split}
\end{equation}
In the pp-channel we define the generalized susceptibility as
\begin{equation}
\begin{split}
\label{Eq:Def_pp}
\chi^{kk^\prime q}_{\pp,\sigma\sigma^\prime}=&\frac{1}{\beta}\int_0^\beta \mathrm{d}\tau_1 \mathrm{d}\tau_2 \mathrm{d}\tau_3 \mathrm{d}\tau_4 \mathrm{e}^{-\i\nu\tau_1} \mathrm{e}^{\i\nu^\prime\tau_2}\mathrm{e}^{-\i(\omega-\nu)\tau_3} \mathrm{e}^{\i(\omega-\nu^\prime)\tau_4}\\
&\frac{1}{N^2}\sum_{1234} \mathrm{e}^{-\i \vk\vR_1} \mathrm{e}^{\i \vk^\prime \vR_2}\mathrm{e}^{-\i(\vq-\vk)\vR_3}\mathrm{e}^{\i(\vq-\vk^\prime)\vR_4} \chi_{\pp,\sigma\sigma^\prime}^{13 |42},\\
\chi_{\pp,\sigma\sigma'}^{13 |42}=&\langle T_{\tau}c_{\mathrm{1},\sigma}^\dagger (\tau_1)c_{\mathrm{2},\sigma}(\tau_2)c^\dagger_{\mathrm{3},\sigma'}(\tau_3)c_{4,\sigma'}(\tau_4)\rangle \\
=&G_{\pp,\sigma\sigma'}^{13 |42},
\end{split}
\end{equation}
where $\sigma\neq \sigma^\prime$ is assumed. Note that this definition of the pp-channel is equivalent to pp-channel in the transversal spin channel in the notation of G. Rohringer in Ref.~\cite{Rohringer2013a,Essl2024}, i.e. $\chi^{kk^\prime q}_{\pp,\sigma\sigma^\prime}=\chi^{kk^\prime q}_{\pp,\overline{\sigma\sigma^\prime},\text{Rohr.}}$.

\section{Symmetry relations for generalized susceptibilities}
From now on, we understand the time ordering operator $T_\tau$ implicitly. For the real space, the following relations hold due to $SU(2)$-spin symmetry, complex conjugation CC and the time ordering convention $T_\tau$ \cite{Rohringer2013a}:
\begin{equation}
\begin{array}{rlc}
    G^{1234}_{\sigma\sigma'}&=G^{1234}_{-\sigma,-\sigma'}& \qquad \mathrm{SU(2)} \\[8pt]
    \left(G^{1234}_{\sigma\sigma'}\right)^*&=G^{-2,-1,-4,-3}_{\sigma\sigma'} & \qquad \mathrm{CC} \\[8pt]
    G^{1234}_{\sigma\sigma'}&=G^{2143}_{\sigma\sigma'} &\qquad T_\tau \\[8pt]
    G^{1234}_{\sigma\sigma'}&=-G^{2341}_{\overline{\sigma\sigma'}} &\qquad T_\tau\\
    G^{1...n} &= G^{*,1...n}\prod_{i=1}^n\e^{\i\vpi\vR_i}& \mathrm{ph sym.}
\end{array}
\end{equation} where $\pm1=(\vR_1,\pm\tau_1)$. Further, we define the real space inversion on a lattice $\vR_1,\vR_2,\vR_3,\vR_4\rightarrow -\vR_1,-\vR_2,-\vR_3,-\vR_4$. 

\paragraph{Consequences for Momentum Space} We will now transfer the real-space symmetries to the momentum space generalized susceptibilities (see also Refs.~\cite{Rohringer2013a,Essl2024}). Note that the \emph{crossing symmetry} and the \emph{complex conjugation}, which are a result of the anticommutation of fermions and the complex conjugation of an expectation value, are strictly speaking no symmetries in the sense that an operator commutes with the Hamiltonian. We remind the reader, that we employ the compact notation $k=(\vk,\nu)$ etc.

For the \emph{crossing symmetry} it holds that
\begin{align}
\label{Eq:cs_sym}
    \chi_{\sigma\sigma^\prime/\overline{\sigma\sigma^\prime},\text{ph}}^{k,k^\prime, q}&= \chi_{\sigma^\prime\sigma/\overline{\sigma^\prime\sigma},\text{ph}}^{k^\prime+q,k+q, -q},\\
\label{Eq:cs_sym_pp}
    \chi_{\sigma\sigma^\prime,\text{pp}}^{k,k^\prime, q}&= \chi_{\sigma^\prime\sigma,\text{pp}}^{q-k,q-k^\prime, q}.
\end{align}
If the Hamiltonian is a real function of $c$ and $c^\dagger$ ($H\in\mathbb{R}$), which implies time reversal symmetry, it holds:
\begin{align}
\label{Eq:H_real}
    \chi_{\sigma\sigma^\prime/\overline{\sigma\sigma^\prime},\text{ph/pp}}^{\vk,\vk^\prime, \vq}(\nu,\nu^\prime,\omega)= \chi_{\sigma^\prime\sigma/\overline{\sigma\sigma^\prime},\text{ph/pp}}^{-\vk^\prime,-\vk, -\vq}(\nu^\prime,\nu,\omega).
\end{align}
For \emph{space inversion symmetry} ($\mathbf{R}=-\mathbf{R}$) it holds:
\begin{align}
\label{Eq:space_inv_sym}
    \chi_{\sigma\sigma^\prime/\overline{\sigma\sigma^\prime},\text{ph/pp}}^{\vk,\vk^\prime, \vq}(\nu,\nu^\prime,\omega)= \chi_{\sigma\sigma^\prime/\overline{\sigma\sigma^\prime},\text{ph/pp}}^{-\vk,-\vk^\prime, -\vq}(\nu,\nu^\prime,\omega).
\end{align}
Combining space inversion \cref{Eq:space_inv_sym} and $H\in\mathbb{R}$ \cref{Eq:H_real} it holds that 
\begin{align}
\label{Eq:space_inv_sym_H_real}
    \chi_{\sigma\sigma^\prime,\text{ph}}^{k,k^\prime, q}= \chi_{\sigma^\prime\sigma,\text{ph}}^{k^\prime,k, q},\;\;
    \chi_{\sigma\sigma^\prime,\text{pp}}^{k,k^\prime, q}= \chi_{\sigma\sigma^\prime,\text{pp}}^{k^\prime,k, q},
\end{align}
from which follows that the all physical channels \cref{Eq:chi_Sz,Eq:chi_Sx,Eq:chi_s_pz,Eq:chi_s_px,Eq:chi_L} are symmetric matrices. 
For complex \emph{complex conjugation} it holds:
\begin{align}
\label{Eq:cc_sym}
    \chi_{\sigma\sigma^\prime/\overline{\sigma\sigma^\prime},\text{ph}}^{k,k^\prime, q}= \left(\chi_{\sigma^\prime\sigma/\overline{\sigma\sigma^\prime},\text{ph}}^{\vk^\prime,\vk, \vq}(-\nu^\prime,-\nu,-\omega)\right)^*\overset{\cref{Eq:space_inv_sym}}{=}\left(\chi_{\sigma^\prime\sigma/\overline{\sigma\sigma^\prime},\text{ph}}^{-k^\prime,-k, -q}\right)^*,\\
\label{Eq:cc_sym_trans}
    \chi_{\sigma\sigma^\prime,\text{pp}}^{k,k^\prime, q}= \left(\chi_{\sigma\sigma^\prime,\text{pp}}^{\vk^\prime,\vk, \vq}(-\nu^\prime,-\nu,-\omega)\right)^*\overset{\cref{Eq:space_inv_sym}}{=}\left(\chi_{\sigma\sigma^\prime,\text{pp}}^{-k^\prime,-k, -q}\right)^*.
\end{align}
Further, for $SU(2)$-symmetry (spin-symmetry), the generalized susceptibilities are invariant under a spin flip in the Bloch sphere's z-direction, i.e.~$(c_{i,\sigma}^\dagger\rightarrow -\sigma c_{i,-\sigma}^\dagger)$ and  $(c_{i,\sigma}\rightarrow -\sigma c_{i,-\sigma})$, which leads to:
\begin{align}
\label{Eq:SU2S_sym}
\begin{split}
    \chi_{\sigma\sigma^\prime/\overline{\sigma\sigma^\prime},\text{ph/pp}}^{k,k^\prime, q}=&\chi_{-\sigma,-\sigma^\prime/\overline{-\sigma,-\sigma^\prime},\text{ph/pp}}^{k,k^\prime, q}.\\
\end{split}
\end{align}
Similarly, they are also invariant under a pseudo-spin flip in z-direction, i.e., $c_{i,\sigma}^\dagger\rightarrow -\mathrm{sign(}\sigma) (-1)^i c_{i,-\sigma}$ and \mbox{$c_{i,\sigma}\rightarrow-\mathrm{sign(}\sigma) (-1)^i c^\dagger_{i,-\sigma}$}, for $SU(2)_P$-symmetry (pseudo-spin-symmetry) which leads to:
\begin{align}
\label{Eq:SU2P_sym_ph}
    \chi_{\sigma\sigma^\prime,\text{ph}}^{k,k^\prime, q}&=\chi_{-\sigma,-\sigma^\prime,\text{ph}}^{-k-\Pi-q,-k^\prime-\Pi-q, q}\overset{\cref{Eq:cs_sym,Eq:cc_sym}}{=}\left(\chi_{-\sigma,-\sigma^\prime,\text{ph}}^{k+\Pi,k^\prime+\Pi, q}\right)^*,\\
\label{Eq:SU2P_sym_pp_trans}
    \chi_{\sigma\sigma^\prime,\text{pp}}^{k,k^\prime, q}&=\chi_{-\sigma^\prime,-\sigma,\text{pp}}^{k^\prime-\Pi-q,k-\Pi-q, -q}\overset{\cref{Eq:cs_sym_pp,Eq:cc_sym_trans}}{=}\left(\chi_{-\sigma,-\sigma^\prime,\text{pp}}^{k+\Pi,k^\prime+\Pi, q}\right)^*,
\end{align}
where Fourier transforming \cref{Eq:SU2P_sym_ph} to real space gives:
\begin{align}
\label{Eq:SU2P_sym_ph_space}
    \chi_{\sigma\sigma^\prime,\text{ph}}^{ij|hl}(\nu,\nu^\prime,\omega)&=(-1)^{i+j+h+l}\chi_{-\sigma,-\sigma^\prime,\text{ph}}^{ji|lh} (-\nu-\omega,-\nu^\prime-\omega,\omega).
\end{align}

\section{Real space derivation}
For $SU(2)_P$-symmetry an arbitrary rotation of the pseudo-spinors \cref{Eq:spinor_rot} does not change the expectation value. Specifically, by rotating by $\pi/2$ around the $y$-axis, we obtain:
\begin{equation}
    \begin{split}
    \label{Eq:pesudospinors}
        \vec{c}\,'\,^\dagger_{i}&=\begin{pmatrix} c'\,^{\dagger}_{i,\uparrow} \\ c'\,^{\dagger}_{i,\downarrow} \end{pmatrix}=\frac{1}{\sqrt{2}}\begin{pmatrix} c^{\dagger}_{i,\uparrow}-\e^{\i\vpi\vR_i}c_{i,\downarrow } \\ c_{i,\uparrow }\e^{\i\vpi\vR_i}+c^{\dagger}_{i,\downarrow}\end{pmatrix},\\
        \vec{c}\,'_{i}&=\begin{pmatrix} c'_{i,\uparrow} \\ c'_{i,\downarrow} \end{pmatrix}=\frac{1}{\sqrt{2}}\begin{pmatrix} c_{i,\uparrow}-\e^{-\i\vpi\vR_i}c^\dagger_{i,\downarrow } \\ c^\dagger_{i,\uparrow }\e^{-\i\vpi\vR_i}+c_{i,\downarrow}\end{pmatrix}.
    \end{split}
\end{equation}
Substituting the fermionic creation- and annihilation operators $c$ and $c^\dagger$ of the definition of the two-particle Green function by the rotated $c'$ and $c'^\dagger$ we obtain in the normal state (particle number conservation):
\begin{equation}
\begin{split}
\label{Eq:realspace_to_pp_Gupdown} 
G^{1234}_{\uparrow\uparrow}=&[G^{1234}_{\uparrow\uparrow}]^\prime\\
    G^{1234}_{\uparrow\uparrow}\overset{T_\tau}{=}&\frac{1}{4}\left[ \langle c^\dagger_{1\uparrow}c_{2\uparrow}c^\dagger_{3\uparrow}c_{4\uparrow}\rangle+\langle c_{1\uparrow}c^\dagger_{2\uparrow}c_{3\uparrow}c^\dagger_{4\uparrow}\rangle\e^{\i\vpi(\vR_1-\vR_2+\vR_3-\vR_4)}\right. \\  &\left. -\e^{\i\vpi(\vR_1-\vR_2)}
    \langle c^\dagger_{2\uparrow}c_{1\uparrow}c^\dagger_{3\downarrow}c_{4\downarrow}\rangle-\e^{\i\vpi(\vR_3-\vR_4)}\langle c^\dagger_{1\uparrow}c_{2\uparrow}c^\dagger_{4\downarrow}c_{3\downarrow}\rangle+\e^{\i\vpi(\vR_3-\vR_2)}
    \langle c^\dagger_{1\uparrow}c_{4\uparrow}c^\dagger_{2\downarrow}c_{3\downarrow}\rangle+\e^{\i\vpi(\vR_1-\vR_4)}\langle c^\dagger_{3\uparrow}c_{2\uparrow} c^\dagger_{4\downarrow}c_{1\downarrow}\rangle\right] 
\end{split}
\end{equation}
This equation holds for systems without SU(2) symmetry. Assuming SU(2) symmetry via $B_z=0$, we can exploit ph-symmetry and complex conjugation \footnote{Note that complex conjugation can not be straightforwardly applied to CDMFT lattice quantities, as they also carry the super-lattice momentum vector $\tilde{\vq}$.}
\begin{equation}
\begin{split}
     G^{1234}_{\uparrow\uparrow}=&\frac{1}{4}\left[ G^{1234}_{\uparrow\uparrow}+G^{1234}_{\uparrow\uparrow}\right.\\&\left. -\e^{\i\vpi(\vR_3-\vR_4)}
    G_{\uparrow\downarrow}^{1243}-\e^{\i\vpi(\vR_3-\vR_4)}G_{\uparrow\downarrow}^{1243}+\e^{\i\vpi(\vR_3-\vR_2)}
    G_{\uparrow\downarrow}^{1423}+\e^{\i\vpi(\vR_3-\vR_2)} G_{\uparrow\downarrow}^{1423}\right],
\end{split}
\end{equation}
exploiting $G_{\uparrow\downarrow}^{1423}=G_{\pp,\uparrow\downarrow}^{1234}$ leads to:
\begin{equation}
\label{Eq:relation_realpsace}
\begin{split}
     2\e^{\i\vpi(\vR_3-\vR_2)} G_{\pp,\uparrow\downarrow}^{1234}=&2G^{1234}_{\uparrow\uparrow}+2\e^{\i\vpi(\vR_3-\vR_4)}G^{1243}_{\uparrow\down}\\
     =&G^{1234}_{\ch}+G^{1234}_{\sp}+\e^{\i\vpi(\vR_3-\vR_4)}\left(G^{1243}_{\ch}-G^{1243}_{\sp}\right).
\end{split}
\end{equation} From this, the expressions for the physical susceptibilities in the main text Fig.~1 can be obtained by Fourier transforming and applying fermionic form factors see next section.

Exploiting ph-symmetry and complex conjugation flips the correlator and reproduces the Matsubara version of Eq.~(\ref{Eq:relation_realpsace}) for the static bosonic frequency $\omega=0$:
\begin{equation}
\begin{split}
     2\e^{\i\vpi(\vR_h-\vR_j)} G_{\pp,\uparrow\downarrow}^{ijhl}(\nu,\nu')=&G^{ijhl}_{\ch}(\nu,\nu')+G^{ijhl}_{\sp}(\nu,\nu')+\e^{\i\vpi(\vR_h-\vR_l)}\left(G^{ijhl}_{\ch}-G^{ijhl}_{\sp}\right)(\nu,-\nu').
\end{split}
\end{equation}
Finally note that the same analysis can be conducted for the disconnected part of $-G^{\mathrm{12}}_{\up}(\tau_1,\tau_2)G^{34}_{\up}(\tau_3,\tau_4)$ where the pseudospin transformation results in
\begin{equation}
\begin{split}
    [G_{disc,\uparrow\uparrow}^{1234}]^\prime=&\frac{1}{4}\left[ \langle c^\dagger_{1\uparrow}c_{2\uparrow}\rangle \langle c^\dagger_{3\uparrow}c_{4\uparrow}\rangle+\langle c_{1\uparrow}c^\dagger_{2\uparrow}\rangle \langle c_{3\uparrow}c^\dagger_{4\uparrow}\rangle\e^{\i\vpi(\vR_1-\vR_2+\vR_3-\vR_4)}\right. \\  &\left. -\e^{\i\vpi(\vR_1-\vR_2)}
    \langle c^\dagger_{2\uparrow}c_{1\uparrow}\rangle \langle c^\dagger_{3\downarrow}c_{4\downarrow}\rangle-\e^{\i\vpi(\vR_3-\vR_4)}\langle c^\dagger_{1\uparrow}c_{2\uparrow}\rangle \langle c^\dagger_{4\downarrow}c_{3\downarrow}\rangle \right.\\   &\left.+\e^{\i\vpi(\vR_3-\vR_2)}
    \langle c^\dagger_{1\uparrow}c^\dagger_{2\downarrow}\rangle \langle c_{3\downarrow}c_{4\uparrow}\rangle+\e^{\i\vpi(\vR_1-\vR_4)}\langle c_{1\downarrow}c_{2\uparrow}\rangle \langle c^\dagger_{3\uparrow}c^\dagger_{4\downarrow}\rangle\right]. \\
\end{split}
\end{equation} The Gor'kov-like (anomalous) correlators are forbidden due to particle number conservation. Hence, one finds that above relation Eq.~(\ref{Eq:relation_realpsace}) holds when inserting the generalized susceptibility for all particle-hole correlators instead of the four-point Green function:
\begin{equation}
\begin{split}
     2\e^{\i\vpi(\vR_3-\vR_2)} \chi_{\pp,\uparrow\downarrow}^{1234}=&2\chi^{1234}_{\uparrow\uparrow}+2\e^{\i\vpi(\vR_3-\vR_4)}\chi^{1243}_{\uparrow\down}\\
     =&\chi^{1234}_{\ch}+\chi^{1234}_{\sp}+\e^{\i\vpi(\vR_3-\vR_4)}\left(\chi^{1243}_{\ch}-\chi^{1243}_{\sp}\right).
\end{split}
\end{equation}

\section{Consequences for physical response functions}\label{sec:phys_respose_degeneracancy}
Analogous to the local spin and pseudo-spin operators \cref{Eq:local_S,Eq:local_S_P} we define the non-local operators by 
\begin{align}
    S_{\alpha}^{12} &= \frac{1}{2}\mathbf{c}^\dagger_1 \sigma_\alpha \mathbf{c}_2,\\
    S_{P,\alpha}^{12} &= \frac{1}{2}\mathbf{c}^\dagger_{P,1} \sigma_\alpha \mathbf{c}_{P,2},
\end{align}
with the space-time index $i=(\tau_i,\vR_i)$. The Fourier transformation ($X^{kk^\prime}=\int_0^\beta \mathrm{d}\tau_1 \mathrm{d}\tau_2\; \mathrm{e}^{-\i\tau_1 \nu}\mathrm{e}^{\i\tau_2 \nu^\prime}\frac{1}{N}\sum_{1,2} \mathrm{e}^{-\i\vk \vR_1} \mathrm{e}^{\i\vk^\prime \vR_2} X^{12}$) reads:
\begin{align}
    S_{x}^{kk^\prime} &= \frac{1}{2}\left(c^\dagger_{k,\uparrow}c_{k^\prime,\downarrow}+c^\dagger_{k,\downarrow}c_{k^\prime,\uparrow}\right),\\
    S_{z}^{kk^\prime} &= \frac{1}{2}\left(c^\dagger_{k,\uparrow}c_{k^\prime,\uparrow}-c^\dagger_{k,\downarrow}c_{k^\prime,\downarrow}\right),\\
    S_{P,x}^{kk^\prime} &= \frac{1}{2}\left(c^\dagger_{k,\uparrow}c^\dagger_{-k^\prime-\Pi,\downarrow}+c_{-k-\Pi,\downarrow}c_{k^\prime,\uparrow}\right),\\
    S_{P,z}^{kk^\prime} &= \frac{1}{2}\left(c^\dagger_{k,\uparrow}c_{k^\prime,\uparrow}+c^\dagger_{-k^\prime-\Pi,\downarrow}c_{-k-\Pi,\downarrow}-1\right),
\end{align}

where the definition is such that under the Shiba transformation $(c^\dagger_{k,\downarrow}\rightarrow c_{-k-\Pi,\downarrow})$ pseudo-spin and spin operators map onto each other $S_\alpha\rightarrow S_{P,\alpha}$.

Further, we define generalized susceptibilities that are related to the (pseudo-)spin operators as:
\begin{align}
\label{Eq:chi_Sz}
    \chi_{S_z}^{k,k^\prime,q}&=\frac{1}{2}\langle S_z^{k,k+q}S_z^{k^\prime+q,k^\prime}\rangle-\frac{1}{2}\langle S_z^{k,k+q}\rangle\langle S_z^{k^\prime+q,k^\prime}\rangle=\frac{1}{2}\left(\chi_{\uparrow\uparrow}+\chi_{\downarrow\downarrow}-\chi_{\uparrow\downarrow}-\chi_{\downarrow\uparrow}\right)^{k,k^\prime,q},\\
\label{Eq:chi_Sx}
    \chi_{S_x}^{k,k^\prime,q}&=\frac{1}{2}\langle S_x^{k,k+q}S_z^{k^\prime+q,k^\prime}\rangle-\frac{1}{2}\langle S_x^{k,k+q}\rangle\langle S_x^{k^\prime+q,k^\prime}\rangle=\frac{1}{2}\left(\chi_{\overline{\uparrow\downarrow}}+\chi_{\overline{\downarrow\uparrow}}\right)^{k,k^\prime,q},\\
\label{Eq:chi_s_pz}
    \chi_{S_{P,z}}^{k,k^\prime,q}&=\frac{1}{2}\langle S_{P,z}^{k,k+q}S_{P,z}^{k^\prime+q,k^\prime}\rangle-\frac{1}{2}\langle S_{P,z}^{k,k+q}\rangle\langle S_{P,z}^{k^\prime+q,k^\prime}\rangle=\frac{1}{2}\left(\chi_{\uparrow\uparrow}^{k,k^\prime,q}+\chi_{\downarrow\downarrow}^{\overline{k},\overline{k^\prime},q}+\chi_{\uparrow\downarrow}^{k,\overline{k^\prime},q}+\chi_{\downarrow\uparrow}^{\overline{k},k^\prime,q}\right),\\
\label{Eq:chi_s_px}
    \chi_{S_{P,x}}^{k,k^\prime,q}&=\frac{1}{2}\langle S_{P,x}^{k,k-q-\Pi}S_{P,x}^{k^\prime-q-\Pi,k^\prime}\rangle-\frac{1}{2}\langle S_{P,x}^{k,k-q-\Pi}\rangle\langle S_{P,x}^{k^\prime-q-\Pi,k^\prime}\rangle=\frac{1}{2}\left(-\chi_{\uparrow\downarrow,\text{pp}}^{k,k^\prime,q}-\left(\chi_{\downarrow\uparrow,\text{pp}}^{k+\Pi,k^\prime+\Pi,q}\right)^*\right),
\end{align}
where the 4 momentum $\overline{k}$ is defined as $\overline{k}=-k-\Pi-q$ with $\Pi=(0,\vpi)$. Note that the vector $\vpi$ has to be modified if another bipartite lattice than the hyper-cubic one is considered. 
Evidently these generalized susceptibilities map onto each other under the Shiba transformation:
\begin{align}
\label{Eq:map_Sz}
    \chi_{S_z;U}^{k,k^\prime,q}=\chi_{S_{P,z};-U}^{k,k^\prime,q},\\
\label{Eq:map_Sx}
    \chi_{S_x;U}^{k,k^\prime,q}=\chi_{S_{P,x};-U}^{k,k^\prime,-q-\Pi},
\end{align}
where again the sub-labels $U$ and $-U$ indicate the parameters of the model that are related via the Shiba transformation, i.e., in the case of the simple square lattice Hubbard model without frustration $(t^\prime=0)$: $U\hat{=}\{\delta\mu,h,U\}$ and $-U\hat{=}\{h,\delta\mu,-U\}$.

To make a connection to more familiar channels \cite{Essl2024,Reitner2025}, namely the (potentially coupled) charge and spin channel
\begin{align}
    \label{Eq:chi_L}
    \begin{split}
    \chi_\text{ch/sp}^{k,k^\prime,q}&=\frac{1}{2}\left(\chi_{\uparrow\uparrow}+\chi_{\downarrow\downarrow}\pm(\chi_{\uparrow\downarrow}+\chi_{\downarrow\uparrow})\right)^{k,k^\prime,q},\\
    \chi_\text{cs/sc}^{k,k^\prime,q}&=\frac{1}{2}\left(\chi_{\uparrow\uparrow}-\chi_{\downarrow\downarrow}\pm(-\chi_{\uparrow\downarrow}+\chi_{\downarrow\uparrow})\right)^{k,k^\prime,q},\\
    \chi_L&=\begin{pmatrix}
        \chi_\text{ch}&\chi_\text{cs}\\
        \chi_\text{sc}&\chi_\text{sp}
    \end{pmatrix},
    \end{split}
\end{align}
with $\chi_\text{cs}=\chi_\text{sc}=0$ for $SU(2)$-symmetry [\cref{Eq:SU2S_sym}]. We now relate $\chi_{S_z}$ and $\chi_{S_{P,z}}$ to charge and spin channel:
\begin{align}
\label{Eq:chi_sz_to_L}
    \chi_{S_z}^{k,k^\prime,q}&=\chi_\text{sp}^{k,k^\prime,q},\\
\label{Eq:chi_spz_to_L}
    \chi_{S_{P,z}}^{k,k^\prime,q}&=\frac{1}{4} \bigg(\chi_\text{ch}^{k,k^\prime,q}+\chi_\text{sp}^{k,k^\prime,q}+\chi_\text{cs}^{k,k^\prime,q}+\chi_\text{sc}^{k,k^\prime,q}
    +\chi_\text{ch}^{\overline{k},\overline{k^\prime},q}+\chi_\text{sp}^{\overline{k},\overline{k^\prime},q}-\chi_\text{cs}^{\overline{k},\overline{k^\prime},q}-\chi_\text{sc}^{\overline{k},\overline{k^\prime},q}\\
    &\phantom{=\frac{1}{4}\bigg(}+\chi_\text{ch}^{k,\overline{k^\prime},q}-\chi_\text{sp}^{k,\overline{k^\prime},q}-\chi_\text{cs}^{k,\overline{k^\prime},q}+\chi_\text{sc}^{k,\overline{k^\prime},q}\nonumber+\chi_\text{ch}^{\overline{k},k^\prime,q}-\chi_\text{sp}^{\overline{k},k^\prime,q}+\chi_\text{cs}^{\overline{k},k^\prime,q}-\chi_\text{sc}^{\overline{k},k^\prime,q}\bigg)\nonumber.
\end{align}
Note, that $\chi_{S_z}$ and $\chi_{S_{P,z}}$ do in general not represent channels with regard to the BSE. When working with the BSE the longitudinal channel $\chi_L$, or the decoupled charge and spin channels in the case of $SU(2)$-symmetry, has to be used.
Further, $\chi_{S_x}$ and $\chi_{S_{P,x}}$ only represent decoupled channels with regards to the BSE for $SU(2)$- respectively $SU(2)_P$-symmetry [this can be seen by using \cref{Eq:SU2S_sym,Eq:SU2P_sym_pp_trans}]. 
If theses symmetries are broken one has to use $\chi_{\overline{\uparrow\downarrow}}$ and $\chi_{\overline{\downarrow\uparrow}}$ (resp. $\chi_{\uparrow\downarrow,\text{pp}}$ and $\chi_{\downarrow\uparrow,\text{pp}}$) separately when working with the BSE.

To relate the physical response function of $S_{P,z}$ with the charge and spin response we introduce form factors $f_r$ for the fermionic (internal) degrees of freedom:
\begin{align}
\label{Eq:phys_response}
\begin{split}
    \chi^{r,q}&=\frac{1}{N\beta^2}\sum_{kk^\prime}f_r(k)\chi^{k,k^\prime,q}f_{r}(k^\prime),\\
    \chi^{(r,r^\prime),q}&=\frac{1}{N\beta^2}\sum_{kk^\prime}f_r(k)\chi^{k,k^\prime,q}f_{r^\prime}(k^\prime),
\end{split}
\end{align}
where we also introduce response functions $\chi^{(r,r')}$ that are of mixed form factor character. Note, that we refer to the geometric sector $r$ of the form factor $f_r$ as ``wave''.
Further, by considering the following relations:
\begin{align}
\label{Eq:parity_sum}
\begin{split}
    \sum_{k^\prime}\chi^{k,\overline{k^\prime},q}f_r(k^\prime) &=    
    \begin{cases}
      \sum_{k}\chi^{k,k^\prime,q}f_r(k^\prime), & \text{if}\ f_r(k)=f_r(\overline{k})\\
      -\sum_{k}\chi^{k,k^\prime,q}f_r(k^\prime), & \text{if}\ f_r(k)=-f_r(\overline{k})
    \end{cases}\\
    \sum_{k}f_r(k)\chi^{\overline{k},k^\prime,q} &=    
    \begin{cases}
      \sum_{k^\prime}f_r(k)\chi^{k,k^\prime,q}, & \text{if}\ f_r(k)=f_r(\overline{k})\\
      -\sum_{k^\prime}f_r(k)\chi^{k,k^\prime,q}, & \text{if}\ f_r(k)=-f_r(\overline{k})
    \end{cases}\\
    \sum_{kk^\prime}f_r(k)\chi^{\overline{k},\overline{k^\prime},q}f_{r^\prime}(k^\prime) &= \sum_{kk^\prime}f_r(k)\chi^{k,k^\prime,q}f_{r^\prime}(k^\prime) \text{ if both $f_r$ and $f_{r^\prime}$ are either ``even" or ``odd" form factors},
\end{split}
\end{align}
where we define a ``$\bar{k}$-parity", i.e.~$f_r(k)=\pm f_r(\overline{k})$, with respect to applying the momentum shift $\overline{k}=-k-\Pi-q$ that depends on both the wave $r$ and the transfer momentum $\vq$. As the momentum-shift originates from the Shiba mapping, its consequences for the fermionic form factors will prove essential to study the consequences of the pseudospin symmetry for the physical susceptibilities.

By using \cref{Eq:chi_s_pz} together with \cref{Eq:parity_sum} it is evident that the physical response function of $S_{P,z}$ is either equal to the charge or the spin response:
\begin{align}
\label{Eq:phys_S_Pz}
    \chi_{S_{P,z}}^{r,q} &=    
    \begin{cases}
      \chi_\text{ch}^{r,q}, & \text{if}\ f_r(k)=f_r(\overline{k})\\
      \chi_\text{sp}^{r,q}, & \text{if}\ f_r(k)=-f_r(\overline{k})
    \end{cases}
\end{align}
if the form factor has a defined $\bar{k}$-parity.

\subsection{Mapping for $SU(2)_P$-symmetry}\label{sec:map_SU2P}
Mapping a system with $SU(2)_P$-symmetry (denoted by the sub-label $U$, where $U\hat{=}\{0,h,U\}$), means that the mapped system of flipped interaction (denoted by where $-U\hat{=}\{h,0,-U\}$) is $SU(2)$-symmetric. Consequently, this allows use to directly relate the charge and spin response to the pp-response.

By using \cref{Eq:SU2P_sym_ph} for the coupled charge-spin channel of \cref{Eq:chi_L} we get 
\begin{align}
\label{Eq:SU2P_c_s}
\chi_\text{ch/sp}^{k,k^\prime,q}&=\chi_\text{ch/sp}^{\overline{k},\overline{k^\prime},q},\\
\label{Eq:SU2P_cs_sc}
\chi_\text{cs/sc}^{k,k^\prime,q}&=-\chi_\text{cs/sc}^{\overline{k},\overline{k^\prime},q}=-\left(\chi_\text{cs/sc}^{k+\Pi,k^\prime+\Pi,q}\right)^*,
\end{align}
which together with \cref{Eq:SU2P_sym_pp_trans} leads to:
\begin{align}
\label{Eq:chi_S_P_z_SU2P}
    \chi_{S_{P,z}}^{k,k^\prime,q}&=\frac{1}{2} \bigg(\chi_\text{ch}^{k,k^\prime,q}+\chi_\text{sp}^{k,k^\prime,q}+\chi_\text{cs}^{k,k^\prime,q}+\chi_\text{sc}^{k,k^\prime,q}+\chi_\text{ch}^{k,\overline{k^\prime},q}-\chi_\text{sp}^{k,\overline{k^\prime},q}-\chi_\text{cs}^{k,\overline{k^\prime},q}+\chi_\text{sc}^{k,\overline{k^\prime},q}\bigg),\\
\label{Eq:chi_S_P_x_SU2P}
    \chi_{S_{P,x}}^{k,k^\prime,q}&=\chi_\pp^{k,k^\prime,q},
\end{align}
where $\chi_\pp=\chi_{\pp,\up\down}$ [Eq.~(4)].

By making use of $SU(2)_P$- and $SU(2)$-symmetry we get a relation between $S_{P,z}$- and pp-channel for the \emph{same} set of system:
\begin{align}
    \chi_{S_{P,z};U}^{k,k^\prime,q}\overset{\cref{Eq:map_Sz}}{=}\chi_{S_{z};-U}^{k,k^\prime,q}\overset{SU(2)}{=}\chi_{S_{x};-U}^{k,k^\prime,q}\overset{\cref{Eq:map_Sx}}{=}\chi_{S_{P,x};U}^{k,k^\prime,-q-\Pi}.
\end{align}
In terms of charge and spin channels [\cref{Eq:chi_S_P_z_SU2P,Eq:chi_S_P_x_SU2P}] this reads 
\begin{align}
\label{Eq:chi_pp_chi_L}
    \chi_\text{pp}^{k,k^\prime,-q-\Pi}=\frac{1}{2}\left(\chi_\text{ch}^{k,k^\prime, q}+\chi_\text{ch}^{k,\overline{k^\prime}, q}+\chi_\text{sp}^{k,k^\prime, q}-\chi_\text{sp}^{k,\overline{k^\prime}, q}+\chi_{\text{sc},U}^{k,k^\prime,q}
    +\chi_{\text{sc},U}^{k,\overline{k^\prime},q}
    +\chi_{\text{cs},U}^{k,k^\prime,q}-\chi_{\text{cs},U}^{k,\overline{k^\prime},q}\right),
\end{align}
which is the $SU(2)$-broken version of Eq.~(7) in the main text.
The physical susceptibility in the pp-channel is degenerate to the charge/spin channel if the form factor has an even/odd ``$\bar{k}$-parity" [\cref{Eq:phys_S_Pz}]
\begin{align}
\label{Eq:phys_pp_ch_sp}
    \chi_{\pp}^{r,-q-\Pi} &=    
    \begin{cases}
      \chi_\text{ch}^{r,q}, & \text{if}\ f_r(k)=f_r(\overline{k})\\
      \chi_\text{sp}^{r,q}, & \text{if}\ f_r(k)=-f_r(\overline{k})
    \end{cases}.
\end{align}
Fourier transforming \cref{Eq:chi_pp_chi_L} to real space leads to 
\begin{align}
\label{Eq:chi_pp_chi_L_space}
\begin{split}
    (-1)^{j+h}\chi_\text{pp}^{ij|hl}(\nu,\nu^\prime,-\omega)=\frac{1}{2}\bigg[&\chi_\text{ch}^{ij|hl}(\nu,\nu^\prime,\omega)+(-1)^{h+l}\chi_\text{ch}^{ij|lh}(\nu,-\nu^\prime-\omega,\omega)+\chi_\text{sp}^{ij|hl}(\nu,\nu^\prime,\omega)-(-1)^{h+l}\chi_\text{sp}^{ij|lh}(\nu,-\nu^\prime-\omega,\omega)\\
    &+\chi_{\text{sc},U}^{ij|hl}(\nu,\nu^\prime,\omega)
    +(-1)^{h+l}\chi_{\text{sc},U}^{ij|lh}(\nu,-\nu^\prime-\omega,\omega)
    +\chi_{\text{cs},U}^{ij|hl}(\nu,\nu^\prime,\omega)-(-1)^{h+l}\chi_{\text{cs},U}^{ij|lh}(\nu,-\nu^\prime-\omega,\omega)\bigg],
\end{split}
\end{align}
which corresponds to the generalized [i.e.,~without requiring $SU(2)$ symmetry] form of Eq.~(6) and Matsubara frequencies instead of imaginary times are used.

As a showcase of this formal development we will now investigate the physical susceptibilities in the bipartite square lattice for transfer momenta $\vq=(0,0),(\pi,\pi),(0,\pi),(\pi,0)$ and the four form factors \cite{Eckhardt18}
\begin{align}
\label{Eq:form_factors}
\begin{split}
    f_s(k)&=1,\\
    f_{p_x}(k)&=\sin \vk_x, \;f_{p_y}=\sin \vk_y,\\
    f_d(k) &= \cos \vk_x - \cos \vk_y,\\
    f_{s_{ext}}(k) &= \cos \vk_x + \cos \vk_y.
\end{split}
\end{align}
Note that all form factors are constant in frequency and have only a momentum structure.

Further, we can see from \cref{Eq:SU2P_cs_sc} that the real part of the physical susceptibility in the cs- and sc-channel is zero for $SU(2)_P$-symmetry, if the $\Pi-$shift has the same impact on the form factors $f_r$ and $f_{r^\prime}$ in terms of $\overline{k}$-parity:
\begin{align}
\label{Eq:cs_sc_zero}
\begin{split}
    \chi_\text{cs/sc}^{(r,r^\prime) ,q}=0 &\text{ if } f_r(k)=f_r(k+\Pi) \text{ and } f_{r^\prime}(k)=f_{r^\prime}(k+\Pi) \\
    &\text{ or }f_r(k)=-f_r(k+\Pi) \text{ and } f_{r^\prime}(k)=-f_{r^\prime}(k+\Pi),
\end{split}
\end{align}
yielding the real part to vanish, while the imaginary part of the physical susceptibility always needs to be zero. For the waves we consider, the only non-zero wave combination is if we combine a $s$-wave form factor with a non $s$-wave form factor. Since we do not consider such wave combinations we will exclude the cs/sc-channel from our discussion of the physical response functions.
We now specify the cases for different transfer momenta:
\begin{itemize}
\item $\vq_{\ph}=(\pi,\pi)$:\\
For this transfer momentum the $\bar{k}$-parity is equivalent to the standard parity since $\vk\rightarrow\overline{\vk}=-\vk-\vpi-\vq=-\vk$ for $\vq=(\pi,\pi)$. 
The parity of the form factors in this case reads
\begin{align}
\label{Eq:parity_q_pi_pi}
\begin{split}
    f_s(k)&=f_s(-k),\\
    f_{p_x}(k)&=-f_{p_x}(-k), f_{p_y}(k)=-f_{p_y}(-k),\\
    f_d(k) &= f_d(-k), \\
    f_{s_{ext}}(k) &= f_{s_{ext}}(-k),
\end{split}
\end{align}
i.e., $s,d$ and $s_{ext}$ are even waves and $p$ form factors are odd parity waves.
According to \cref{Eq:phys_S_Pz}, $\chi_\text{pp}^{r,-q-\Pi}=\chi_\text{ch}^{r,q}$ for $s$, $d$ and $s_{ext}$ form factors and $\chi_\text{pp}^{r,-q-\Pi}=\chi_\text{sp}^{r,q}$ for $p$ form factors hold.

\item $\vq_{\ph}=(0,0)$:\\
The ``$\bar{k}$-parity" of the waves under the transformation $\vk\rightarrow-\vk-\vpi$ is even for $s$- and $p$-waves, while it is odd for $d$ and $s_{ext}$-waves. 
Therefore, $\chi_\text{pp}^{r,-q-\Pi}=\chi_\text{ch}^{r,q}$ for $s$ and $p$ form factors while $\chi_\text{pp}^{r,-q-\Pi}=\chi_\text{sp}^{r,q}$ holds for $d$- and $s_{ext}$ form factors. 

\item $\vq_\ph=(\pi,0)$:\\
The ``$\bar{k}$-parity'' of the waves under the transformation $\vk\rightarrow-\vk-(0,\pi)$ is even for $s$- and $p_{y}$-waves while it is odd for $p_x$-waves. For these waves with defined ``$\bar{k}$-parity" we can proceed as in the previous cases. Further we find that $d$- and $s_{ext}$-waves have no defined ``$\bar{k}$-parity" for $\vq=(\pi,0)$, hence some extra work is needed. However, since $f_d(k)=f_{s_{ext}}(\overline{k})$ [resp. $f_d(k)=-f_{s_{ext}}(-k+\Pi+q_{\pp})$] we can relate the physical response functions of these two waves. 
Specifically, it holds for $SU(2)_P$-symmetry [\cref{Eq:SU2P_sym_ph,Eq:SU2P_sym_pp_trans}] that 
\begin{align}
    \chi_\eta^{d,q}&=\chi_\eta^{s_{ext},q} \text{ and } \chi_\eta^{(d,s_{ext}),q}=\chi_\eta^{(s_{ext},d),q} \text{ for $\eta=$ch,sp, $S_x$ and pp},
\end{align}
and we remind the reader that $q_{\pp}=-q_{\ph}-\Pi$ for the pp-channel. For the pp-channel we further used that the imaginary part of physical susceptibility needs to vanish. 
Since the physical susceptibilities in cs- and sc-channel vanish [\cref{Eq:cs_sc_zero}] we only need to consider the ch/sp-channel in \cref{Eq:chi_S_P_z_SU2P}. Considering $\frac{1}{N\beta^2}\sum_{kk^\prime}f_d(k)\chi_\text{ch/sp}^{k,\overline{k^\prime},\vq=(\pi,0)}f_d(k^\prime)=\chi_\text{ch/sp}^{(d,s_{ext}),\vq=(\pi,0)}$, we find
\begin{align}
\label{Eq:phys_pp_q0pi}
    \chi_\text{pp}^{d,-q-\Pi}=\chi_\text{ch}^{d,q}+\chi_\text{sp}^{d,q}+\chi_\text{ch}^{(d,s_{ext}),q}-\chi_\text{sp}^{(d,s_{ext}),q}=\chi_{\pp}^{s_{ext},-q-\Pi},
\end{align}
which shows that here the pp-channel is not degenerate with either the ch- or sp-channel but rather a mixture of both. 
\item $\vq_\ph=(0,\pi)$:\\
This case is analogous to the case of  $\vq_{\ph}=(\pi,0)$ with the only differences that $s$- and $p_{x}$-waves are even while the $p_y$-wave is odd and that $f_d(k)=-f_{s_{ext}}(\overline{k})$ while $f_d(k)=f_{s_{ext}}(-k+\Pi+q_{\pp})$. Since for the square lattice the response for $\vq=(0,\pi)$ and $\vq=(\pi,0)$ is degenerate also a superposition between these two cases can appear.
\end{itemize}
The findings of this investigation are summarized in Fig.~1 in the main text. 

Finally, we want to emphasize that this investigation can easily be extended to other transfer momenta and form factors, where a degeneracy between pp-channel and ch- or sp-channel will appear if one considers combination of transfer momenta and form factors that exhibit a defined ``$\bar{k}$-parity", i.e., $f_r(k)=\pm f_r(\overline{k})$.

\section{Degeneracy for embedding methods}
Remarkably, the results of \cref{sec:phys_respose_degeneracancy} also hold for various embedding methods, namely dynamical mean field theory (DMFT) \cite{Georges1992} and its cluster extensions, dynamical cluster approximation (DCA) \cite{Jarrell2001} and the cellular dynamical mean field theory (CDMFT) \cite{,Kotliar2000}.
In this section we analytically prove the relations found in \cref{sec:phys_respose_degeneracancy} for these embedding theories. Since the dual BSE (DBSE) \cite{vanLoon2024-2} represents a crucial ingredient for our proof we will restrict ourselves to the case without magnetic field, i.e., $SU(2)_P$- \emph{and} $SU(2)$-symmetry which is equal to ph-symmetry, in order to avoid having to deal with the coupled longitudinal channel \cref{Eq:chi_L}. 
The crucial ingredient of the dual BSE is the full vertex $F$ of the impurity therefore we now show that \cref{Eq:chi_pp_chi_L,Eq:chi_pp_chi_L_space} also hold for the full vertex, by using the relation 
\begin{align}
\label{Eq:F_chi}
    F_\eta = \chi_{0,\eta}^{-1}(\chi_{0,\eta}-\chi_\eta)\chi_{0,\eta}^{-1}\quad \eta=\text{ch, sp, pp},
\end{align}
where we omitted indices such that the quantities may represent either matrices in momentum \cite{Maier2005} or real space \cite{Meixner2026a}.
Further, applying the ph-transformation [$c_i\rightarrow (-1)^ic_i^\dagger$] to the Green function \cref{Eq:Def_G} gives 
\begin{align}
\label{Eq:ph_G_momentum}
    G_\vk(\nu)&=-G_{-\vk-\vpi}(-\nu),\\
\label{Eq:ph_G_space}
    G_{ij}(\nu)&=-(-1)^{i+j}G_{ji}(-\nu),
\end{align}
in momentum and real space, respectively. Substituting one Green function in the bubbles  \cref{Eq:Def_chi0_ph,Eq:Def_chi0_pp} gives:
\begin{align}
\label{Eq:bubble_ph_sym_momentum}
    &\chi_{0,\ph}^{kk^\prime q}=\chi_{0,\pp}^{k,k^\prime,-q-\Pi},\\
\label{Eq:bubble_ph_sym_space}
    &\chi_{0,\ph}^{ij|hl}(\nu,\nu^\prime,\omega)=(-1)^{j+h}\chi_{0,\pp}^{ij|hl}(\nu,\nu^\prime,-\omega),
\end{align}
for momentum and real space formulations, respectively.

Combining \cref{Eq:chi_pp_chi_L} [or \cref{Eq:chi_pp_chi_L_space}] with \cref{Eq:F_chi} we can show that \cref{Eq:chi_pp_chi_L} (or \cref{Eq:chi_pp_chi_L_space}) also directly holds for $F$, i.e.,
\begin{align}
\label{Eq:F_pp_chi_L_momentum}
    F_\text{pp}^{k,k^\prime,-q-\Pi}&=\frac{1}{2}\bigg(F_\text{ch}^{k,k^\prime, q}+F_\text{ch}^{k,\overline{k^\prime}, q}+F_\text{sp}^{k,k^\prime, q}-F_\text{sp}^{k,\overline{k^\prime}, q}\bigg)=F_{S_{P,z}}^{k,k^\prime, q},\\
\label{Eq:F_pp_chi_L_space}
\begin{split}
    (-1)^{j+h}F_\text{pp}^{ij|hl}(\nu,\nu^\prime,-\omega)&=\frac{1}{2}\bigg(F_\text{ch}^{ij|hl}(\nu,\nu^\prime,\omega)+(-1)^{h+l}F_\text{ch}^{ij|lh}(\nu,-\nu^\prime-\omega,\omega)\\
    &\phantom{=\frac{1}{2}\bigg(}+F_\text{sp}^{ij|hl}(\nu,\nu^\prime,\omega)-(-1)^{h+l}F_\text{sp}^{ij|lh}(\nu,-\nu^\prime-\omega,\omega)\bigg)=F_{S_{P,z}}^{ij|hl}(\nu,\nu^\prime,\omega),
\end{split}
\end{align}
for both momentum and real space, respectively. 

Since DMFT and its cluster extensions exactly solve an (cluster-)impurity problem, \cref{Eq:bubble_ph_sym_momentum,Eq:bubble_ph_sym_space,Eq:F_pp_chi_L_momentum,Eq:F_pp_chi_L_space} hold for the cluster quantities of these embedding theories if the bath is also ph-symmetric. 

In the following we will denote $F,\chi$ and $\chi_0$ without a $q$ index as the impurity quantities while $\chi^q$ and $\chi_0^q$ will denote the lattice quantities of the embedding theory and $\tilde\chi_{0}^q=\chi_{0}^q-\chi_{0}$ is the dual bubble.

The dual BSE (DBSE), accounting for the different signs of $\Gamma$ and $F$ in comparison to van Loon et al., Ref.~\cite{vanLoon2024-2}, reads
\begin{align}
\label{Eq:d-BSE}
    \chi^q_\eta&=\chi_\eta+\bar\chi^q_\eta=\chi_\eta+L_\eta\tilde\chi^q_\eta L_\eta,\\
    &\text{with }\tilde\chi^q_\eta=\tilde\chi_{0,\eta}^q-\tilde\chi_{0,\eta}^q\sum_{n=1}^\infty \left(F_\eta\tilde\chi_{0,\eta}^q\right)^n,\nonumber\\
    &\phantom{\text{with }}L_\eta=1-\chi_{0,\eta}F_\eta,\nonumber\\
    &\phantom{\text{with }}\tilde\chi_{0,\eta}^q=\chi_{0,\eta}^q-\chi_{0,\eta},\nonumber
\end{align}
where we again omitted the specific indices which have to be added if DMFT, DCA or CDMFT is investigated.

\subsection{Degeneracy for DMFT}\label{sec:dmft}
As a first pedagogical example we \sout{will }start with DMFT \cite{Georges1992}, where only the s-wave form factor is relevant because of the local nature of DMFT. 
In DMFT the lattice Green function, lattice bubbles and impurity bubbles read (see Ref.~\cite{georges:1996})
\begin{align}
\label{Eq:G_lattice_DMFT}
    G_k&=(i\nu+\mu-\varepsilon_{\vk}-\Sigma_\nu)^{-1},\\
\label{Eq:chi_0q_ph_DMFT}
    \chi_{0,\ph}^{\nu\nu^\prime\omega,\vq}&=- \frac{\beta}{N}\sum_{\vk} G_{k}G_{k+q}\delta_{\nu\nu^\prime},\\
\label{Eq:chi_0q_pp_DMFT}
    \chi_{0,\pp}^{\nu\nu^\prime\omega,\vq}&=\frac{\beta}{N}\sum_{\vk} G_{k}G_{q-k}\delta_{\nu\nu^\prime},\\
\label{Eq:chi_0_ph_DMFT}
    \chi_{0,\ph}^{\nu\nu^\prime\omega}&= -\beta G_{\nu}G_{\nu+\omega}\delta_{\nu\nu^\prime},\\
\label{Eq:chi_0_pp_DMFT}
    \chi_{0,\pp}^{\nu\nu^\prime\omega}&=\beta G_{\nu}G_{\omega-\nu}\delta_{\nu\nu^\prime},
\end{align}
where $G_\nu$ and $\Sigma_\nu$ are the Green function and self-energy of the impurity and $\varepsilon_\vk$ is the lattice dispersion. 
Due to ph-symmetry, the self-energy (excluding the Hartree term) obeys $\Sigma_\nu=-\Sigma_{-\nu}$ while the dispersion $\varepsilon_{\vk}=-2t(\cos k_x+\cos k_y)$ satisfies $\varepsilon_{\vk}=-\varepsilon_{-\vk-\vpi}$. Consequently, the relation \cref{Eq:ph_G_momentum} also holds for the lattice Green function of DMFT. Therefore, \cref{Eq:bubble_ph_sym_momentum} also holds for the lattice bubble of DMFT.

Further, we define the operator $J^{\nu\nu^\prime\omega}=\delta_{\nu,-\nu^\prime-\omega}$ and view all two-particle quantities as matrices in the fermionic Matsubara frequencies with fixed parameters $\omega$ and $\vq$. 
We will now show that, for ph-symmetry, $J$ commutes with every matrix that appears in the DBSE \cref{Eq:d-BSE} for charge and spin channel. Since for the local case ($\vk,\vk'$ independent), \cref{Eq:SU2P_c_s} can be expressed as $J\chi_\text{ch/sp}J=\chi_\text{ch/sp}$ in this matrix notation, J obviously commutes with the impurity generalized susceptibility. 
As $G_\nu=-G_{-\nu}$ [\cref{Eq:ph_G_momentum}], also the impurity bubble commutes with $J$, it follows from \cref{Eq:F_chi} that $[J,F_\text{ch,sp}]=0$ must also hold. 
For the lattice bubble $[J,\chi_{0,\ph}^{q}]=0$ can be shown by considering
\begin{align}
    J\chi_{0,\ph}^{q}J&=-\frac{\beta}{N}\sum_{\vk}G_{\vk}(-\nu-\omega)G_{\vk+\vq}(-\nu)\overset{\cref{Eq:ph_G_momentum}}{=}-\frac{\beta}{N}\sum_{\vk}G_{-\vk-\vpi}(\nu+\omega)G_{-\vk-\vpi-\vq}(\nu)\\
    &=-\frac{\beta}{N}\sum_{\vk^\prime=-\vk-\vpi-\vq}G_{\vk^\prime+\vq}(\nu+\omega)G_{\vk^\prime}(\nu)=\chi_{0,\ph}^{q}.\nonumber
\end{align}
Together with \cref{Eq:F_pp_chi_L_momentum} for the local impurity vertex we now have all ingredients to show that the $s$-wave DMFT lattice susceptibility fulfills
\begin{align}
    f_s\chi_\pp^{-q-\Pi}f_s=f_s\chi_\text{ch}^{q}f_s \quad\forall q \text{ in DMFT},
\end{align}
as it is also found for the exact lattice solution in \cref{sec:phys_respose_degeneracancy}.

Evidently, this is true for the first term of the DBSE \cref{Eq:d-BSE} as this is the exact solution of a specific (ph-symmetric) impurity problem and therefore the derivation done in \cref{sec:phys_respose_degeneracancy} holds.
Since the DBSE also contains a geometric series in the impurity vertex $F$ we aim to prove by induction, that  
\begin{align}
\label{Eq:induction_hypothesis}
    (F_{S_{P,z}}\tilde\chi_{0,\ph}^q)^n L_{S_{P,z}}f_s=(F_\text{ch}\tilde\chi_{0,\ph}^q)^n L_\text{ch}f_s\quad n\in\mathbb{N},
\end{align} 
where the form factors are viewed as vectors ($f_s$ is just a vector of all ones). To that aim, we start the induction
\begin{align}
\label{Eq:induction_start_DMFT}
\begin{split}
    n=1:&\\
    &F_{S_{P,z}}\tilde\chi_{0,\ph}^q L_{S_{P,z}}f_s = \frac{1}{2}\left(F_\text{ch}+F_\text{ch}J+F_\text{sp}-F_\text{sp}J\right)\tilde\chi_{0,\ph}^q\left(1-\chi_{0,\ph}\frac{1}{2}(F_\text{ch}+F_\text{ch}J+F_\text{sp}-F_\text{sp}J)\right)f_s\\
    &\phantom{F_{S_{P,z}}\tilde\chi_{0,\ph}^q L_{S_{P,z}}f_s}=F_\text{ch}\tilde\chi_{0,\ph}^q L_\text{ch}f_s,
\end{split}
\end{align}
where in the second line we used that $[J,\tilde\chi_{0,\ph}^q]=[J,\chi_{0,\ph}]=[J,F_\text{ch/sp}]=0$ and that $Jf_s=f_s$.
\begin{align}
\label{Eq:induction_step_DMFT}
\begin{split}
    n\rightarrow n+1:&\\
    &(F_{S_{P,z}}\tilde\chi_{0,\ph}^q)^{n+1} L_{S_{P,z}}f_s = \frac{1}{2}\left(F_\text{ch}+F_\text{ch}J+F_\text{sp}-F_\text{sp}J\right)\tilde\chi_{0,\ph}^q(F_\text{ch}\tilde\chi_{0,\ph}^q)^n L_\text{ch}f_s\\
    &\phantom{(F_{S_{P,z}}\tilde\chi_{0,\ph}^q)^{n+1} L_{S_{P,z}}f_s}=(F_\text{ch}\tilde\chi_{0,\ph}^q)^{n+1} L_\text{ch}f_s \quad\square,
\end{split}
\end{align}
which completes the induction proof. Putting all this together and starting from the DBSE for the pp-channel we get:
\begin{align}
\label{Eq:dmft_final}
\begin{split}
    f_s\chi^{-q-\Pi}_\pp f_s&=f_s\left[\chi_\pp+L_\pp\left(\tilde\chi_{0,\pp}^{-q-\Pi}-\tilde\chi_{0,\pp}^{-q-\Pi}\sum_{n=1}^\infty \left(F_\pp\tilde\chi_{0,\pp}^{-q-\Pi}\right)^n\right)L_\pp\right]f_s\\
    &=f_s\left[\chi_{S_{P,z}}+L_{S_{P,z}}\left(\tilde\chi_{0,\ph}^{q}-\tilde\chi_{0,\ph}^{q}\sum_{n=1}^\infty \left(F_{S_{P,z}}\tilde\chi_{0,\ph}^{q}\right)^n\right)L_{S_{P,z}}\right]f_s\\
    &=f_s\chi^{q}_\text{ch}f_s
\end{split}
\end{align}
where \cref{Eq:bubble_ph_sym_momentum,Eq:F_pp_chi_L_momentum} are used in the second line and \cref{Eq:induction_hypothesis} in the third line.

\subsection{Degeneracy for DCA}
To investigate non-uniform form factors we consider DCA \cite{Hettler2000} which is a cluster extension of DMFT in momentum space. DCA tiles the Brillouin zone into patches of constant self-energy, resulting in a momentum-dependent self-consistency equation, in contrast to DMFT. These patches are represented by a specific patch momentum vector. For the mapping to be applicable, we require the vector $\vk_\mathbf{P}=\vpi$ when added to a patch momentum vector to always result in another patch momentum vector of the DCA tiling. This results in bipartite boundary conditions for the impurity.

In DCA the lattice Green function, lattice bubbles and impurity bubbles read (see Ref.~\cite{Maier2005}):
\begin{align}
\label{Eq:G_lattice_DCA}
        G_\nu^{\vk+\tilde\vk}&=(i\nu+\mu-\varepsilon_{\vk+\tilde\vk}-\Sigma_\nu^{\vk})^{-1},\\
\label{Eq:chi_0q_ph_DCA}
    \chi_{0,\ph}^{\vk,\vk^\prime,\vq+\tilde\vq}(\nu,\nu^\prime,\omega)&=-\frac{\beta}{N_\vtk}\sum_{\tilde\vk} G_{\nu}^{\vk+\tilde\vk}G_{\nu+\omega}^{\vk+\tilde\vk+\vq+\tilde\vq}\delta_{\nu\nu^\prime}\delta_{\vk\vk^\prime},\\
\label{Eq:chi_0q_pp_DCA}
    \chi_{0,\pp}^{\vk,\vk^\prime,\vq+\tilde\vq}(\nu,\nu^\prime,\omega)&=\frac{\beta}{N_\vtk}\sum_{\tilde\vk} G_{\nu}^{\vk+\tilde\vk}G_{\omega-\nu}^{\vq+\tilde\vq-\vk-\tilde\vk}\delta_{\nu\nu^\prime}\delta_{\vk\vk^\prime},\\
\label{Eq:chi_0_ph_DCA}
    \chi_{0,\ph}^{\vk,\vk^\prime,\vq}(\nu,\nu^\prime,\omega)&= -\beta G_{\nu}^{\vk}G_{\nu+\omega}^{\vk+\vq}\delta_{\nu\nu^\prime}\delta_{\vk\vk^\prime},\\
\label{Eq:chi_0_pp_DCA}
    \chi_{0,\pp}^{\vk,\vk^\prime,\vq}(\nu,\nu^\prime,\omega)&= \beta G_{\nu}^{\vk}G_{\omega-\nu}^{\vq-\vk}\delta_{\nu\nu^\prime}\delta_{\vk\vk^\prime},\\
\end{align}
where $G_\nu^{\vk}$ and $\Sigma_\nu^{\vk}$ are the Green function and self-energy of the momentum cluster impurity with the cluster momentum $\vk$. Further, $\tilde\vk, \tilde\vq$ denote the momenta of the reduced Brillouin-zone (RBZ) of the super-lattice and $N_\vtk$ are the number of points in the RBZ. Moreover, $G_\nu^{\vk+\tilde\vk}$ represents the lattice Green function. Generally, all quantities that depend only on the cluster momenta are cluster quantities, while the lattice quantities also carry a dependence on the momenta of the super-lattice that we denote with a tilde. Again we will consider all two-particle quantities as matrices in the space of fermionic Matsubara frequencies and fermionic cluster momenta, with $\omega,\vq$ and $\tilde\vq$ as fixed parameters. In the DBSE \cref{Eq:d-BSE} we then have to replace all $\vq$ by $\vq+\tilde\vq$ while also the impurity quantities will only depend on the cluster momenta $\vq$.

Again, \cref{Eq:ph_G_momentum} also holds for the DCA lattice Green function, since the self-energy is local (independent of $\tilde{\vk}$), $\Sigma_\nu^{\vk}=-\Sigma_{-\nu}^{-\vk-\vpi}$. Therefore, \cref{Eq:bubble_ph_sym_momentum} also holds for the lattice bubble of DCA.

Analogously to the previous section we define the operator $J^{\vk,\vk^\prime,\vq}(\nu,\nu^\prime,\omega)=\delta_{\nu,-\nu^\prime-\omega}\delta_{\vk,-\vk^\prime-\vpi-\vq}$ and by using \cref{Eq:SU2P_sym_ph,Eq:ph_G_momentum} we show again that the lattice and impurity bubble as well as the full vertex of the impurity commute with $J$ for ph-symmetry, i.e. $[J,\chi_{0,\ph}^{\vq+\tilde\vq}]=[J,\chi_{0,\ph}^{\vq}]=[J,F_\text{ch/sp}^{\vq}]=0$. 
By using induction together with these commutation relations we can prove that 
\begin{align}
\label{Eq:induction_hypothesis_DCA}
    (F^{\vq}_{S_{P,z}}\tilde\chi_{0,\ph}^{\vq+\tilde\vq})^n L_{S_{P,z}}^{\vq}f_r &=
    \begin{cases}
        (F_\text{ch}^{\vq}\tilde\chi_{0,\ph}^{\vq+\tilde\vq})^n L_\text{ch}^{\vq}f_r & \text{if}\ f_r=Jf_r\\
        (F_\text{sp}^{\vq}\tilde\chi_{0,\ph}^{\vq+\tilde\vq})^n L_\text{sp}^{\vq}f_r & \text{if}\ f_r=-Jf_r
    \end{cases}\quad n\in\mathbb{N}
\end{align}
holds, which covers all summands of the dual BSE.
Analogously to \cref{sec:dmft} we get the same relation as obtained for the exact lattice solution \cref{Eq:phys_pp_ch_sp}, i.e.
\begin{align}
\label{Eq:phys_pp_ch_sp_DCA}
    f_r\chi_\pp^{-\vq-\tilde\vq-\vpi}(-\omega)f_r &=
    \begin{cases}
        f_r\chi_\text{ch}^{\vq+\tilde\vq}(\omega)f_r & \text{if}\ f_r=Jf_r\\
        f_r\chi_\text{sp}^{\vq+\tilde\vq}(\omega)f_r & \text{if}\ f_r=-Jf_r,
    \end{cases}
\end{align}
by using \cref{Eq:induction_hypothesis_DCA}, \cref{Eq:F_pp_chi_L_momentum} and the DBSE \cref{Eq:d-BSE}.
The lattice momentum is replace by the sum of cluster momentum $\vq$ and the super-lattice momentum $\tilde\vq$.
The more complicated equations for stripe order \cref{Eq:phys_pp_q0pi} can be proven analogously by using that $f_d=Jf_{s_{ext}}$ for $\vq=(\pi,0)$ while $f_d=-Jf_{s_{ext}}$ for $\vq=(0,\pi)$.

\subsection{Degeneracy for CDMFT}
Finally, we also investigate CDMFT \cite{Kotliar2001}, which is a cluster extension of DMFT in real space, and is therefore also able to capture non-uniform form factors.

In CDMFT the lattice Green function, lattice bubbles and impurity bubbles read (see Ref.~\cite{Meixner2026a})
\begin{align}
\label{Eq:G_lattice_CDMFT}
        G_{\nu,\tilde\vk}^{ij}&=\left(i\nu+\mu-\varepsilon_{\tilde\vk}^{ij}-\Sigma_\nu^{ij}\right)^{-1},\\
\label{Eq:chi_0q_ph_CDMFT}
    \chi_{0,\ph}^{ij|hl,\tilde\vq}(\nu,\nu^\prime,\omega)&=-\frac{\beta}{N_\vtk}\sum_{\tilde\vk} G_{\nu,\tilde\vk}^{li}G_{\nu+\omega,\tilde\vk+\tilde\vq}^{jh}\delta_{\nu\nu^\prime},\\
\label{Eq:chi_0q_pp_CDMFT}
    \chi_{0,\pp}^{ih|jl,\tilde\vq}(\nu,\nu^\prime,\omega)&=\frac{\beta}{N_\vtk}\sum_{\tilde\vk} G_{\nu,\tilde\vk}^{li}G_{\omega-\nu,\tilde\vq-\tilde\vk}^{jh}\delta_{\nu\nu^\prime},\\
\label{Eq:chi_0_ph_CDMFT}
    \chi_{0,\ph}^{ij|hl}(\nu,\nu^\prime,\omega)&= -\beta G_{\nu}^{li}G_{\nu+\omega}^{jh}\delta_{\nu\nu^\prime},\\
\label{Eq:chi_0_pp_CDMFT}
    \chi_{0,\pp}^{ih|jl}(\nu,\nu^\prime,\omega)&= \beta G_{\nu}^{li}G_{\omega-\nu}^{jh}\delta_{\nu\nu^\prime},\\
\end{align}
where $G_{\nu,\tilde\vk}^{ij}$ and $\Sigma_\nu^{ij}$ are the Green function and self-energy of the real space cluster impurity with the cluster lattice sites $i,j$. Further, $\tilde\vk, \tilde\vq$ denote the momenta of the reduced Brillouin-zone (RBZ) of the super-lattice and $\varepsilon_{\tilde\vk}^{ij}$ is the dispersion in the unit cell of the super-lattice. All quantities that depend only on the cluster sites are cluster quantities, while the lattice quantities also obtain a dependence on the momenta of the super-lattice that we again denote with a tilde. Again we will consider all two-particle quantities as matrices in the space of fermionic Matsubara frequencies and cluster sites (see Ref.~\cite{Meixner2026a} for the convention), with $\omega$ and $\tilde\vq$ as fixed parameters. In the DBSE \cref{Eq:d-BSE} we than have to replace all $\vq$ by $\tilde\vq$ while the impurity quantities will now depend on four cluster indices.

Since $\Sigma_\nu^{ij}=-(-1)^{i+j}\Sigma_{-\nu}^{ji}$ and $\varepsilon_{\tilde\vk}^{ij}=-(-1)^{i+j}\varepsilon_{-\tilde\vk}^{ji}$, the lattice Green function obeys 
\begin{align}
\label{Eq:ph_G_CDMFT}
    G_{\nu,\tilde\vk}^{ij}=-(-1)^{i+j}G_{-\nu,-\tilde\vk}^{ji},
\end{align}
which is a mixed representation of \cref{Eq:ph_G_momentum,Eq:ph_G_space}. From this it follows that the lattice bubble of CDMFT satisfies
\begin{align}
\label{Eq:bubble_phy_sym_CDMFT}
    \chi_{0,\ph}^{ij|hl,\tilde\vq}(\nu,\nu^\prime,\omega)=(-1)^{j+h}\chi_{0,\pp}^{ij|hl,-\tilde\vq}(\nu,\nu^\prime,-\omega),
\end{align}
which is a mixed representation of \cref{Eq:bubble_ph_sym_momentum,Eq:bubble_ph_sym_space}. We now want to show that the $SU(2)$-symmetric version of \cref{Eq:chi_pp_chi_L_space,Eq:relation_realpsace},i.e. Eq.~(6) of the main text, which holds for the impurity, can be transferred to the lattice via the DBSE. 

For the CDMFT BSE-formalism all quantities are viewed as matrices in the real space indices and the fermionic Matsubara frequencies by introducing combined cluster real-space and frequency indices \cite{Musshoff2021,Meixner2026a}. 
Within this framework a matrix multiplication is defined as $(AB)^{ij|hl}(\nu,\nu^\prime,\omega)=A^{ij|mn}(\nu,\nu_1,\omega)B^{nm|hl}(\nu_1,\nu^\prime,\omega)$ where we sum over repeated indices (see Ref.~\cite{Meixner2026a} for details).

We define the operator $J^{ij|hl}(\nu,\nu^\prime,\omega)=(-1)^{i+j}\delta_{\nu,-\nu^\prime-\omega}\delta_{i,h}\delta_{j,l}$ and by using \cref{Eq:SU2P_sym_ph_space,Eq:ph_G_space,Eq:ph_G_CDMFT} we show again that the lattice and impurity bubble as well as the full vertex of the impurity commute with $J$ in the case of ph-symmetry, i.e. $[J,\chi_{0,\ph}^{\tilde\vq}]=[J,\chi_{0,\ph}]=[J,F_\text{ch/sp}]=0$.

We now derive the relationship between the pp- and the $S_{P,z}$-channels via the DBSE \cref{Eq:d-BSE}. The essential step is that the phase factor appearing in \cref{Eq:F_pp_chi_L_space} vanishes for contracted indices due to the structure of the matrix multiplication, i.e. $(-1)^{j+m}A^{ij|mn}(\nu,\nu_1,\omega)\,(-1)^{m+h}B^{nm|hl}(\nu_1,\nu^\prime,\omega)=(-1)^{j+h}\,A^{ij|mn}(\nu,\nu_1,\omega)B^{nm|hl}(\nu_1,\nu^\prime,\omega)$. This phase factor vanishes also for the identity matrix, i.e. $\mathbb{1}^{ij|hl}(-1)^{j+h}=\delta_{il}\delta_{jh}(-1)^{j+h}=\mathbb{1}^{ij|hl}$. Therefore, the phase factors do only appear for non-contracted indices for all terms in the DBSE.
By applying the ph-symmetries for the impurity susceptibility, bubble, and vertex [\cref{Eq:SU2P_sym_ph_space,Eq:bubble_ph_sym_space,Eq:F_pp_chi_L_space}] alongside the lattice bubble symmetry [\cref{Eq:bubble_phy_sym_CDMFT}], the DBSE yields 
\begin{align}
\label{Eq:DBSE_CDMFT}
\begin{split}
    (-1)^{j+h}\chi_\pp^{ij|hl,-\tilde\vq}(\nu,\nu^\prime,-\omega)&=\chi_{S_{P,z}}^{ij|hl,\tilde\vq}(\nu,\nu^\prime,\omega)\\
    &=\frac{1}{2}\bigg(\chi_\text{ch}^{ij|hl,\tilde\vq}(\nu,\nu^\prime,\omega)+(-1)^{h+l}\chi_\text{ch}^{ij|lh,\tilde\vq}(\nu,-\nu^\prime-\omega,\omega)\\
    &\phantom{=\frac{1}{2}\bigg(}+\chi_\text{sp}^{ij|hl,\tilde\vq}(\nu,\nu^\prime,\omega)-(-1)^{h+l}\chi_\text{sp}^{ij|lh,\tilde\vq}(\nu,-\nu^\prime-\omega,\omega)\bigg).
\end{split}
\end{align}
As a postprocessing after the BSE it is useful to map the lattice sites to cluster momenta \cite{Meixner2026a} via
\begin{align}
\label{Eq:CDMFT_FT_ph}
    \chi_{\ph}^{\vk,\vk^\prime,\vq;\tilde\vq}(\nu,\nu^\prime,\omega)=\frac{1}{N_c^2}\sum_{ijhl}\mathrm{e}^{-i\vk\vR_i} \mathrm{e}^{\i(\vk+\vq)\vR_j}\mathrm{e}^{-\i(\vk^\prime+\vq)\vR_h}\mathrm{e}^{\i\vk^\prime\vR_l} \chi_\ph^{ij|hl,\vtq}(\nu,\nu^\prime,\omega),\\
\label{Eq:CDMFT_FT_pp}
    \chi_{\pp}^{\vk,\vk^\prime,\vq;\tilde\vq}(\nu,\nu^\prime,\omega)=\frac{1}{N_c^2}\sum_{ijhl}\mathrm{e}^{-i\vk\vR_i} \mathrm{e}^{\i\vk^\prime\vR_j}\mathrm{e}^{-\i(\vq-\vk)\vR_h}\mathrm{e}^{\i(\vq-\vk^\prime)\vR_l} \chi_\pp^{ih|lj,\vtq}(\nu,\nu^\prime,\omega),
\end{align}
where $N_c$ is the number of cluster lattice sites. Note that this is not a Fourier transformation since CDMFT breaks translation invariance and therefore this transformation with three independent momenta is not invertible by a Fourier transform.
Fourier transforming Eq.~(\ref{Eq:DBSE_CDMFT}) results in a periodized momentum space equation
\begin{equation}
\label{Eq:DBSE_CDMFT_MOMENTUM}
\begin{split}
    \chi_\text{pp}^{k,k^\prime,-q-\Pi;-\vtq}=\frac{1}{2}\left(\chi_\text{ch}^{k,k^\prime, q;\vtq}+\chi_\text{ch}^{k,\overline{k^\prime}, q;\vtq}+\chi_\text{sp}^{k,k^\prime, q;\vtq}-\chi_\text{sp}^{k,\overline{k^\prime}, q;\vtq}\right).
\end{split}
\end{equation}

Contracting this with the form factors $f_r(k)$ gives us a physical susceptibility which depends on the cluster momentum $\vq$ and the super-lattice momentum $\tilde\vq$, $\vq$ gives the modulation inside a cell of the CDMFT superlattice while $\tilde \vq$ gives the modulation across different superlattice cells.

The show the effect of the operator $J$ it is useful to consider
\begin{align}
\label{Eq:J_cdmft}
\begin{split}
    \sum_{\vk^\prime,mnhl}X^{ij|nm}J^{mn|hl} \mathrm{e}^{-\i(\vk^\prime+\vq)\vR_h}\mathrm{e}^{\i\vk^\prime\vR_l} f_r(\vk^\prime)&=\sum_{\vk^\prime,mn}X^{ij|nm} (-1)^{n+m} \mathrm{e}^{-\i(\vk^\prime+\vq)\vR_m} \mathrm{e}^{\i\vk^\prime\vR_n} f_r(\vk^\prime)\\
    &=\sum_{\vk_1=-\vk^\prime-\vpi-\vq,mn}X^{ij|nm} \mathrm{e}^{-\i(\vk_1+\vq)\vR_n} \mathrm{e}^{\i\vk_1\vR_m} f_r(-\vk_1-\vpi-\vq),
\end{split}
\end{align}
where we omitted both, the frequency arguments, and $\tilde\vq$, and where $X$ is a place holder for some term in the DBSE of the $S_{P,z}$-channel. Evidently, if  $f_r(-\vk-\vpi-\vq)=\pm f_r(\vk)$, then the spin or charge part of the $S_{P,z}$-channel cancel. 
Using \cref{Eq:J_cdmft} and the fact that $[J,\chi_{0,\ph}^{\tilde\vq}]=[J,\chi_{0,\ph}]=[J,F_\text{ch/sp}]=0$ we can prove
\begin{align}
\label{Eq:induction_hypothesis_CDMFT}
\begin{split}
    \sum_{\vk^\prime,hl}&\left[(F_{S_{P,z}}\tilde\chi_{0,\ph}^{\tilde\vq})^n L_{S_{P,z}}\right]^{ij|hl}\mathrm{e}^{-\i(\vk^\prime+\vq)\vR_h}\mathrm{e}^{\i\vk^\prime\vR_l} f_r(\vk^\prime) \\
    &=\begin{cases}
         \sum_{\vk^\prime,hl}\left[(F_\text{ch}\tilde\chi_{0,\ph}^{\tilde\vq})^n L_\text{ch}\right]^{ij|hl}\mathrm{e}^{-\i(\vk^\prime+\vq)\vR_h}\mathrm{e}^{\i\vk^\prime\vR_l} f_r(\vk^\prime) & \text{if}\ f_r(\vk)=f_r(-\vk-\vpi-\vq)\\
         \sum_{\vk^\prime,hl}\left[(F_\text{sp}\tilde\chi_{0,\ph}^{\tilde\vq})^n L_\text{sp}\right]^{ij|hl}\mathrm{e}^{-\i(\vk^\prime+\vq)\vR_h}\mathrm{e}^{\i\vk^\prime\vR_l} f_r(\vk^\prime) & \text{if}\ f_r(\vk)=-f_r(-\vk-\vpi-\vq)
    \end{cases}\quad n\in\mathbb{N}
\end{split}
\end{align}
by induction. Finally, by ``Fourier transforming'' \cref{Eq:DBSE_CDMFT} with \cref{Eq:CDMFT_FT_ph} we show that
\begin{align}
\label{Eq:phys_pp_ch_sp_CDMFT}
    f_r\chi_\pp^{-\vq-\vpi;-\tilde\vq}(-\omega)f_r &=
    \begin{cases}
        f_r\chi_\text{ch}^{\vq;\tilde\vq}(\omega)f_r & \text{if}\ f_r(\vk)=f_r(-\vk-\vpi-\vq)\\
        f_r\chi_\text{sp}^{\vq;\tilde\vq}(\omega)f_r & \text{if}\ f_r(\vk)=-f_r(-\vk-\vpi-\vq)
    \end{cases}
\end{align}
by using \cref{Eq:induction_hypothesis_CDMFT}.
The only difference to the relation for the exact lattice solution \cref{Eq:phys_pp_ch_sp} is that the susceptibility depends on the cluster and the super-lattice momenta. 
The more complicated equations for stripe order \cref{Eq:phys_pp_q0pi} can again be proven analogous.

With this we accomplished our goal to show that the degeneracy found for the exact lattice solution [\cref{sec:phys_respose_degeneracancy}] essentially also hold for DMFT and its cluster extensions. With the only difference that the transfer momentum $\vq$ has to be split into cluster momentum $\vq$ and super-lattice momentum $\tilde\vq$, which can be summed in DCA but remain two distinct momenta in CDMFT.

\subsection{Numerical fulfillment for CDMFT calculations} 
\begin{figure}
    \centering
    \includegraphics[width=0.8\linewidth]{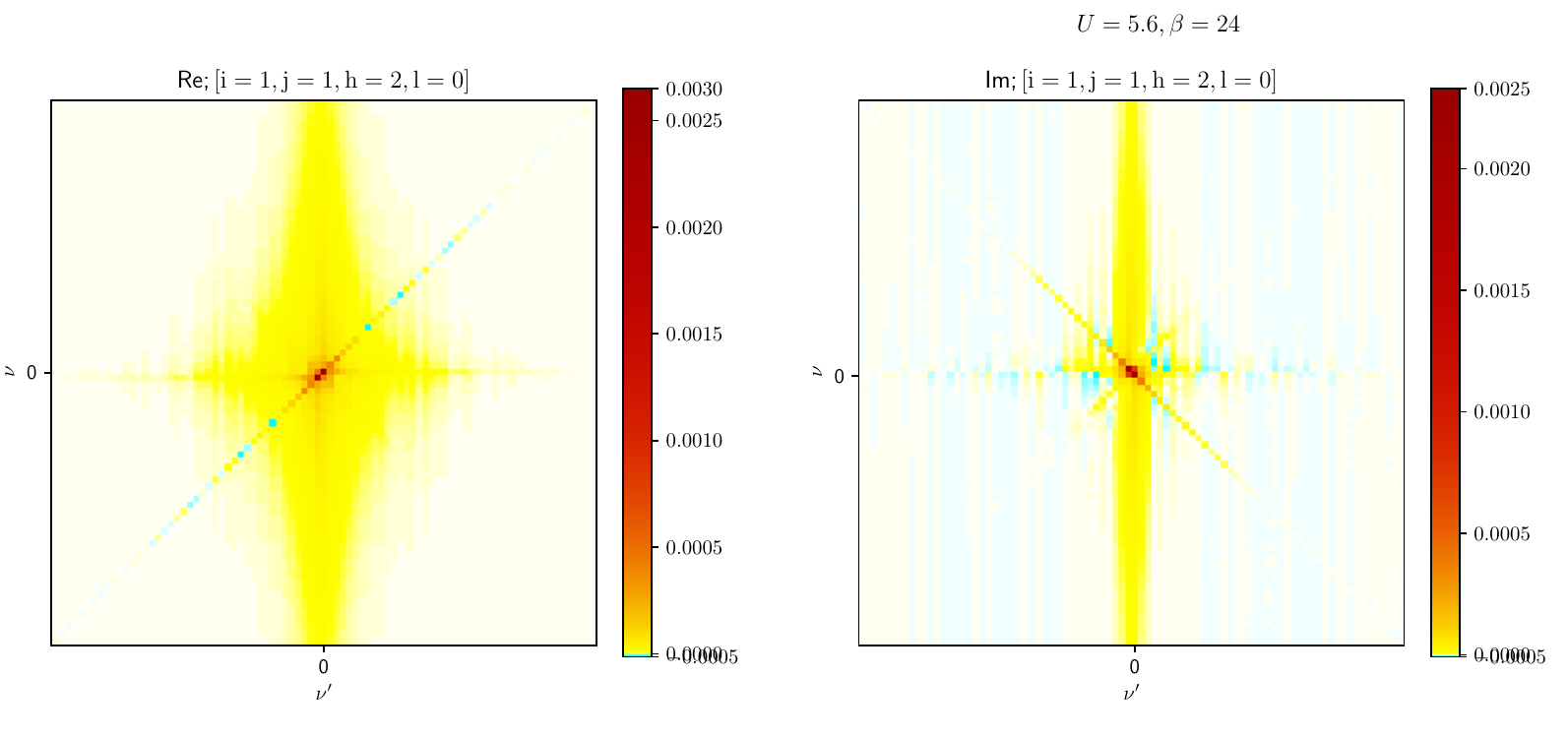}
    \includegraphics[width=0.8\linewidth]{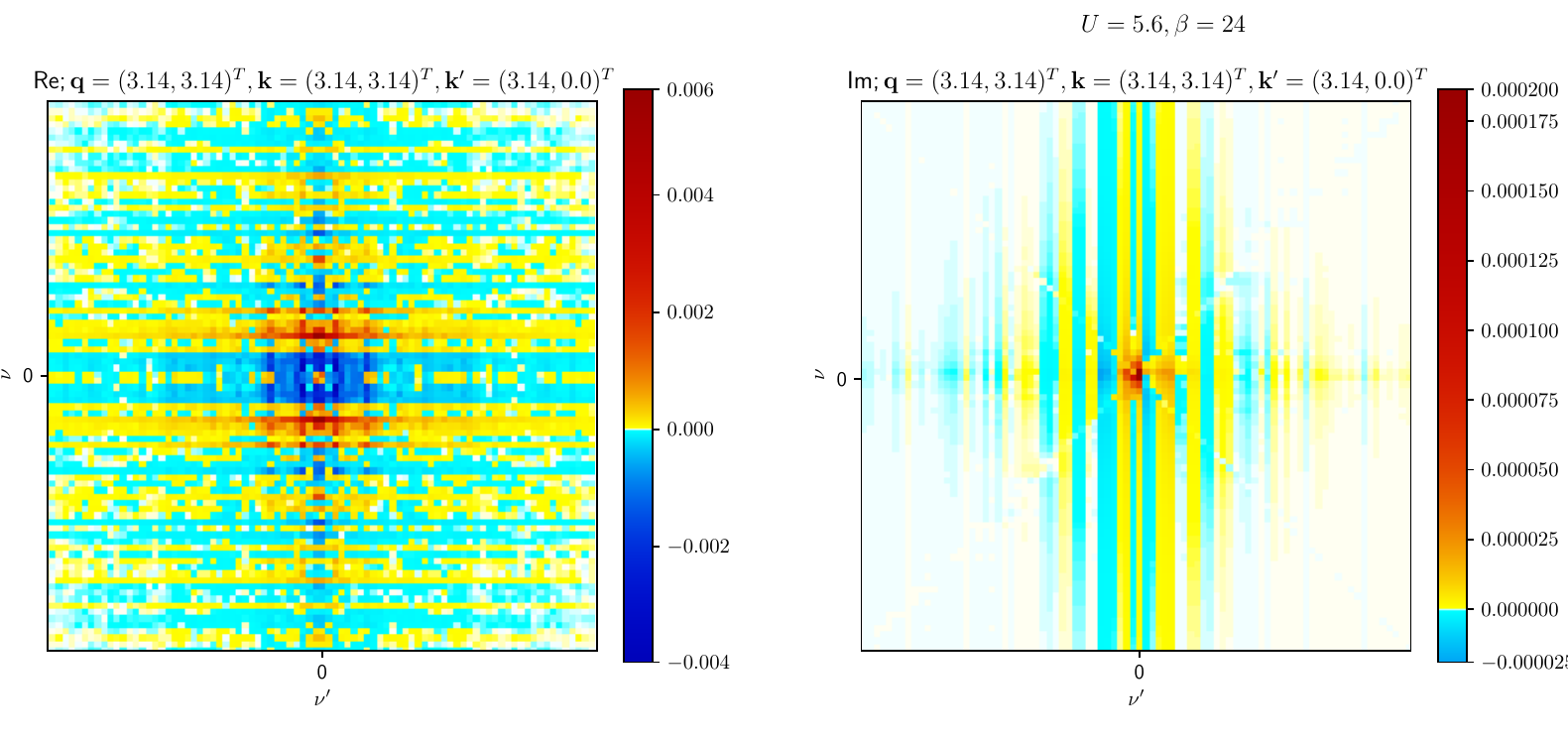}
    \caption{Benchmark for the identity for CDMFT. Absolute difference between the left hand side and right hand side real and imaginary parts of real-space and momentum-space identities for the generalized susceptibility Eq.~(\ref{Eq:DBSE_CDMFT}) (upper row) and Eq.~(\ref{Eq:DBSE_CDMFT_MOMENTUM}) (lower row) at $U=5.6t$, $T=t/24$ and the super lattice vector $\vtq=(0.3,0.3)^T$. The data is selected from the component which has the largest relative Error.}
    \label{fig:Appendix_Fulfillment_CDMFT-lattice}
\end{figure}

\begin{figure}
    \centering
    \includegraphics[width=0.5\linewidth]{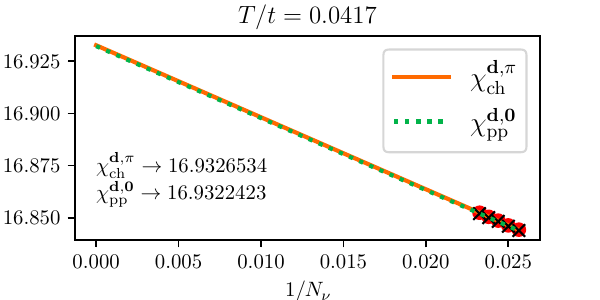}
    \caption{Extrapolation in the number of fermionic frequencies $N_\nu$ of the BSE Eq.~(A1) (dashed and dotted lines) from finite grid data points of the physical susceptibilities for $d$-DW $\chi_\ch^{d,\vq=\vpi}$ ($\circ$-marker) and $\pp$ d-wave $\chi_\pp^{d,\vq=\vnull}$ ($\times$-marker) for $U=5.6t,\,\vtq=\vnull$ at $T/t=1/24$, short before the instability. The linear fit considers the last three available frequency points.}
    \label{fig:Extrapolate_BSE}
\end{figure}

For a nontrivial superlattice vector of $\vtq=(0.3,0.3)^T$ and $U=5.6t,\,T=t/24$, short before the critical temperature, we analyze the fulfillment of Eq.~(6) and Eq.~(7) in the main text, or to be more precise, the respective CDMFT version Eq.~(\ref{Eq:DBSE_CDMFT},\ref{Eq:DBSE_CDMFT_MOMENTUM}). The largest error in the real space Eq.~(6), given in Fig.~\ref{fig:Appendix_Fulfillment_CDMFT-lattice} in the upper row, corresponds to a relative error of $0.001$. The largest component after Fourier transforming, resulting in Eq.~(\ref{Eq:DBSE_CDMFT_MOMENTUM}) has a similar relative error, given in the lower row of Fig.~\ref{fig:Appendix_Fulfillment_CDMFT-lattice}. We regard this error as very low, considering that the computation relies on Monte Carlo sampling. Further, we stress that the BSE computation is in a linear regime in the number of fermionic frequencies $N_\nu$, which allows an extrapolation to $N_\nu\rightarrow \infty$ for the physical response functions, see Fig.~\ref{fig:Extrapolate_BSE}. For these, a relative error of $0.000023$ is achieved between $\chi_\pp^{d,\vnull}$ and $\chi_\ch^{d,\vpi}$ for $\vtq=\vnull$.

\section{Benchmark} The critical temperatures in the pairing channel of our numeric implementation agreeare within errorbars with the published data in \cite{Walsh2023} at $U=6.2t$ and $n=0.98$ filling, and \cite{Harland2019} at $U=8t$, $t'=-0.3t$ and $n=0.95$. Further, for the half-filled case, data is publicly available \cite{Fratino2016} for $U=5.5t,\, T_c\approx0.038t$ and $U=5.7t,\, T_c\approx 0.0444t$. This allows to interpolate $U=5.6t,\,T_c\approx0.0414t$ and hence $\beta_c\approx24.14/t$. We find a critical temperature of $\beta_c t\in \left[ 24.1, 24.25\right]$.
\section{SBE analysis for spin, charge and singlet exchange bosons}
\begin{figure}
    \centering
    \includegraphics[width=0.46\linewidth]{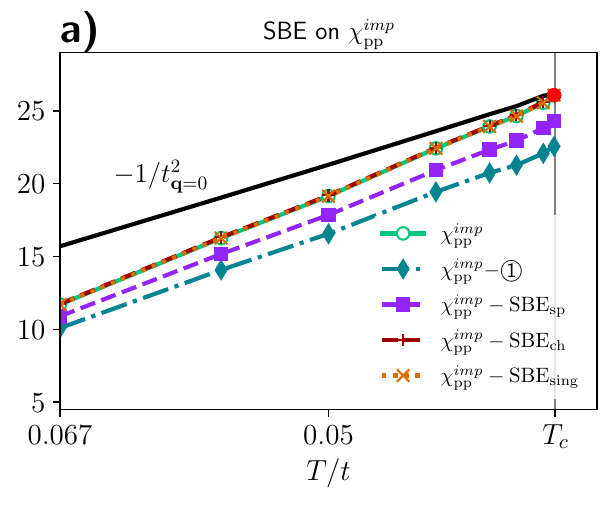}
    \includegraphics[width=0.46\linewidth]{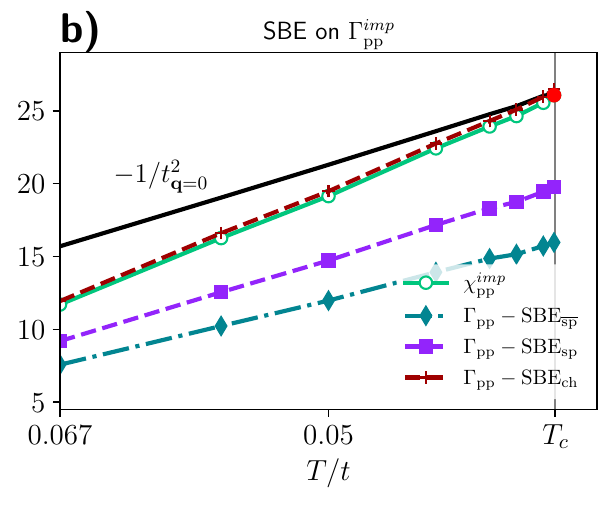}
    \caption{a) SBE analysis as in the main text [Fig.~2b)], on the level of the impurity susceptibility $\chi_\pp^{imp}$, including the horizontal diagrams and b) SBE analysis as in the main text [Fig.~2b)], solely on the level of $\Gamma_\pp^{imp}$, including the horizontal diagrams.}
    \label{fig:Appendix_SBE_Analysis}
\end{figure}

Similar to Fig.~2b) of the main text, we present the impact of the horizontal SBE charge (ch), singlet (pp) and spin (sp) diagrams of $\chi_\pp^{imp}$ [first summand in Fig.~2c)]. We here contrast it to the subtraction of the $\overline{\sp}$ SBE diagram on the level of the generalized impurity susceptibility \cite{Krien2019c} Fig~\ref{fig:Appendix_SBE_Analysis}a). We note that a subtraction of the horizontal singlet and charge SBE diagram do hardly change the contribution of $\chi^{imp}_\pp$ to the instability. The horizontal spin diagram (square markers) is slightly more important, while the transversal spin diagram $\overline{\sp}$ is the most important. Similarly, a subtraction of the SBE diagrams on the level of $\Gamma_\pp$ is undertaken \cite{Krien2020b,Meixner2026a} in Fig~\ref{fig:Appendix_SBE_Analysis}b). Here, the picture remains the same. Note, that the subtraction of the spin transverse diagrams $\Gamma_\pp - \mathrm{SBE}_{\overline{\sp}}$. equaling $\chi_\pp-$\numcircled{2} of Fig.~2b) of the main text, which results in a change of slope, thereby completely omitting a transition for low temperatures.  Note, that the subtraction of the horizontal spin diagram has the second largest impact, yielding an impurity contribution which is as good as parallel to the impurity threshold in this temperature regime.

\section{Eigenvalue analysis of the BSE}
The Bethe-Salpeter equation of the real-space cell for $\omega=0$ reads \cite{Meixner2026a}
\begin{equation}
    \label{Eq:eigenvector_pp_diverence}
\left(\left[\chi^{imp}_\pp\right]^{-1}+t^2_{\vtq=0}\right)^{-1}=\chi_\pp^{\vtq=0}\approx V^{\text{crit}} \lambda_{\text{crit}} V^{\text{crit},-1}.
\end{equation} For an orthonormal basis we study the effect of the impurity quantities in the direction of the leading superlattice eigenvector. Specifically, for a diverging $ \lambda_{\infty}\coloneq \lambda_{\text{crit}}(T_c)$ and using the notation of the scalar product $\langle.,.\rangle$ \footnote{We use the right- and left-hand side eigenvectors for the respective sides of the scalar product.}, we can write Eq.~(\ref{Eq:eigenvector_pp_diverence}):
\begin{equation}
\label{Eq:leading_Eigenvector_pp}
    0=\frac{1}{\lambda_\infty}=\langle V^{\infty},\left(\left[\chi^{imp}_\pp\right]^{-1}+t^2_{\vtq=0}\right)V^{\infty}\rangle,
\end{equation}
which gives a necessary condition for the divergence of the lattice charge response.
\begin{figure}
    \centering
    \includegraphics[width=0.9\linewidth]{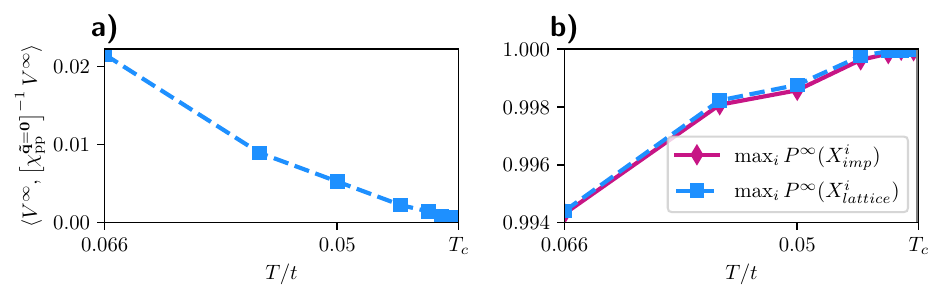}
    \caption{Projection of a) the full lattice susceptibility $\langle V^{\infty},\left[\chi^{\vtq=\vnull}_{\mathrm{pp}}\right]^{-1}V^{\infty}\rangle$ and b) the temperature dependent eigenvectors of $\chi_\pp^{imp}$ and $\chi_\pp^{\vq}$ to the lattice eigenvector $V^\infty$ taken from the instability.}
    \label{fig:Projection_Lattice_impurity}
\end{figure}
When changing the temperature away from the critical point, the left hand side remains of very small value $\langle V^{\infty},\left[\chi^{\vtq=\vnull}_{\mathrm{pp}}\right]^{-1}V^{\infty}\rangle\approx 0$, see Fig.~\ref{fig:Projection_Lattice_impurity}a), and we remark, that $V^{\infty}$ remains almost parallel to an eigenvector of $\chi^{\vtq=\vnull}_{\mathrm{pp}}$ when the temperature changes. Together with the spectral decomposition of the impurity susceptibility
\begin{equation}
\chi_{\mathrm{ch}}=\sum_i X^{i}E_i X^{i,-1}
\end{equation} we get
\begin{equation}
\begin{split}
\label{Eq:overlapp_impurity_pp_respone}
    \langle V^{\infty}, -t^2_{\vq=0} V^{\infty}\rangle&\approx\langle V^{\infty},\left[\chi^{imp}_{\mathrm{pp}}\right]^{-1}V^{\infty}\rangle\\
    &\approx\sum_i\langle V^{\infty},X^{i}\rangle\frac{1}{E_i}\langle X^{i},V^{\infty}\rangle\\
    &\approx\sum_iP^\infty(X^i)\frac{1}{E_i},
\end{split}
\end{equation} where we introduced the projection $P^\infty(X^i)$ of the eigenvector $X^i$ of the impurity pp susceptibility onto the leading eigenvector of the lattice pp response $V^\infty$. In the analyzed temperature range $T\in\left[0.0666,T_c\right]$ we have $\underset{i}{\text{max }} P^\infty(X^i)^2>0.994$ and increases towards $1$ when approaching $T_c$, see Fig.~\ref{fig:Projection_Lattice_impurity}b).

 Hence $V^\infty$ is as good as an eigenvector of $\chi_\pp^{imp}$. This results in a diagonal block structure of the BSE where the sum over the impurity eigenvectors can be omitted and we obtain
\begin{equation}
\begin{split}
\label{Eq:overlapp_impurity_pp_respone_final}
    1/\langle V^{\infty}, -t^2_{\vq=0} V^{\infty}\rangle\approx E_{max} \qquad \text{for}\qquad P^\infty(X^{max})=1,
\end{split}
\end{equation} which justifies the analysis in the main text Fig.~2b).
\bibliography{supplemental}